\documentclass[twocolumn]{aastex631}

\let\textbf\relax
\newcounter{chem}
\newcounter{temp}
\newenvironment{chequation}{%
  \setcounter{temp}{\value{equation}}%
  \setcounter{equation}{\value{chem}}%
  \renewcommand{\theequation}{R\arabic{equation}}%
}{%
  \setcounter{chem}{\value{equation}}%
  \setcounter{equation}{\value{temp}}%
}

\usepackage[version=4]{mhchem}
\usepackage{multirow}
\usepackage{xcolor}
\usepackage{newtxtext}
\usepackage[most]{tcolorbox}
\usetikzlibrary{patterns}
\pgfdeclarepatternformonly{mystrikeout}{\pgfqpoint{-1pt}{-1pt}}{\pgfqpoint{11pt}{11pt}}{\pgfqpoint{10pt}{10pt}}%
{
  \pgfsetlinewidth{0.4pt}
  \pgfpathmoveto{\pgfqpoint{0pt}{0pt}}
  \pgfpathlineto{\pgfqpoint{10.1pt}{10.1pt}}
  \pgfusepath{stroke}
}
\newtcolorbox{tcbstrikeout}{breakable,
 enhanced jigsaw,
 opacityback=0,
 parbox=false,
 boxrule=0mm,
 top=0mm,bottom=0pt,left=0pt,right=0pt,
 boxsep=0pt,
 frame hidden,
 finish={\fill[pattern=mystrikeout] (frame.north west) rectangle (frame.south east);}
}

\begin{document}

\title{Formation of the interstellar sugar precursor, (Z)-1,2-ethenediol, through radical reactions on dust grains}

\author[0000-0002-3251-3594]{Juan Carlos del Valle}
\affiliation{Institute for Theoretical Chemistry,
University of Stuttgart,
Pfaffenwaldring 55, D-70569 
Stuttgart, Germany}

\author[0000-0001-7876-4818]{Pilar Redondo}
\affiliation{Computational Chemistry Group, Departamento de Química Física y Química Inorg\'anica, Facultad de Ciencias, Universidad de Valladolid, E-47011 Valladolid,
Spain}

\author[0000-0001-6178-7669]{Johannes K\"astner}
\affiliation{Institute for Theoretical Chemistry,
University of Stuttgart,
Pfaffenwaldring 55, D-70569 
Stuttgart, Germany}

\author[0000-0001-8803-8684]{Germ\'an Molpeceres}
\affiliation{Departamento de Astrofísica Molecular, Instituto de Física Fundamental (IFF-CSIC), C/ Serrano 121, E-28006 Madrid, Spain}

\correspondingauthor{Germ\'an Molpeceres}
\email{german.molpeceres@iff.csic.es}

\correspondingauthor{Johannes K\"astner}
\email{kaestner@theochem.uni-stuttgart.de}



\begin{abstract}

In recent years, the continued detection of complex organic molecules of prebiotic interest has refueled the interest on a panspermic origin of life. \textbf{The prebiotic molecule} glyceraldehyde is proposed to be formed from (Z)-1,2-ethenediol, a molecule recently detected towards the G+0.693-0.027 molecular cloud at the galactic center. In this work, we computationally simulate the formation of (Z)-1,2-ethenediol from vinyl alcohol on the surface of amorphous solid water in a two-step synthesis involving a OH addition and a H abstraction reaction. In total, we considered all reaction possibilities of the 1,1 and 1,2-OH addition to vinyl alcohol followed by H-abstraction or H-addition reactions \textbf{on the resulting radicals.} The combination of these reactions is capable of explaining the formation of (Z)-1,2-ethenediol provided a suprathermal diffusion of OH. \textbf{We also conclude that our proposed formation pathway is not selective and also yields other abstraction and addition products.} Key in our findings is the connection between the adsorption modes of the reactants and intermediates and the stereoselectivity of the reactions. 

\end{abstract}



\section{Introduction} \label{sec:intro}

The formation and detection of complex organic molecules (COMs), molecules with 6 or more atoms of which at least one is carbon \citep{Herbst2009}, is one of the most active areas of research in modern astrochemistry. For example, one of the most striking facts in the update from the 2018 census on interstellar molecules \citep{mcguire_2018_2018} to the 2021 one \citep{McGuire2022} is the number of complex molecules involved. From all these molecules, two families are especially relevant. First, aromatic molecules, \citep[][to provide just some examples]{c6h5cn, pah,Cernicharo2021,fulvenallene} and second, prebiotic molecules \citep{belloche_re-exploring_2019, Rodriguez-Almeida2021, Rivilla2020, Rivilla2021, rivilla_precursors_2022, andres_first_2024}. Prebiotic COMs garner the attention of the astrochemical community due to the possibility of establishing a connection between the abiotic chemical space and the biotic one. In this regard, the astronomical detection of biomolecular intermediates and the elucidation of chemical pathways leading to them is especially important. So far, intermediates in the synthesis of aminoacids and peptides \citep{Rivilla2020, Rodriguez-Almeida2021,belloche08,Zeng_2021}, lipids \citep{Rivilla2021,npropanol}, and finally sugars \citep{rivilla_precursors_2022,zeng19} have been detected.

\begin{figure}[hbt] 
    \centering
    \includegraphics[width=0.20\linewidth]{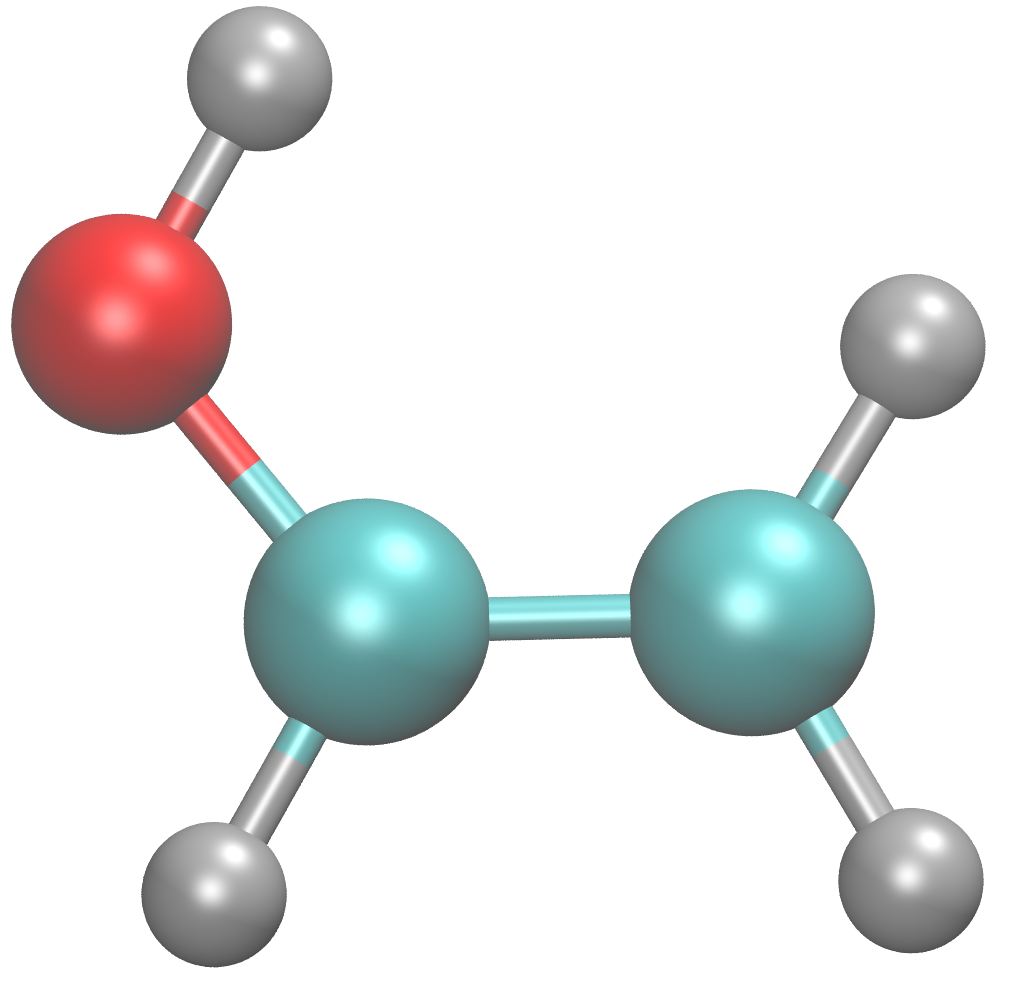} 
    \includegraphics[width=0.22\linewidth]{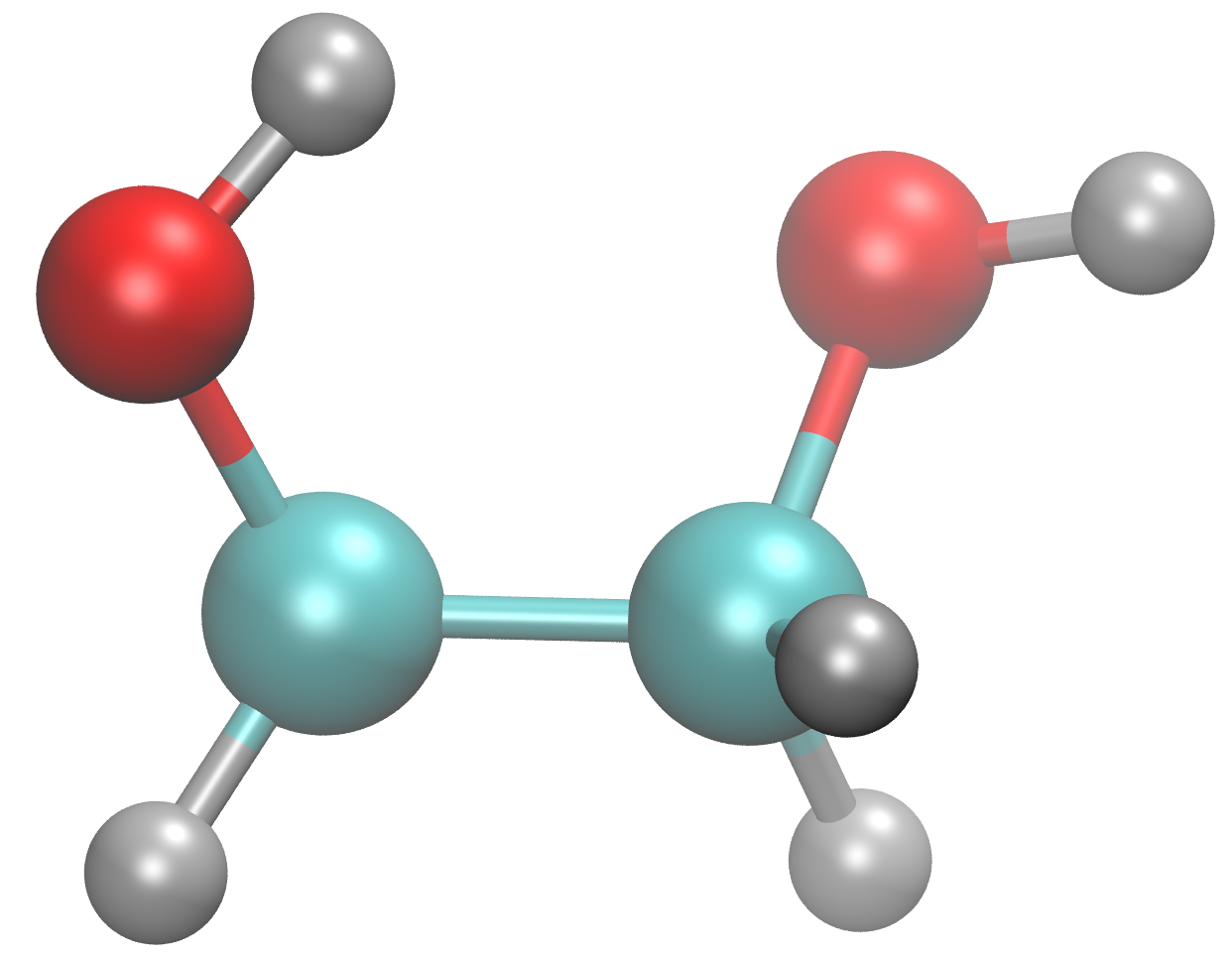} 
    \includegraphics[width=0.20\linewidth]{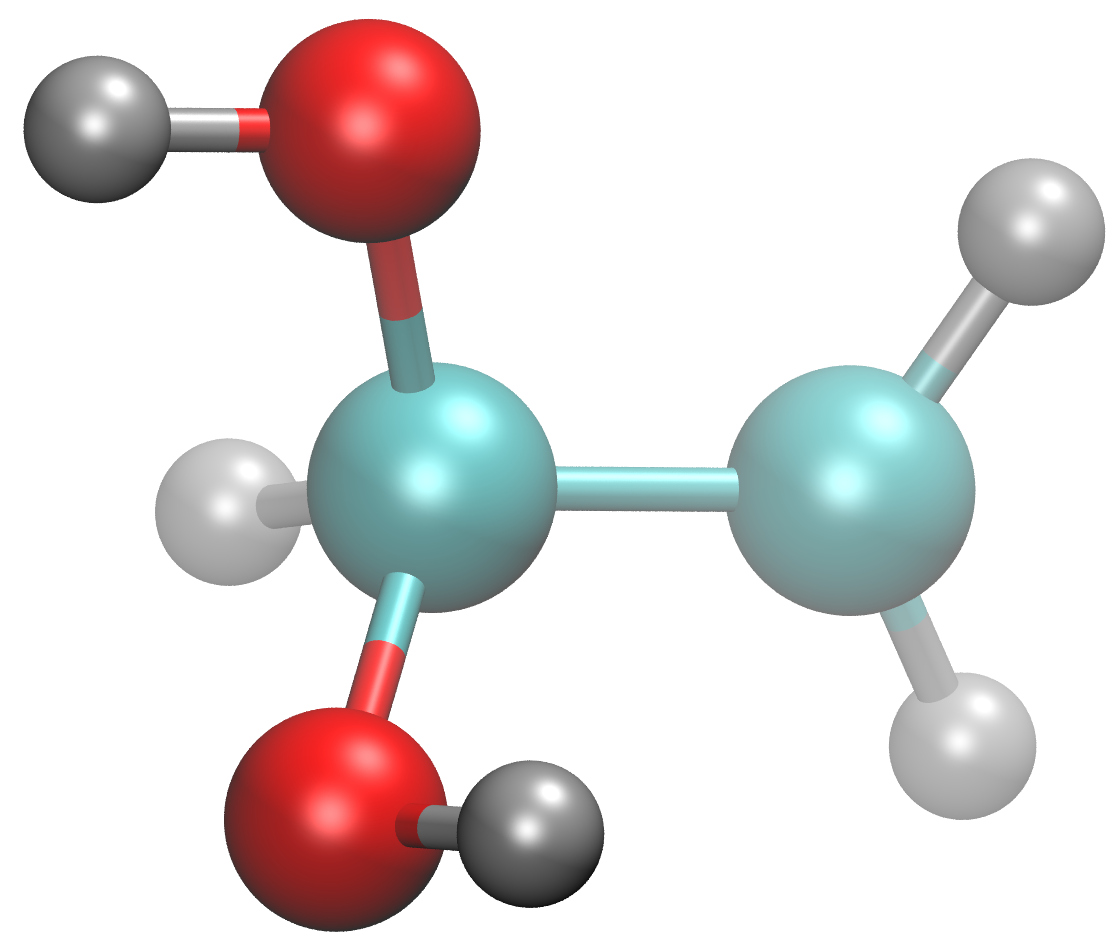}
    \includegraphics[width=0.19\linewidth]{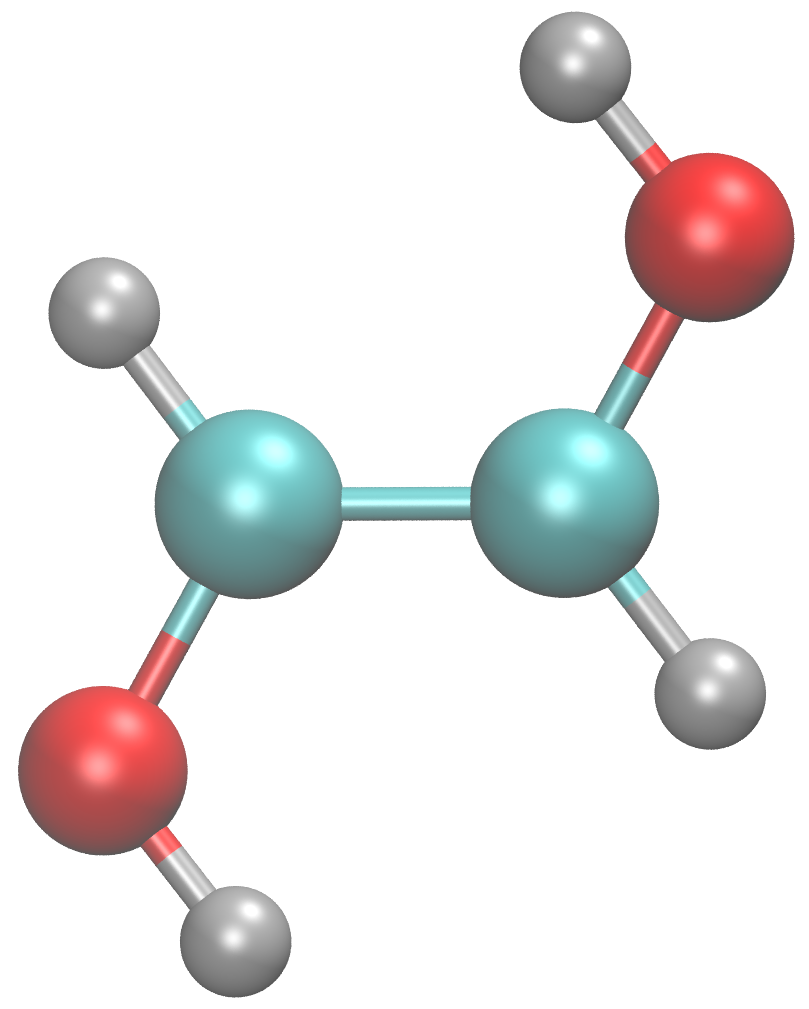}
    \vspace{0.10cm}
    \begin{tabular}{@{}c}
    \quad \quad \ce{C2H3OH}  \quad   \ce{HOC2H3OH} \quad \ce{C2H3(OH)2} \ce{(E)-HOC2H2OH} \\
    \end{tabular} \\ 
    \includegraphics[width=0.21\linewidth]{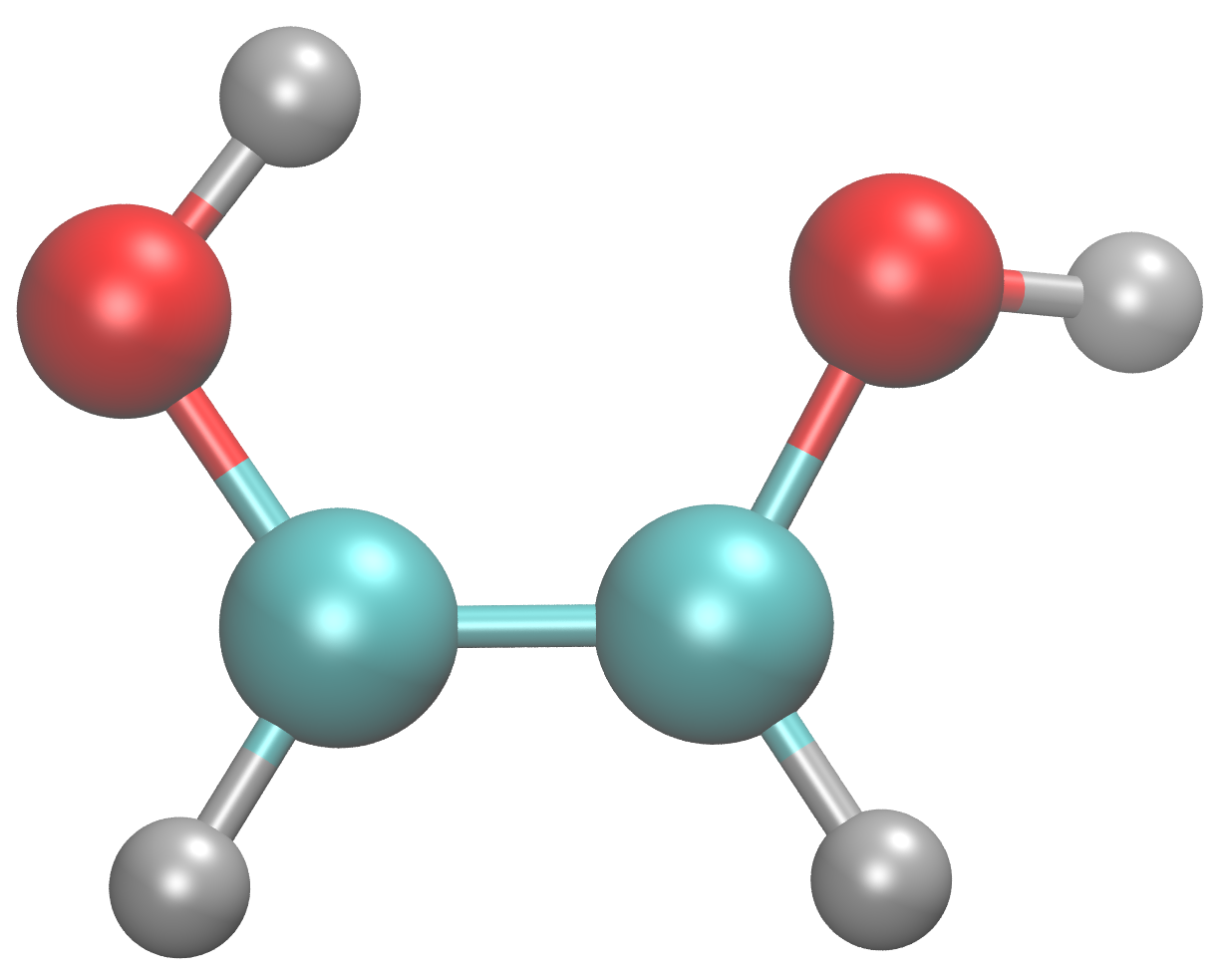} 
    \includegraphics[width=0.21\linewidth]{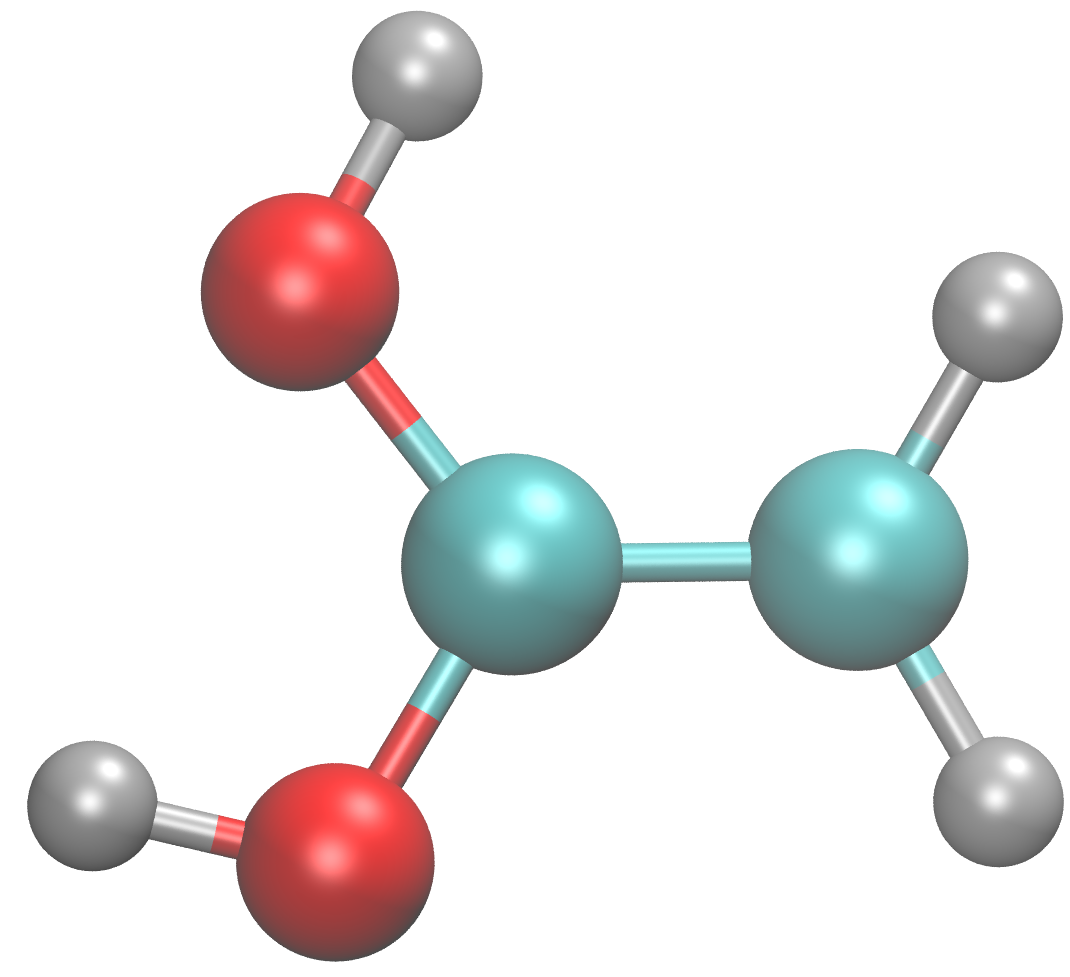} 
  \includegraphics[width=0.19\linewidth]{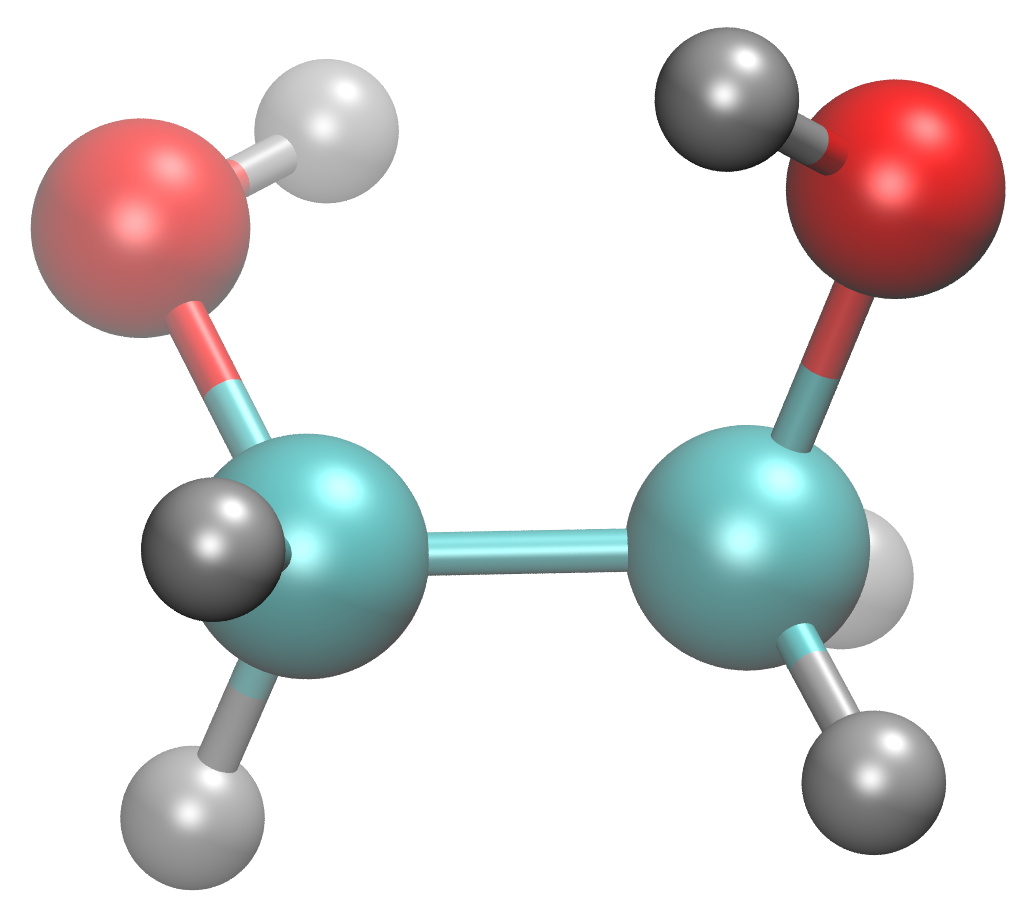}
    \includegraphics[width=0.20\linewidth]{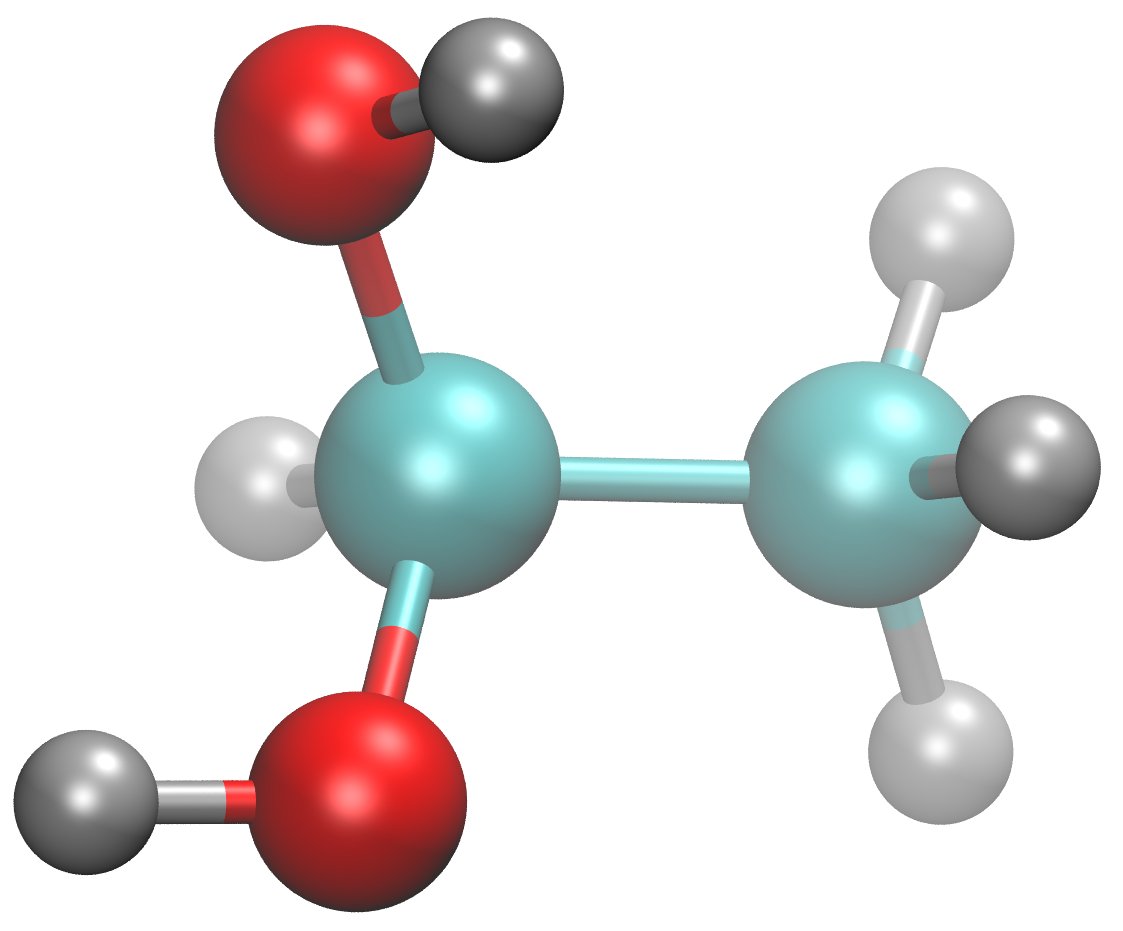}
    \vspace{0.10cm}
    \hskip-0.75cm \begin{tabular}{@{}c}
    \ce{(Z)-HOC2H2OH}  \quad \ce{C2H2(OH)2}  \quad \ce{HOC2H3OH}  \quad \ce{C2H2(OH)2} \\
    \end{tabular} \\
    \caption{Gas phase structures of the molecular species involved in the reaction schemes.}
    \label{fig:gasphase}
\end{figure}

Focusing on the detection of the sugar precursor, our molecule of interest in this work is (Z)-ethenediol, \ce{(Z)-HOC2H2OH}, (bottom leftmost structure in Figure \ref{fig:gasphase}), derived from vinyl alcohol, \ce{C2H3OH}, (top leftmost structure in Figure \ref{fig:gasphase}). This molecule is the so-called enol form of a more stable isomer, which under Earth conditions and in equilibrium will dominate, glycolaldehyde \citep{Etim_Gorai_Das_Arunan_2017}. However, under ISM conditions the thermal energy is not enough to reach thermodynamic equilibrium, \textbf{stimulating the spectroscopic characterization of (Z)-\ce{HOC2H2OH} in the laboratory \citep{melosso},} ultimately leading to its detection in the interstellar medium (ISM) \citep{rivilla_precursors_2022}. \textbf{Recently, the authors of the spectroscopic characterization of (Z)-\ce{HOC2H2OH} have completed the equivalent study on the molecule's isotopologues \citep{nonne_tracing_2024}} The importance of intermediates like \ce{(Z)-HOC2H2OH} in the interstellar sugar synthesis picture lies in its importance in the formose reaction, one of the oldest and most well-established chemical routes for the formation of sugars, i.e. ribose \citep{butlerow_bildung_1861, shapiro_prebiotic_1988}. It has been proposed that \ce{(Z)-HOC2H2OH} is an ideal candidate for an equivalent of such reaction under ISM conditions, through a reaction with formaldehyde (\ce{H2CO}) \textbf{\citep{eckhardt_gas-phase_2018,melosso}}. The formation of \ce{(Z)-HOC2H2OH} on interstellar ices was recently proposed through electron irradiation of \ce{CH3OH}/CO ices \citep{kleimeier_identification_2021}. For such a route to be operative, the ice abundance of the precursors needs to be substantial, which is an easily satisfied condition in prestellar cores, where CO is completely depleted onto thick ices \citep{Caselli1999}. However, information on additional formation routes of sugar precursors in less-evolved molecular clouds, e.g. on purely polar amorphous solid water (ASW) ices is required to feed astrochemical models and promote further detections.

In this work, we computationally simulate the formation on interstellar dust grain analogs of (Z)-1,2-ethenediol, (\ce{(Z)-HOC2H2OH}), from vinyl alcohol, (\ce{C2H3OH}), a molecule detected in young interstellar objects, e.g. TMC-1 \citep{agundez_o-bearing_2021}, in a two-step synthesis involving a first OH-addition followed by a H-abstraction reaction. For the first reaction, we expect the 1,2 and 1,1 addition products:

\begin{chequation}
\begin{alignat}{2}
&\ce{C2H3OH + OH &->& HOC2H3OH} \label{eq:OH1}\\
&\ce{C2H3OH + OH &->& C2H3(OH)2} \label{eq:OH2}
\end{alignat}
\end{chequation}

These reactions have been studied in the gas phase \citep{Ballotta23}. In their study, the authors find the adducts \ce{HOC2H3OH} and \ce{C2H3(OH)2} submerged with respect to reactants and forming without barrier. However, subsequent evolution is either endoergic (formation of (Z)-1,2-ethenediol) or have significantly emerged barriers (formation of glycolaldehyde). Additionally, the authors study H-abstraction reactions (\ce{C2H3OH + OH -> C2H2OH + H2O}) finding barriers higher than in the case of the OH addition, and reducing their importance for surface chemistry. Therefore, accounting for the differences in gas phase and surface chemistry on ices, we consider that formation on the grain should be feasible through hydrogen abstraction after a OH addition. Thanks to the third body effect promoted by the interstellar ice, on the surface the addition product can stabilize and further react. We consider reactions with hydrogen for this subsequent step. Hydrogen abstraction reactions are performed on the C-H moiety, yielding (E)-1,2-ethenediol, (Z)-1,2-ethenediol and 1,1-ethenediol:

\begin{chequation}
\begin{alignat}{2}
&\ce{HOC2H3OH + H &->& (E)-HOC2H2OH + H2} \label{eq:H1} \\
&\ce{HOC2H3OH + H &->& (Z)-HOC2H2OH + H2} \label{eq:H2} \\
&\ce{C2H3(OH)2 + H &->& C2H2(OH)2 + H2} \label{eq:H3}
\end{alignat}
\end{chequation}

\textcolor{black}{Hydrogen addition reactions have been also considered, leading to the formation of 1,2-ethanediol \textbf{(ethylene glycol, a molecule detected in the ISM \citep{hollis_interstellar_2002}}), and 1,1-ethanediol:}

\begin{chequation}
\begin{alignat}{2}
&\ce{HOC2H3OH + H &->&  HOC2H4OH}\label{eq:H4} \\
&\ce{C2H3(OH)2 + H &->& C2H4(OH)2} \label{eq:H5}
\end{alignat}
\end{chequation}

The previous species can be found in Figure \ref{fig:gasphase}. In our chemical route we propose that OH and H are the mobile reagents on the surface, meeting the other reaction partners. 

In a Langmuir--Hinshelwood mechanism on interstellar dust grains, the reaction rate constant $k_{\textrm{t}}$, where t stands for thermal, for such a process reads \citep{hasegawa_three-phase_1993, ruaud_gas_2016}:

\begin{equation} \label{eq:thermal}
    k_{\textrm{t}} = \kappa_{ij} \left( \frac{1}{\tau\left(i\right)} \text{+} \frac{1}{\tau\left(j\right)} \right) \dfrac{1}{N_{\text{Site}}n_{\text{Dust}}},
\end{equation}

\noindent with 

\begin{equation} \label{eq:kappa}
    \kappa_{ij} = \dfrac{k_{r}}{k_{r} + k_\text{diff}(i) + k_\text{diff}(j) + k_\text{des}(i) + k_\text{des}(j) }.
\end{equation}

In the above equations, $\tau(i,j)$ are the hopping (single diffusion step) times for reactants $i$ and $j$, $N_{\text{site}}$ is the number of adsorption sites on a dust grain, $n_{\text{Dust}}$ is the number density of dust grains and $k_{r}$, $k_{\text{diff}}$ and $k_{\text{des}}$ are the rate constants for reaction, diffusion and desorption of the species involved. From a quick inspection of equation \ref{eq:thermal}, it is clear that at least one of the reactants needs to diffuse to have a large enough rate constant for the process. This condition is satisfied for H atoms. For OH radicals, diffusion starts to be noticeable above 36 K \citep{miyazaki_direct_2022}, so reactions \ref{eq:OH1} and \ref{eq:OH2} must be too slow at 10 K, as a characteristic temperature of a molecular cloud. However, in recent years experiments have shown that, in addition to thermal diffusion of OH, non-thermal diffusion and reaction is an important promoter of interstellar chemistry \citep{Ishibashi2021}. Non-thermal diffusion and reaction can be triggered by a variety of actors, like photodissociation \citep{Ishibashi2021} or oxygen atom hydrogenation \citep{Garrod2011, Ioppolo2020}. Both mechanisms generate a translationally excited OH radical (suprathermal) that can shortly overcome kinetic barriers for diffusion or reaction. The formalism for suprathermal chemistry has been developed for cosmic ray and photon initiated chemistry \citep{shingledecker_cosmic-ray-driven_2018, shingledecker_simulating_2019, mullikin_new_2021}, but it holds for any mechanism capable of inoculating sufficient kinetic energy to freely roam the surface for a short time, like chemical energy \citep{molpeceres_reaction_2023}. The rate constant $k_{\textrm{st}}$ (with st referring to suprathermal) for such a process is given by \textbf{\citep{shingledecker_cosmic-ray-driven_2018}}:

\begin{equation} \label{eq:suprathermal}
    k_{\textrm{st}} = f\left[ \dfrac{\nu^{i} + \nu^{j}}{N_{\text{Site}}n_{\text{Dust}}} \right]
\end{equation}

\noindent where $f$ is the branching ratio, $\nu^{i,j}$ is the characteristic frequency of diffusion of the partners involved. \textbf{The formation of a suprathermal radical that can quickly roam the surface and react for a short time is essentially equivalent to the non-diffusive mechanism postulated in \citet{Chang2018, Jin2020, Garrod2022}.} It is important to note that the time to freely roam and react is short, because the molecule relaxes (quenches) on the surface with a rate constant \textbf{equivalent to the characteristic vibrational frequency of the suprathermal species, \citep{Landau1976Mechanics,shingledecker_cosmic-ray-driven_2018}}:

\begin{equation} \label{eq:quench}
    k_{\textrm{q}} = \sqrt{\dfrac{2N_{\text{Site}}E_{d}}{\pi^{2} m}},
\end{equation}

\noindent Where the only non-defined quantities are $E_{d}$ and $m$ are the diffusion energy and mass of the diffusing partner, OH in this case. \textbf{We note that the characteristic frequencies are on the order of 1-3 ps for molecules the size of typical interstellar molecules; and such, the quenching times are coherent with dissipation times of excess energy reported in the literature \citep{Pantaleone_2020, Pantaleone2021, ferrero_where_2023,  molpeceres_reaction_2023}.} Although suprathermal OH can roam and react, a necessary condition for the application of equation \ref{eq:suprathermal} in astrochemical models, is that the inoculated energy is sufficiently large to consider the otherwise activated processes effectively barrierless. This is a trivial condition for cosmic ray induced chemistry, as the deposited energy is enormous. For photoprocesses and chemical energy, some limit cases may exist, and computational simulations such as the ones in this work are pivotal to identify them.

This paper is divided as follows. In Section \ref{sec:methods} we discuss the methods that we use to construct our theoretical ice models, and how we study the subsequent reactions. In Section \ref{sec:results} we present the the results of our work. In Section \ref{sec:discussion} we discuss our findings and contextualize the results, focusing on the isomerism of the reactions and the possible astrochemical avenues that our study opens. We conclude our work with a very brief summary of our findings.

\section{Methods} \label{sec:methods}

We simulate amorphous solid water (ASW) using two cluster models, one consisting of 18 \ce{H2O} molecules, and another one consisting on 33 \ce{H2O} molecules featuring a pocket site \citep{Rimola2018, Perrero2022}. Several binding sites are available in these models, for example, dangling oxygen (dO) sites, which are those where the oxygen atom from the water ice acts as hydrogen acceptor, and dangling hydrogen (dH) where water acts as the H-donor. Finally, the pocket sit refers to a shallow depression on the ice surface.

We characterized the stationary points in the potential energy surfaces by using density functional theory (DFT) calculations with ORCA software \citep{Neese2020}. First, we studied the OH radical addition step with the MPWB1K exchange and correlation functional \citep{Zhao2005}, and the triple-zeta basis-set def2-TZVP (MPWB1K/def2-TZVP) \citep{Weigend2005} including the D3BJ \citep{Grimme_Ehrlich_Goerigk_2011} dispersion correction. DFT calculations were sped up by introducing the Split-RI-J variant of the density fitting approximation for the calculation of the Coulomb part of the Fock matrix and the "chain-of-spheres" algorithm for the exchange part in the so called RIJCOSX procedure \citep{NEESE200998}. Artificial overbinding effects that may arise from the basis set superposition error (BSSE) are corrected with the geometrical Counterpoise Correction (gCP) \citep{Kruse_Grimme_2012}. Therefore the method used in this work can be abbreviated as MPWB1K-D3(BJ)-gCP/def2-TZVP. The choice of this functional is justified on the basis of a benchmark testing several exchange and correlation functionals compared against the high-level results in the gas phase of \citet{Ballotta23}. This benchmark can be found in Appendix \ref{sec:appendix_methods}.

%\begin{equation}
%E_{\text{total}} = E_{\text{DFT}} + E_{\text{gCP}}
%\end{equation}

Vinyl alcohol (\ce{C2H3OH}) has two possible rotamers, syn and trans vinyl alcohol. Early spectroscopic studies acknowledge \ce{syn-C2H3OH} as the most stable rotamer \citep{RODLER198523,SAITO1976399}. Therefore, we proceed with this conformation during our study and we will refer to it from now on as simply \ce{C2H3OH}. A first optimization step for both 18 \ce{H2O} and 33 \ce{H2O} clusters was carried out. Afterwards, we placed \ce{C2H3OH} on the surface and reoptimized this new structure: we assess the adsorption on both dH and dO sites on the 18 \ce{H2O} cluster, leaving the pocket sampling for the 33 \ce{H2O} cluster. The pocket is described as a confinement on the surface where both adsorbates can be inserted in, either forming dO or dH interactions with the surrounding water molecules. Finally, we add the OH radical onto the surface near the adsorbate and re-optimize the geometry of this guess at an internuclear distance of $\sim$3.5~\AA~in all cases. Using these three resulting conformations as input, we carried out relaxed geometry scans along the C-O distances to produce either $\ce{HOC2H3OH}$ or $\ce{C2H3(OH)2}$ radicals. \textbf{Because the initial guess for the reactants is placed at a relatively short distance, the barriers emerging in our scans are purely reaction barriers, and we avoid the sampling of diffusion barriers. Hence, the activation energies reported in this paper for reactions \ref{eq:OH1} and \ref{eq:OH2}, the ones reported in \citet{Ballotta23}, should be obtained taking as reference their structure Prc1, instead of the separated reactants, namely 1.0 kcal mol$^{-1}$ and 1.9 kcal mol$^{-1}$, for reactions \ref{eq:OH1} and \ref{eq:OH2}}. The stationary points along the scan trajectory are then further optimized: minima as either reactants or products and the maxima as transition states (TS). Vibrational frequencies were calculated for all optimized geometries to validate them as well as to provide the zero point vibrational energy (ZPVE).

Later, we test  H-addition and H-abstraction reactions on the products of the first OH addition. For these calculations, we adopted a broken symmetry formalism. \textbf{We note that the inherent (methodological) error may be larger in the case of these reactions than in the first OH addition, due to the multireference nature of the system.} Considering the amount of reactions, we opted for a double-zeta basis-set instead (def2-SVP) to lower the computational cost while achieving reasonable accuracy. The method for the second set of reactions is therefore MPWB1K-D3(BJ)-gCP/def2-SVP. We follow the same procedure from the first reaction step, though this time producing C-H distance relaxed geometry scans for the H-addition reactions and H-H scans for the H-abstraction reaction. We opted to fix the atom positions of the ice slab for the reactions on the pocket to ease geometry optimizations.

Finally, Binding Energies (BE) are defined as the attractive interactions between an adsorbate and a surface. In our case, the different chemical species involved in the reaction schemes and our ice slabs accounting for the respective binding sites. 
%- \Delta H^{\circ}_\text{bind} =
\begin{equation}
\label{eq:bindenergies}
\text{BE} =  H^{\circ}_\text{bind, Ice + Adsorbate} - \left(H^{\circ}_\text{bind, Ice} + H^{\circ}_\text{bind, Adsorbate}\right)    
\end{equation}
We provide individual binding energies for each molecule and binding site. These are calculated as the difference between the electronic energies including ZPVE of the cluster formed by the ice slab and the adsorbate in a given binding site, the ice slab, and the adsorbate in gas phase.

\section{Results} \label{sec:results}

\subsection{Hydroxyl addition to \ce{C2H3OH}}
\label{section:OHadd}

\begin{figure*}[hbt] 
    \centering
    \hrulefill \\[1ex]
    {\large \bfseries \ce{C2H3OH} (Vinyl alcohol)} \\
    \includegraphics[width=0.30\linewidth]{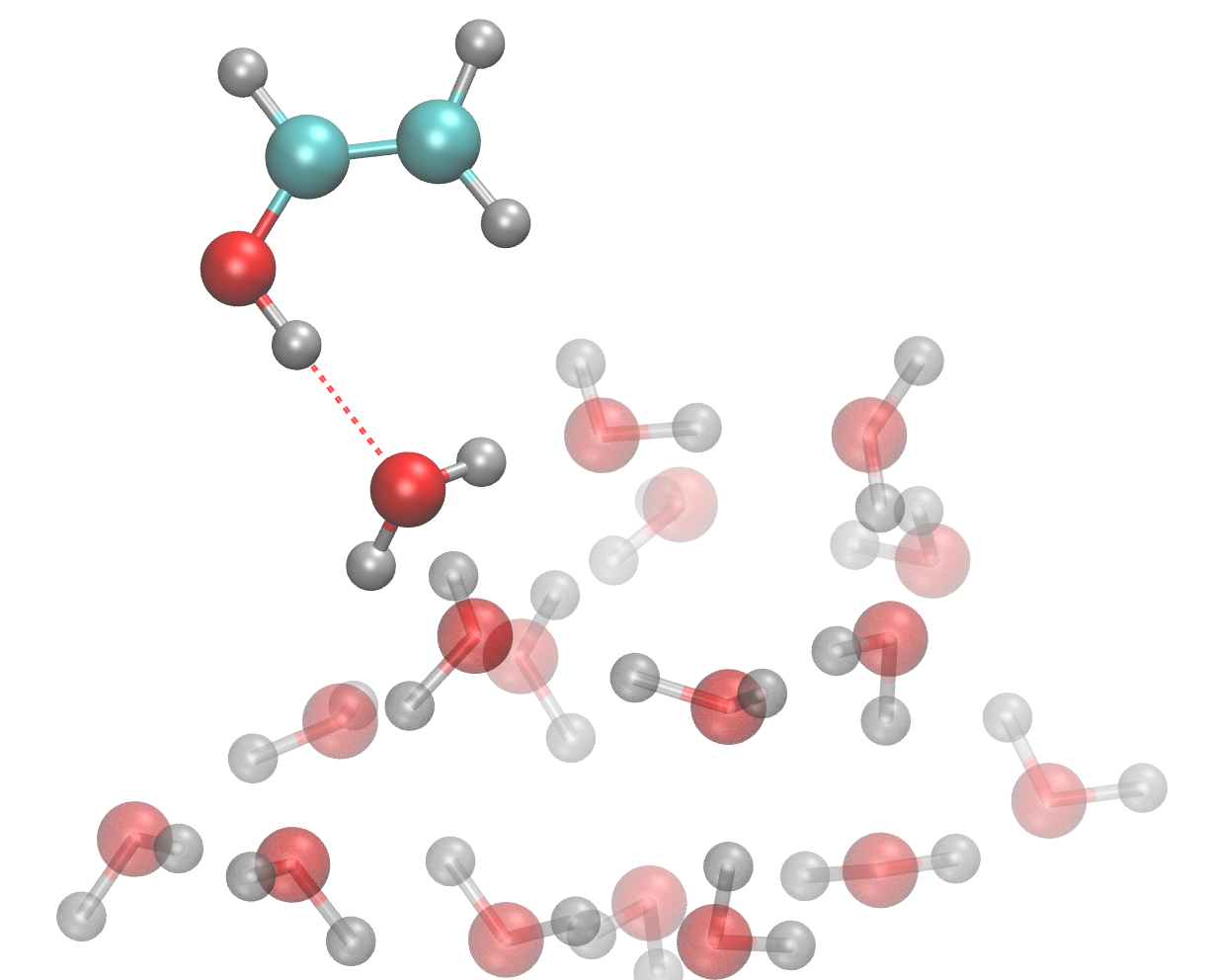} 
    \includegraphics[width=0.30\linewidth]{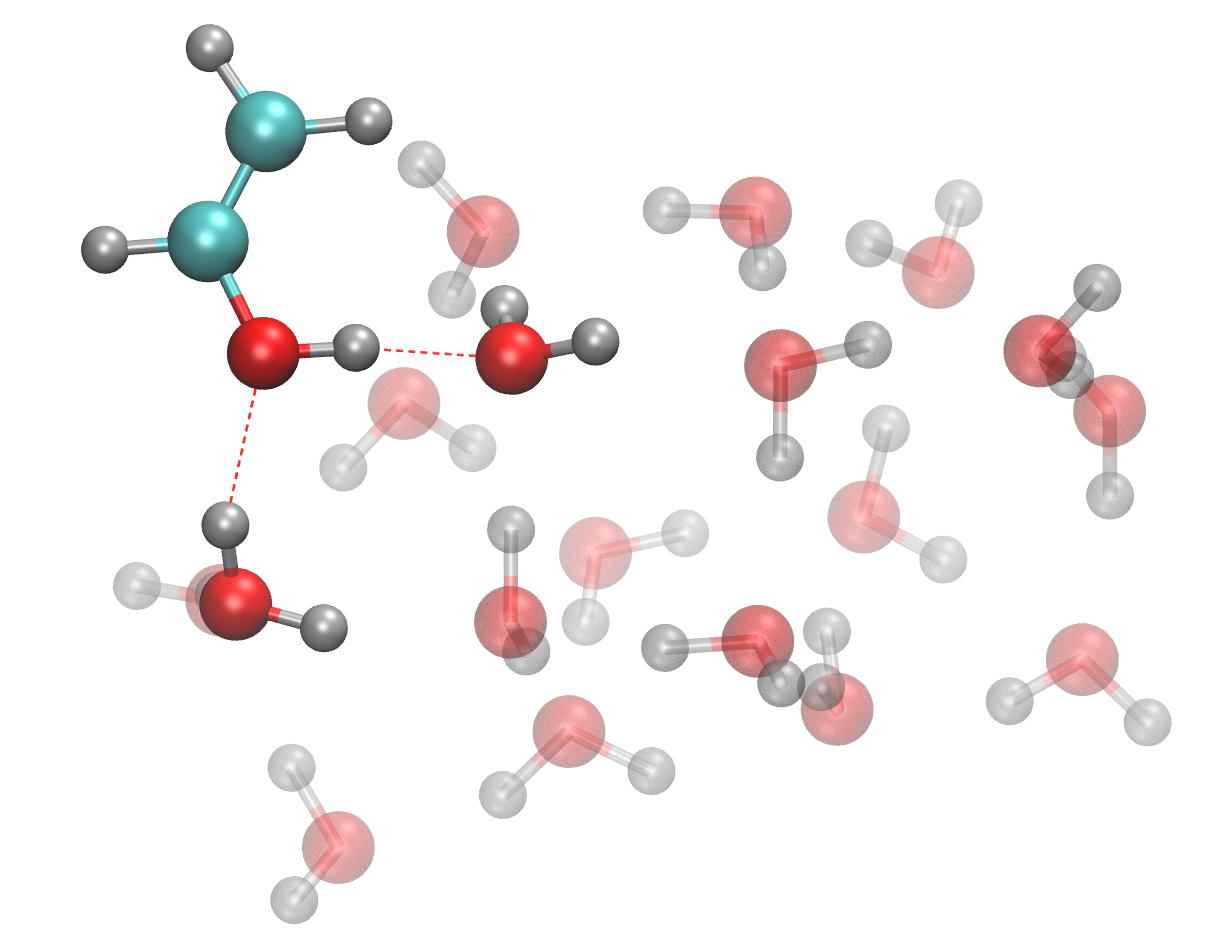}
    \includegraphics[width=0.30\linewidth]{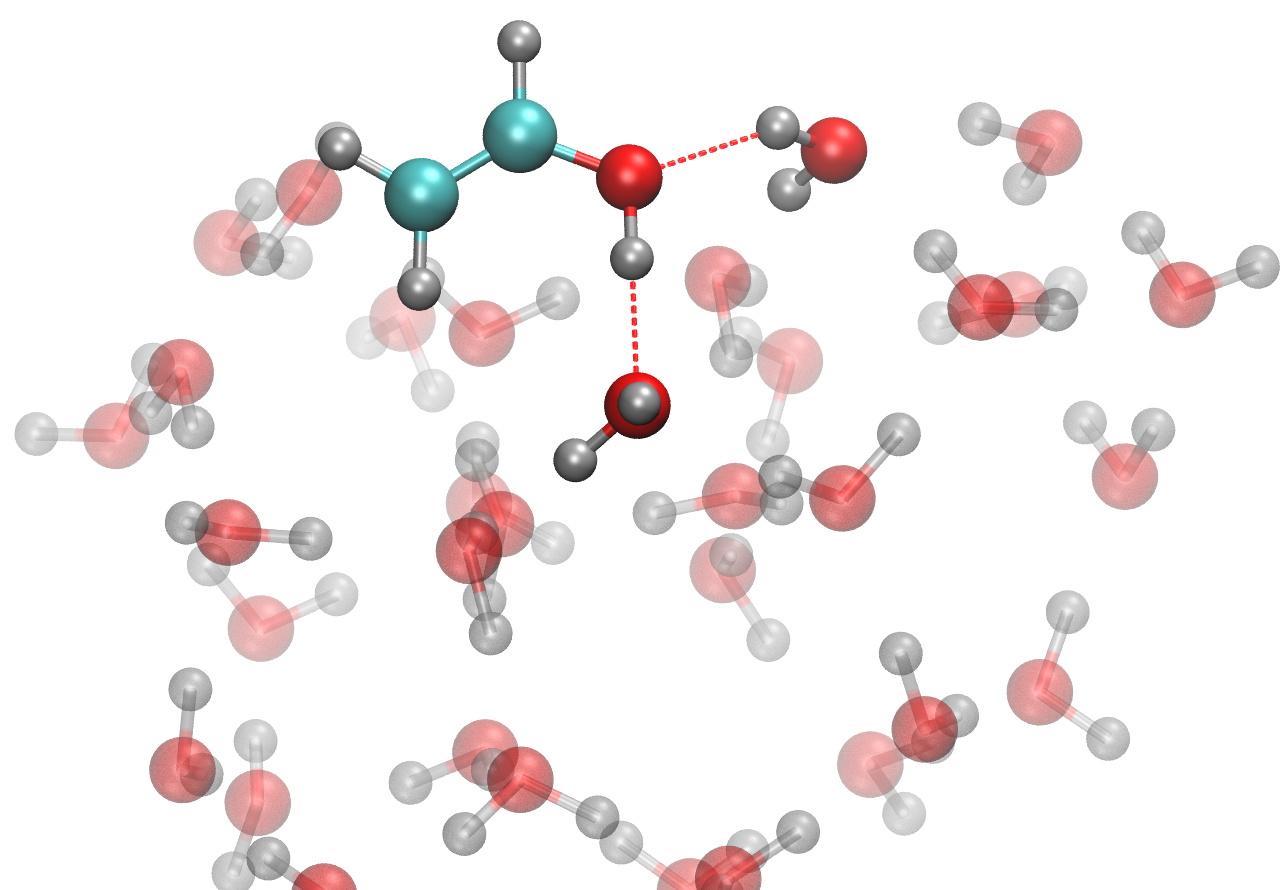} \\
    \vspace{0.1cm}
    \begin{tabular}{lll}
    dO  & \hspace{0.3\linewidth}  dH  & \hspace{0.3\linewidth}  Pocket \\
    \end{tabular} \\
    \hrulefill \\[1ex]
    {\large \bfseries \ce{HOC2H3OH} (1,2 Addition Product)} \\
    \includegraphics[width=0.30\linewidth]{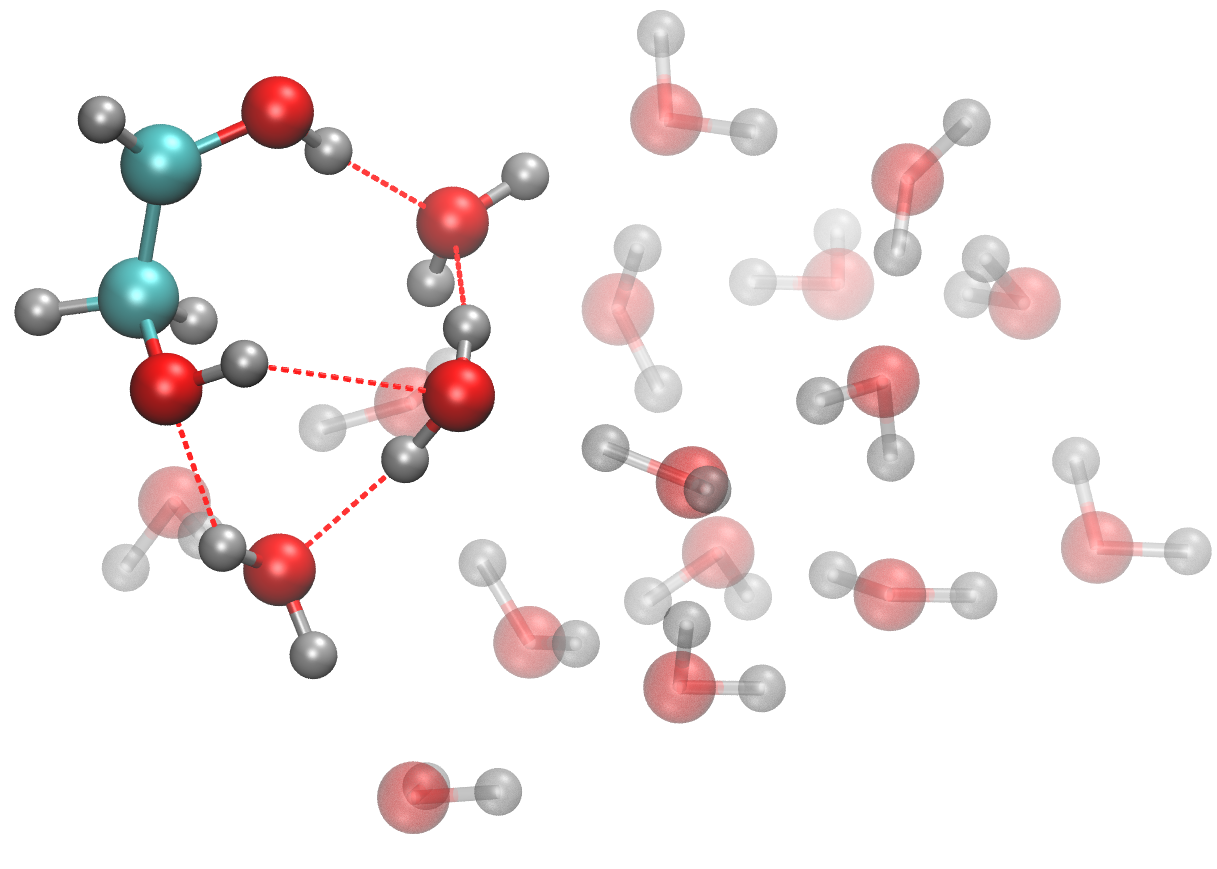} 
    \includegraphics[width=0.30\linewidth]{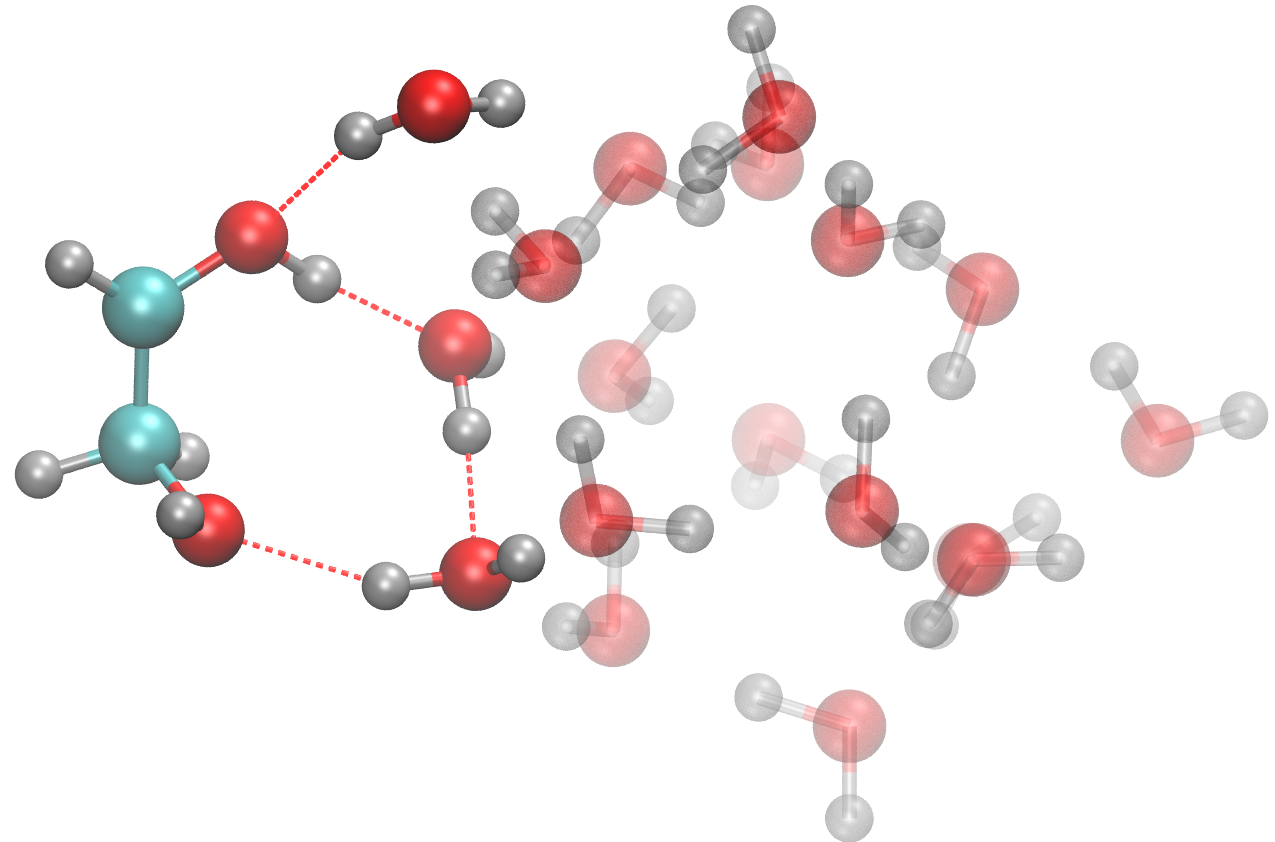}
    \includegraphics[width=0.30\linewidth]{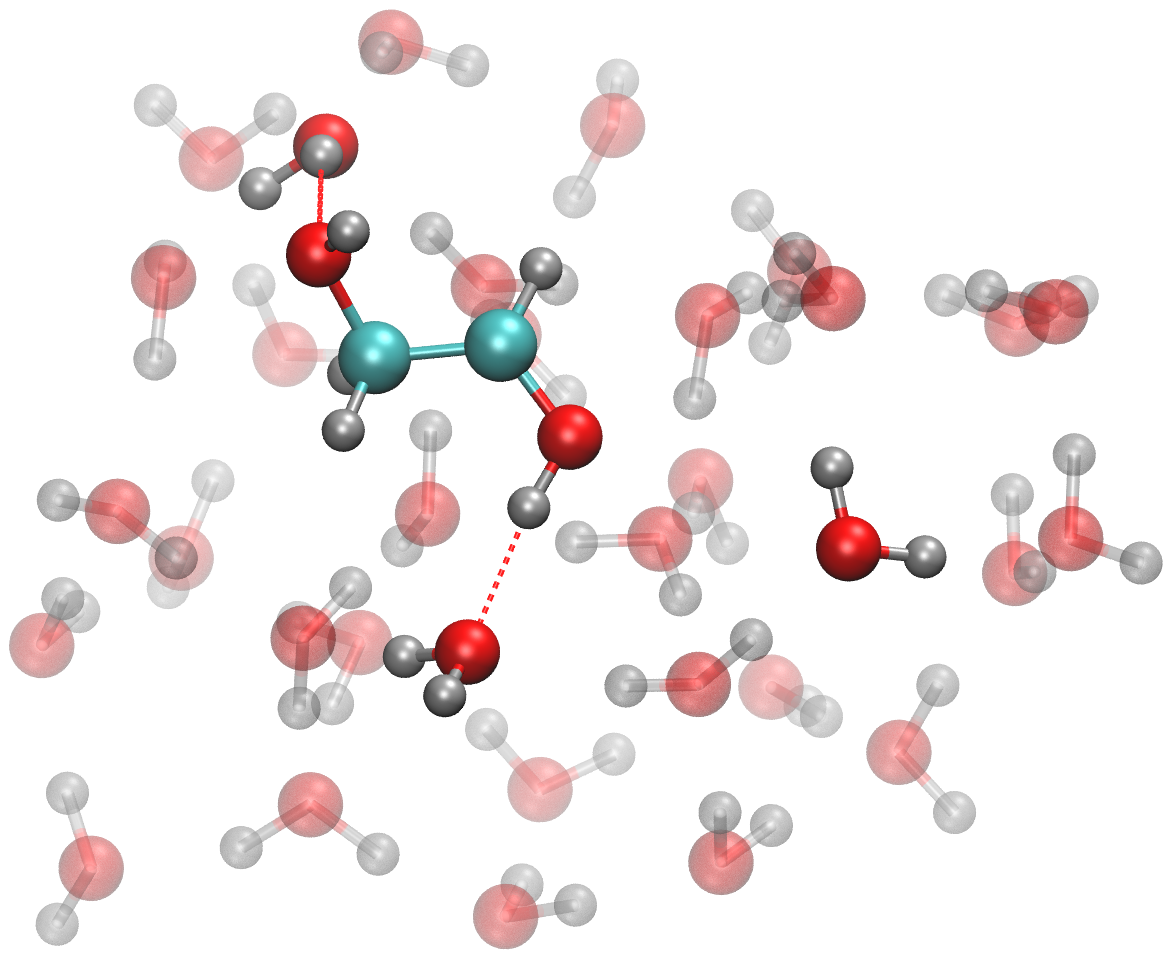} \\
    \vspace{0.1cm}
    \begin{tabular}{lll}
    dO  & \hspace{0.3\linewidth}  dH  & \hspace{0.3\linewidth}  Pocket \\
    \end{tabular} \\
    \hrulefill \\[1ex]
    {\large \bfseries \ce{C2H3(OH)2} (1,1 Addition Product)} \\
    \includegraphics[width=0.30\linewidth]{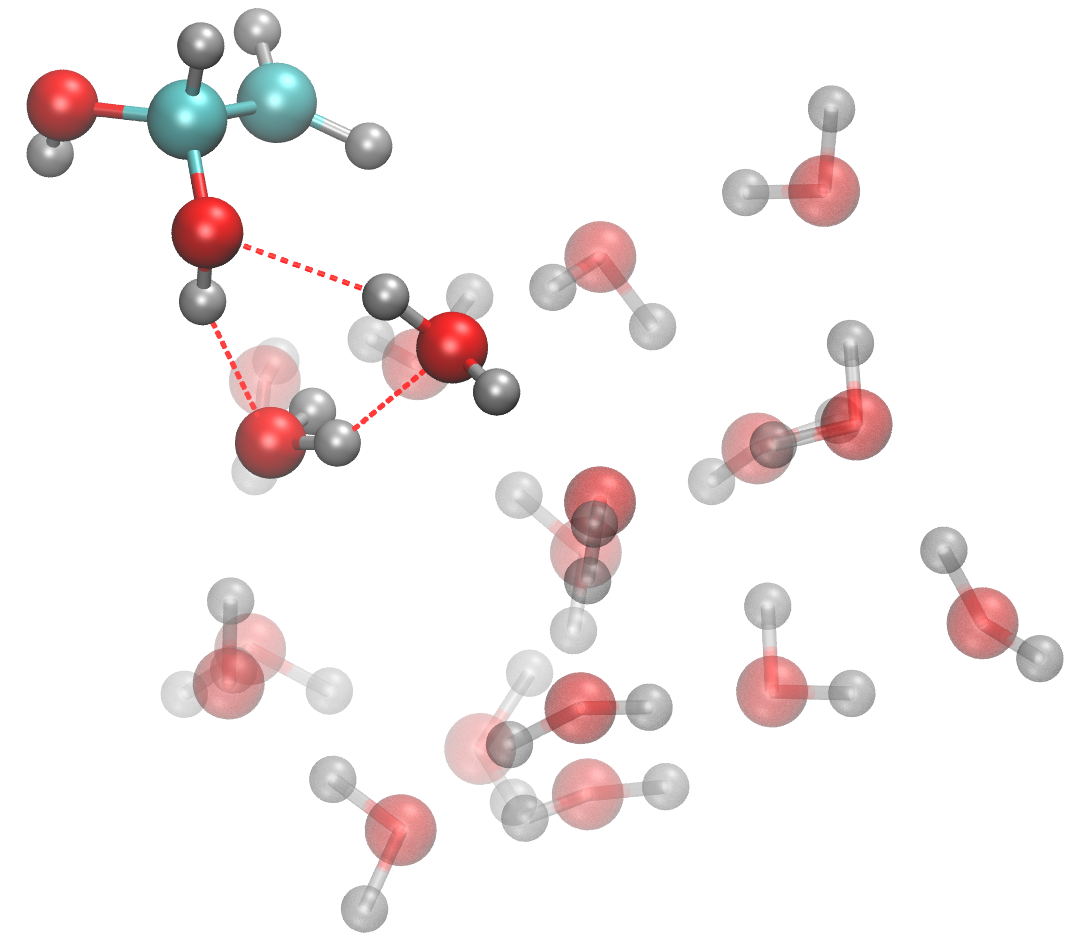} 
    \includegraphics[width=0.30\linewidth]{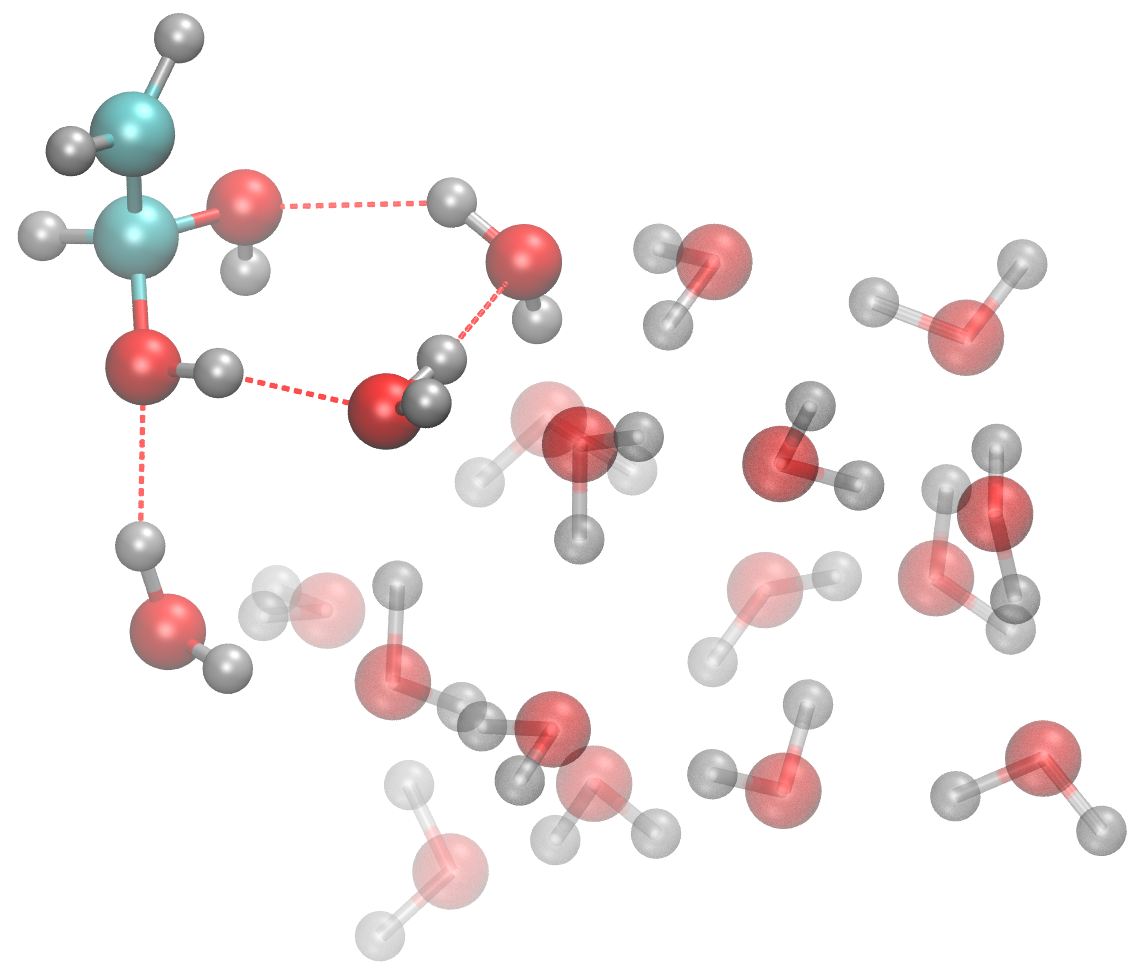}
    \includegraphics[width=0.30\linewidth]{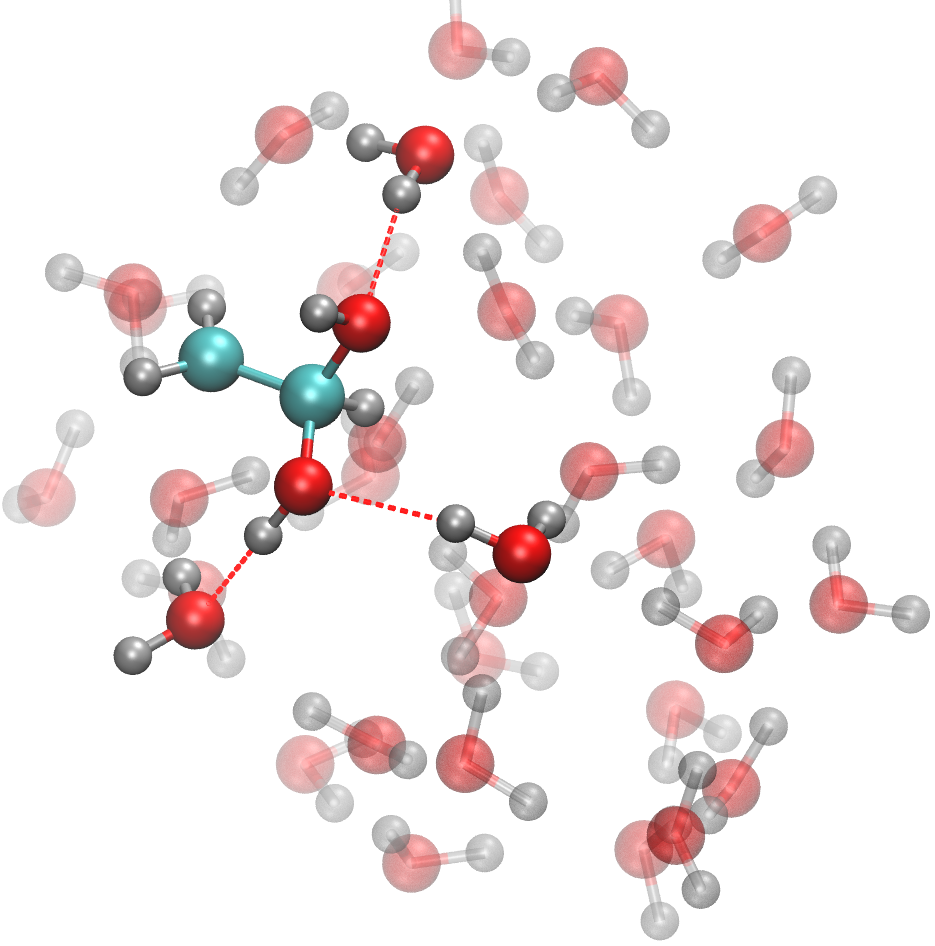
    } \\
    \vspace{0.1cm}
    \begin{tabular}{lll}
    dO  & \hspace{0.3\linewidth}  dH  & \hspace{0.3\linewidth}  Pocket \\
    \end{tabular} \\
    \hrulefill \\[2ex]
    \caption{Adsorption geometries of vinyl alcohol (\ce{C2H3OH}) (Top panel) and its OH addition products (\ce{HOC2H3OH}) (Middle panel) and (\ce{C2H3(OH)2}) (Bottom panel). From left to right in each panel we show the different binding sites considered in this work, namely dO, dH and the pocket site. The highlighted atoms in the figure indicate the adsorbate atoms and directly interacting water molecules.}
    \label{fig:OHadditionproducts}
\end{figure*}

The adsorption geometries of \ce{C2H3OH} and the products of the reaction with OH (Reactions \ref{eq:OH1} and \ref{eq:OH2}) are shown in Figure \ref{fig:OHadditionproducts}, and BE values are gathered in Table \ref{tab:BE}. 

\begin{deluxetable}{lc|c}[bt]
\label{tab:BE}
\tablecaption{Binding energy (BE) in kcal mol$^{-1}$ (in parenthesis in K) of the species and binding sites considered in this study.}
\tablehead{
\colhead{Species} &
\colhead{ Binding Site } &
\colhead{ BE } 
}
\startdata
\multirow{3}{*}{\ce{C2H3OH}} & dO & 5.6 (2808)  \\
& dH & 7.8 (3950)  \\
& Pocket & 12.5 (6315)  \\
\hline
\multirow{3}{*}{\ce{HOC2H3OH}} & dO & 10.3 (5190)   \\
& dH & 12.5 (6272)   \\
& Pocket & 8.6 (4341)  \\
\hline
\multirow{3}{*}{\ce{C2H3(OH)2}} & dO & 7.4 (3721)   \\
& dH & 10.7 (5408)   \\
& Pocket & 11.0 (5550)  \\
\hline
\multirow{2}{*}{\ce{(Z)-HOC2H2OH}} & dO & 11.4 (5733)   \\
& dH & 9.3 (4671)   \\
\hline
\multirow{3}{*}{\ce{HOC2H4OH}} & dO & 10.9 (5466)   \\
& dH & 14.0 (7044)   \\
& Pocket & 15.4 (7762)  \\
\hline
\multirow{3}{*}{\ce{C2H4(OH)2}} & dO & 8.3 (4193)   \\
& dH & 14.9 (7505)   \\
& Pocket & 17.5 (8793)  \\
\hline
\multirow{3}{*}{\ce{C2H2(OH)2}} & dO & 12.4 (6218)   \\
& dH & 9.2 (4653)   \\
& Pocket & 16.3 (8199)  \\
\hline
\multirow{2}{*}{\ce{(E)-HOC2H2OH}} & dO & 5.4 (2728) \\
& Pocket & 16.1 (8115) \\
\enddata
\vspace{-2em}
\end{deluxetable}

\begin{figure*}[hbt] 
    \centering
    \includegraphics[width=0.8\linewidth]{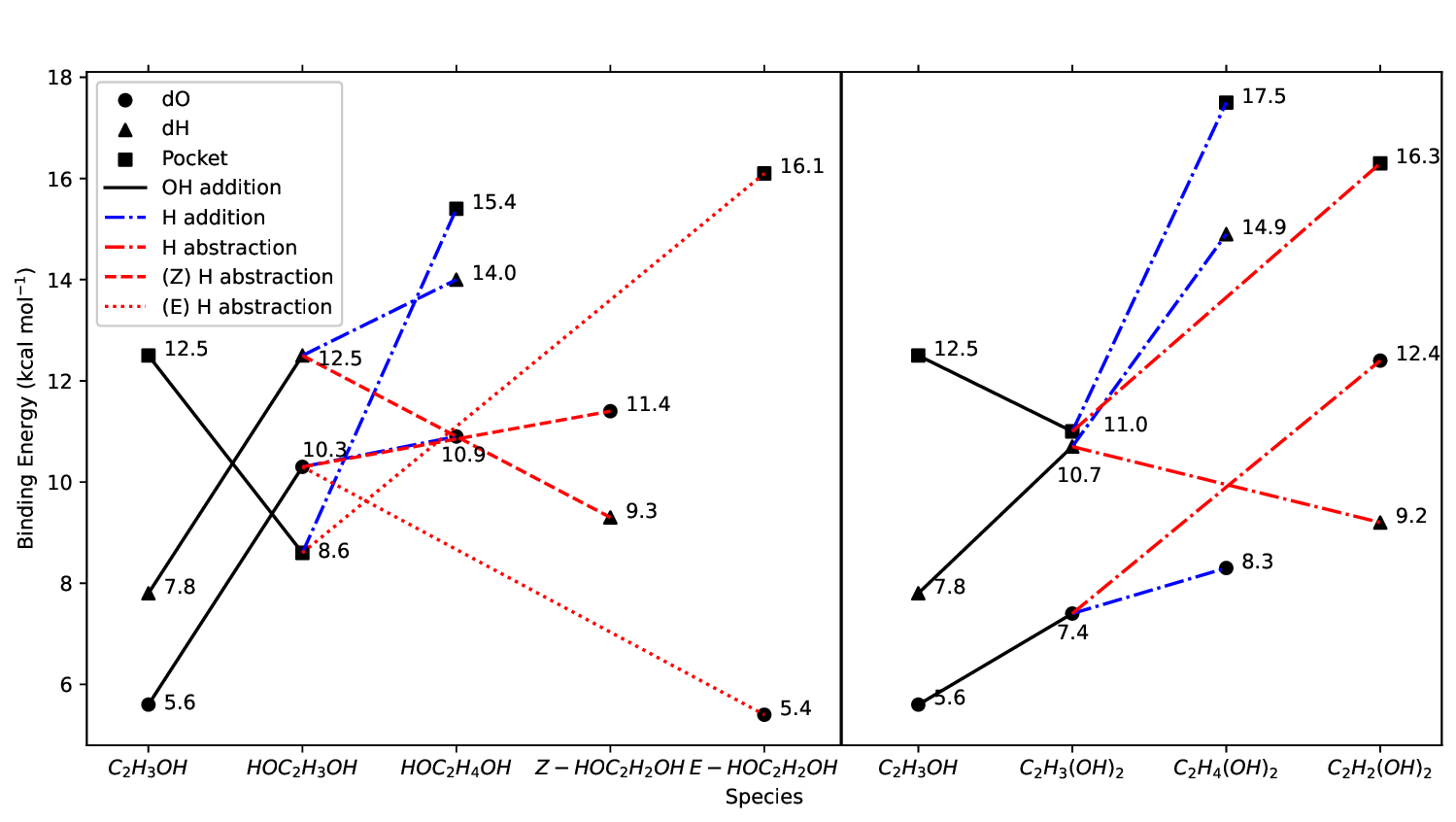} 
    \caption{Evolution of the binding energies of the species on the different binding sites. The left-hand side of the figure features the species evolved from the 1,2-addition of OH to \ce{C2H3OH} and the right-hand side those evolved from the 1,1-addition.}
    \label{fig:BEgraph}
\end{figure*}

\begin{figure*}[hbt] 
    \centering
    \hrulefill \\[1ex]
    {\large \bfseries \ce{(Z)-HOC2H2OH} ((Z)-1,2-ethenediol)} \\
    \includegraphics[width=0.40\linewidth]{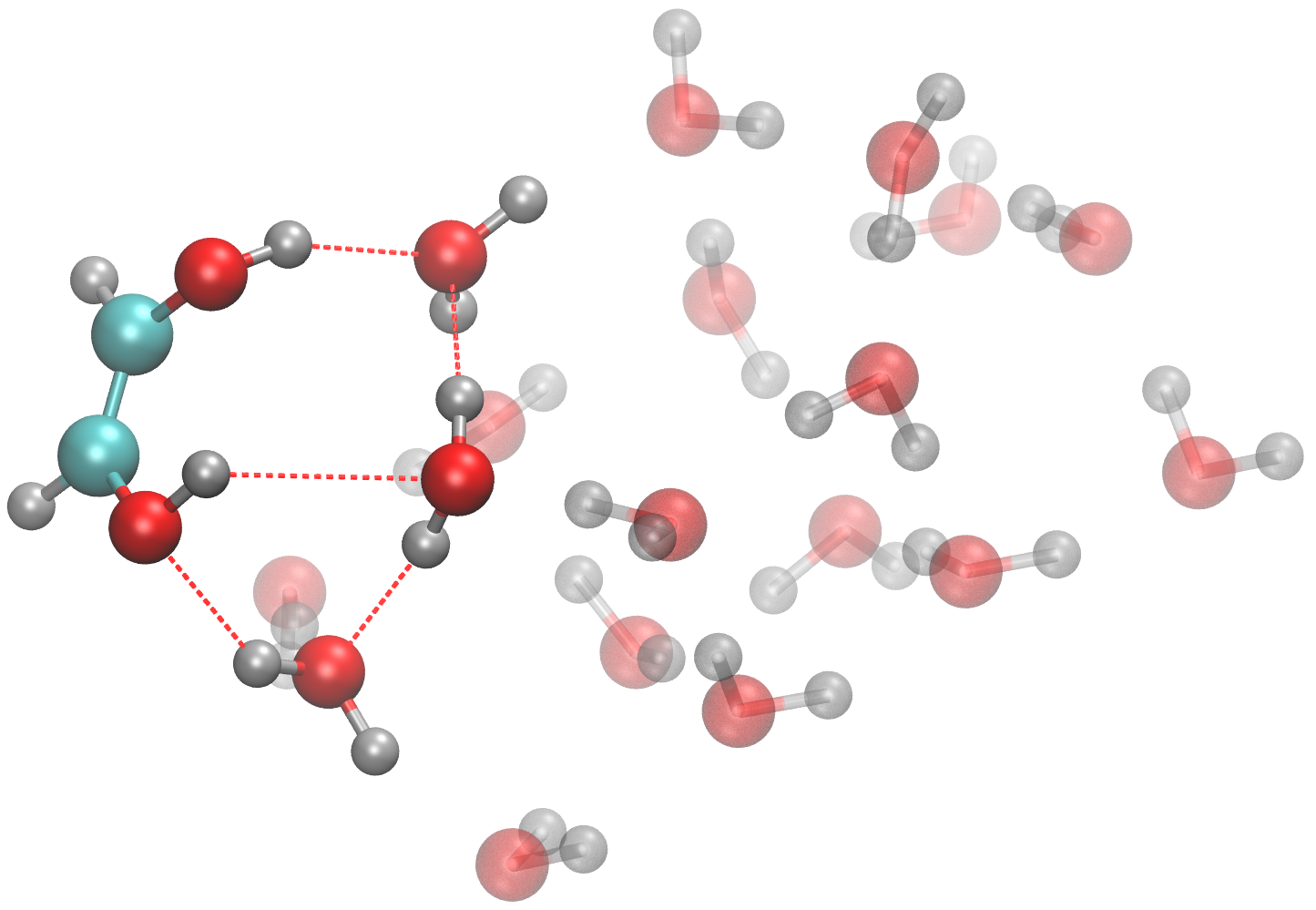} 
    \includegraphics[width=0.40\linewidth]{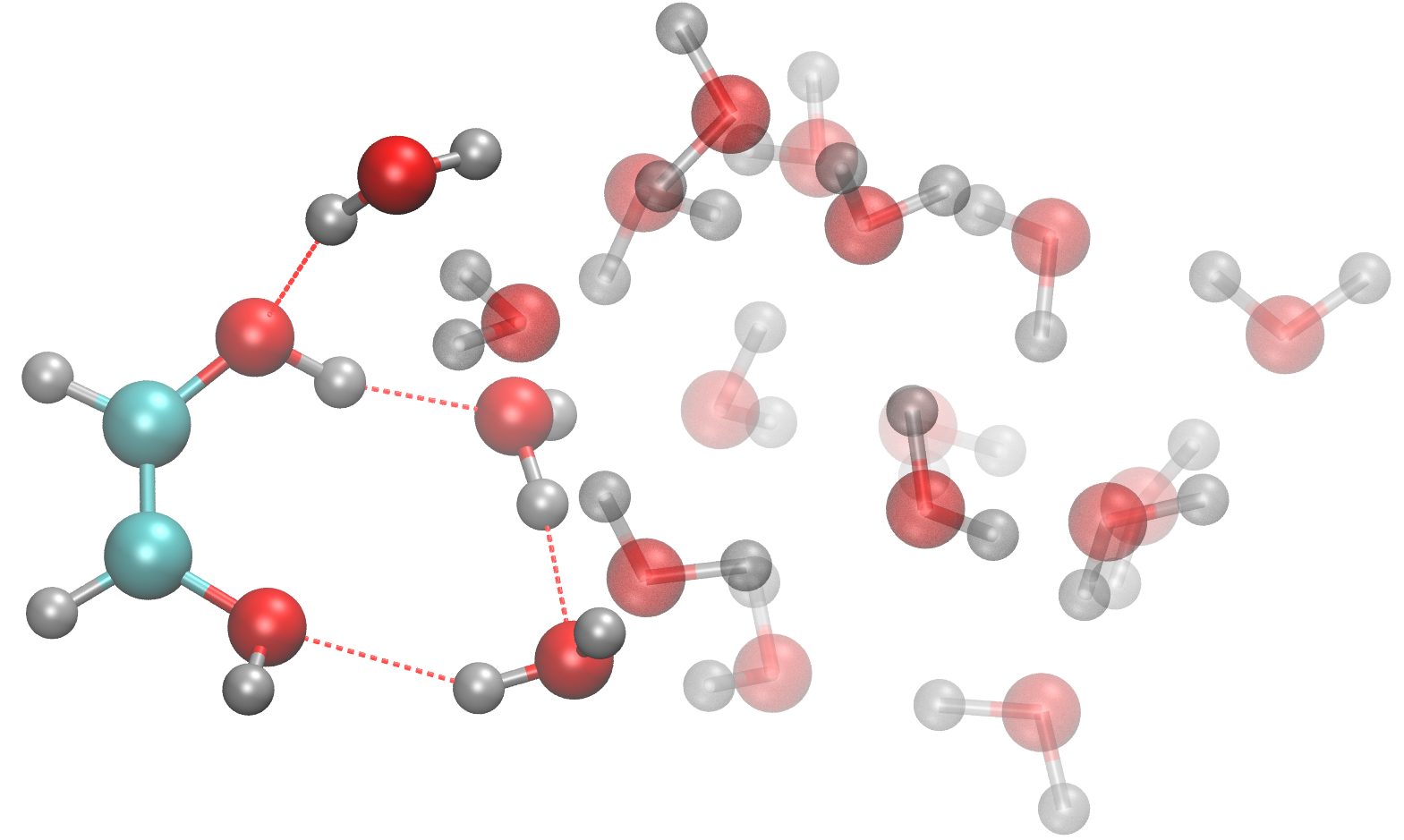} \\
    \vspace{0.1cm}
    \begin{tabular}{lll}
    dO  & \hspace{0.4\linewidth}  dH \\
    \end{tabular} \\
    \hrulefill \\[1ex]
    {\large \bfseries \ce{(E)-HOC2H2OH} ((E)-1,2-ethenediol)} \\
    \includegraphics[width=0.40\linewidth]{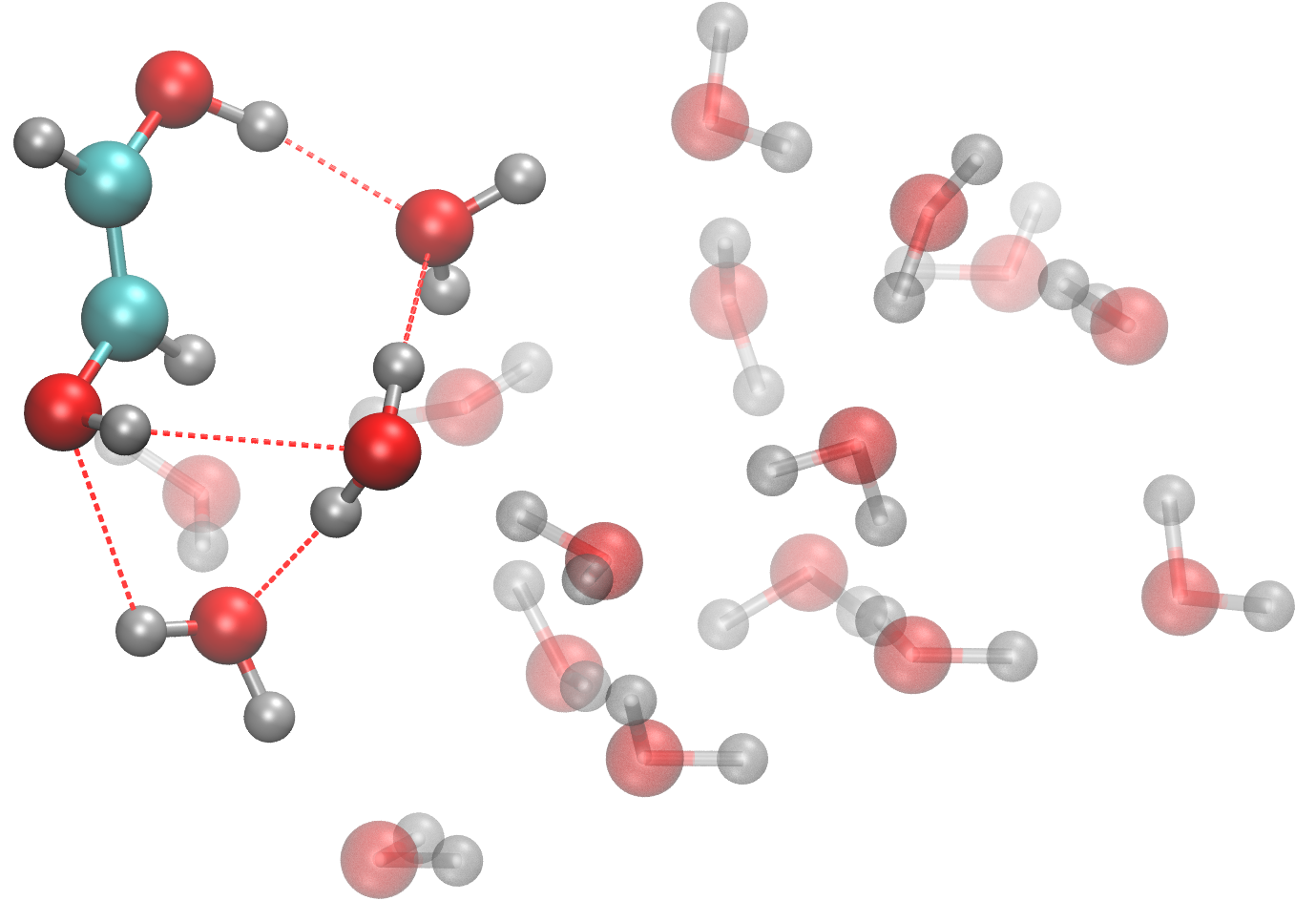} 
    \includegraphics[width=0.40\linewidth]{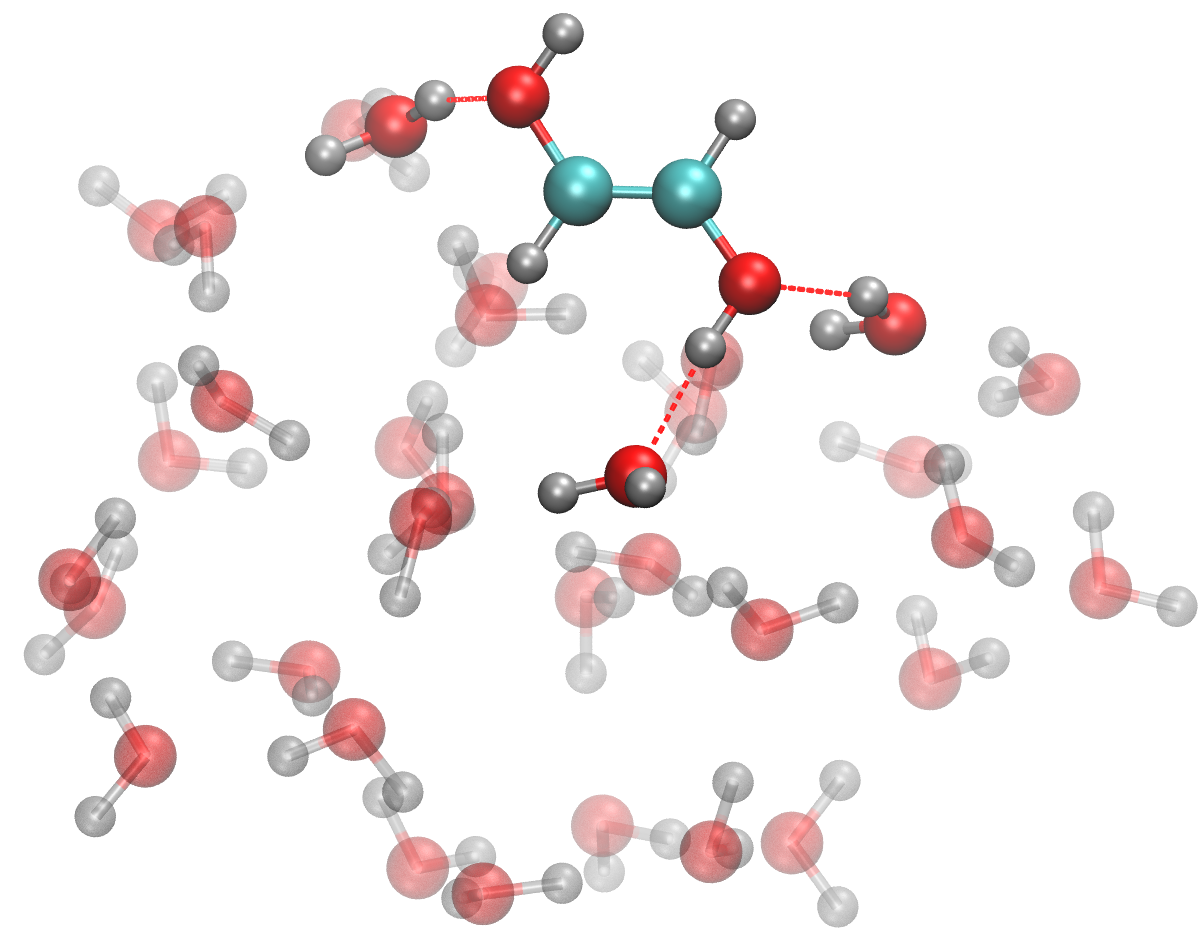} \\
    \vspace{0.1cm}
    \begin{tabular}{lll}
    dO  & \hspace{0.4\linewidth}  Pocket \\
    \end{tabular} \\
    \hrulefill \\[1ex]
    {\large \bfseries \ce{C2H2(OH)2} (1,1-ethenediol)} \\
    \includegraphics[width=0.30\linewidth]{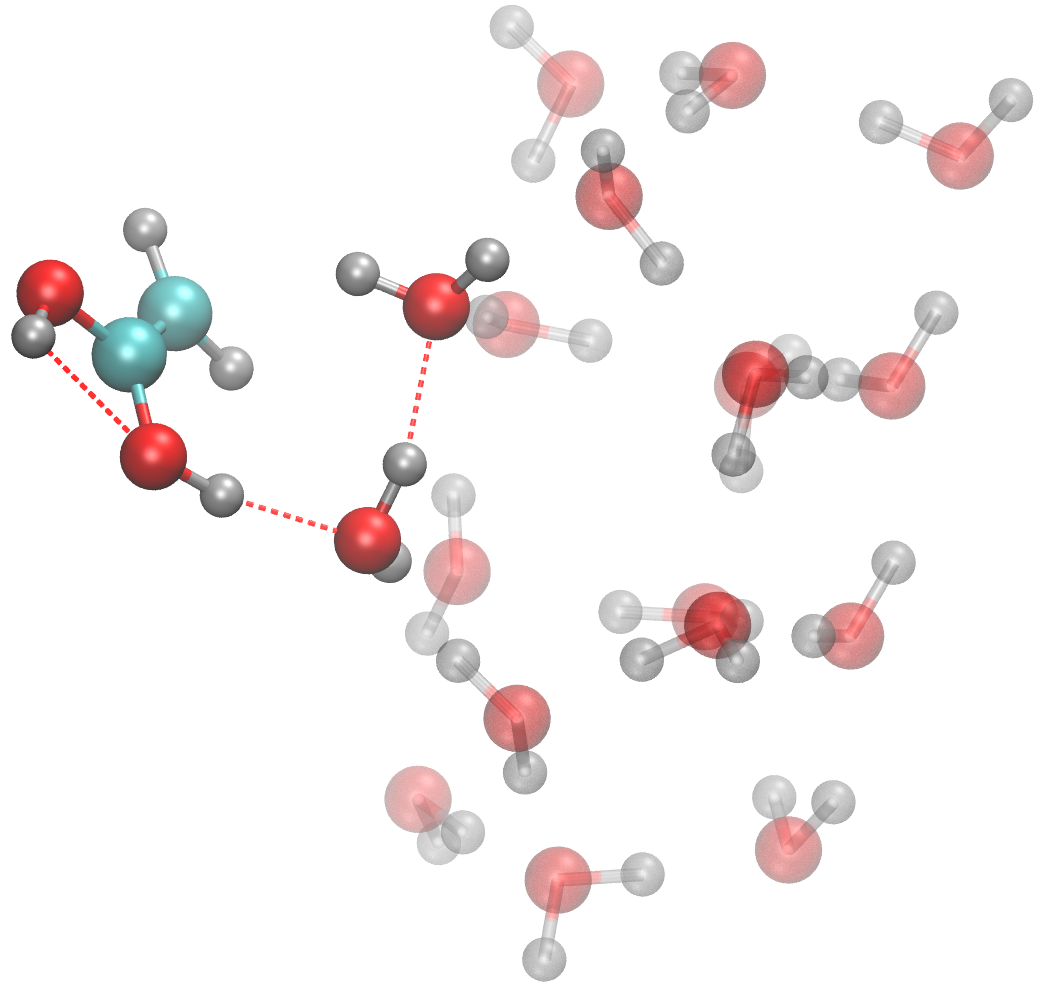} 
    \includegraphics[width=0.30\linewidth]{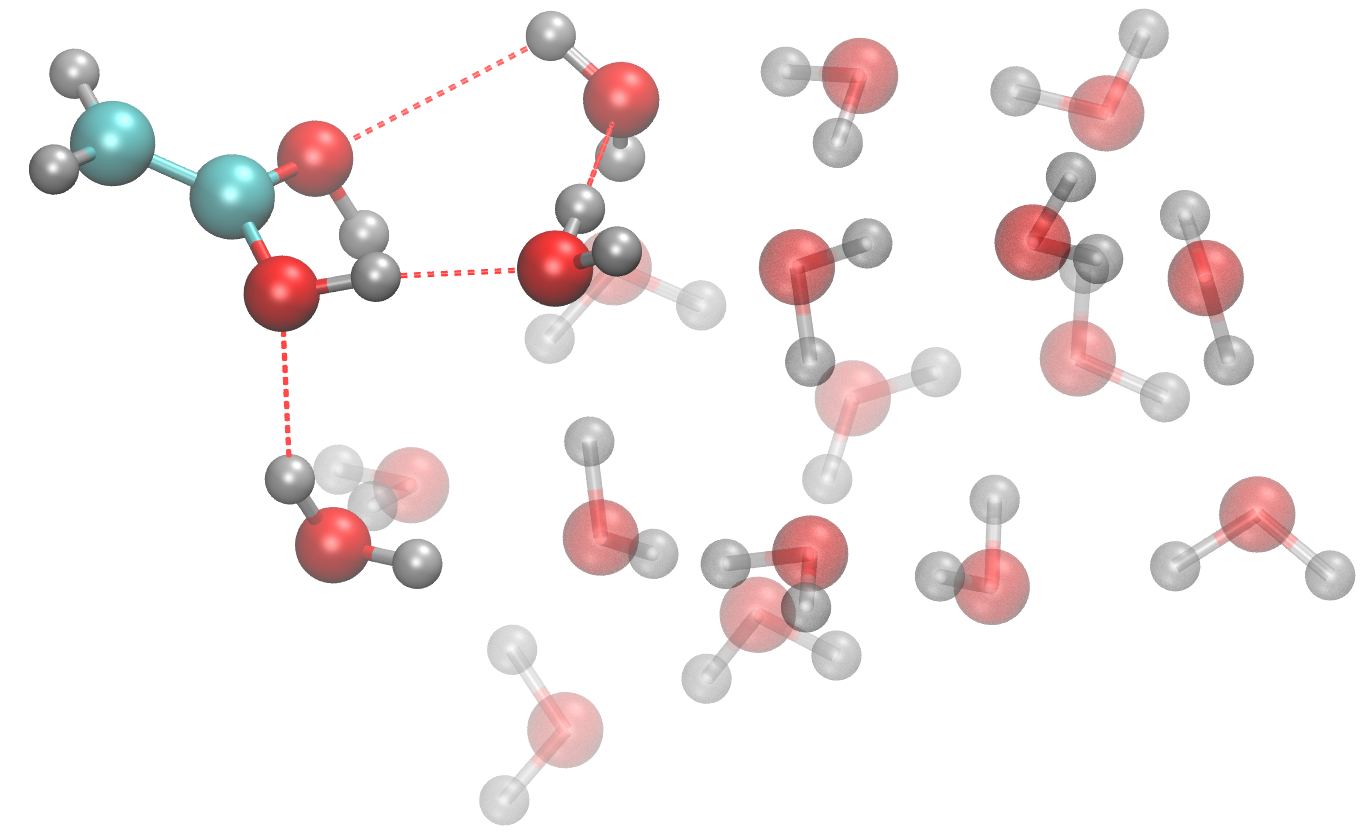}
    \includegraphics[width=0.30\linewidth]{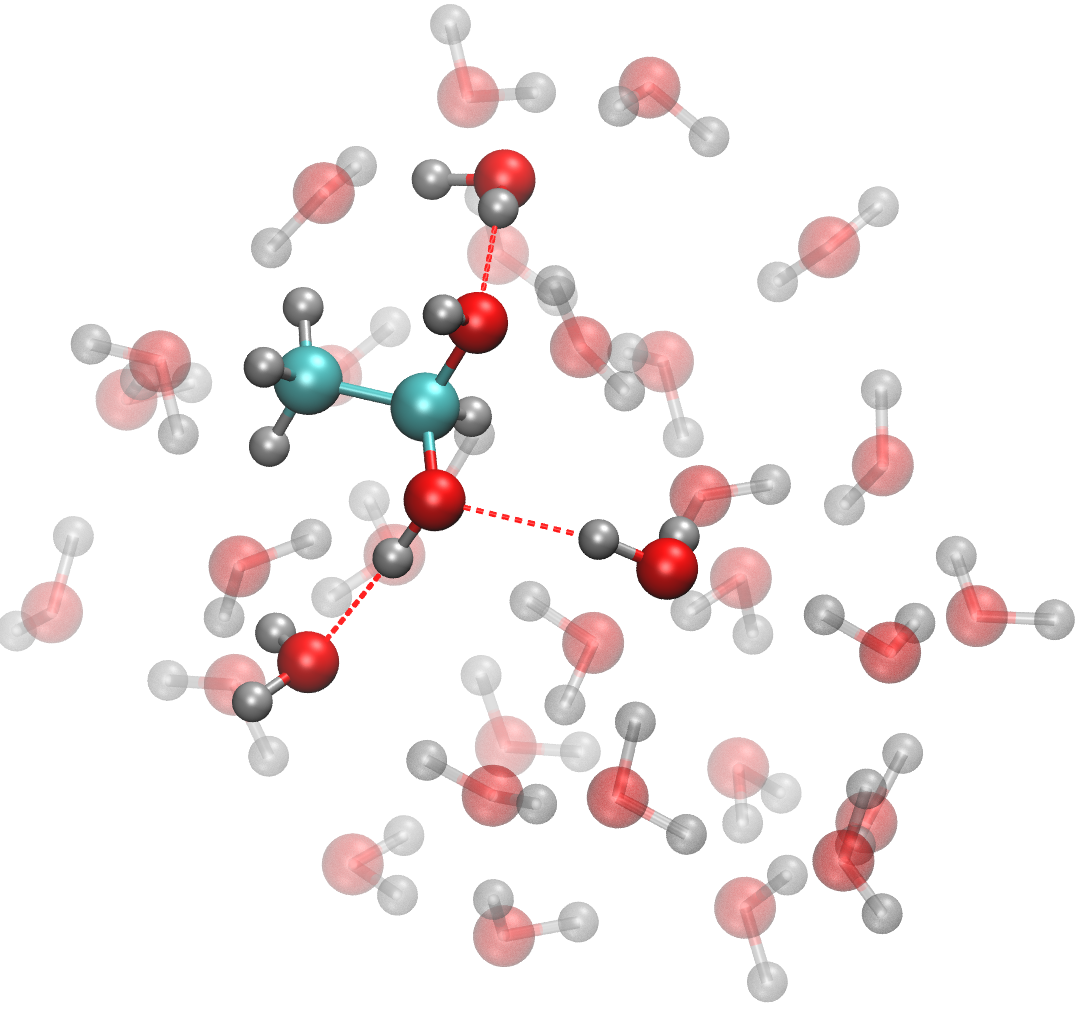}\\
    \vspace{0.1cm}
    \begin{tabular}{lll}
    dO  & \hspace{0.3\linewidth}  dH  & \hspace{0.3\linewidth}  Pocket \\
    \end{tabular} \\
    \hrulefill \\[2ex]
    \caption{Adsorption geometries of (Z)-1,2-ethenediol (\ce{(Z)-HOC2H2OH}) (Top panel), (E)-1,2-ethenediol (\ce{(E)-HOC2H2OH}) (Middle panel) and 1,1-ethenediol (\ce{C2H2(OH)2}) (Bottom panel). From left to right in each panel we show the different binding sites considered in this work, namely dO, dH and the pocket site. The highlighted atoms in the figure indicate the adsorbate atoms and directly interacting water molecules.}
    \label{fig:Habstractionproducts}
\end{figure*}

\begin{figure}[hbt] 

    \centering
    \hrulefill \\[1ex]
    { \bfseries \ce{(Z)-HOC2H2OH} ((Z)-1,2-ethenediol)} \\
    \includegraphics[width=0.40\linewidth]{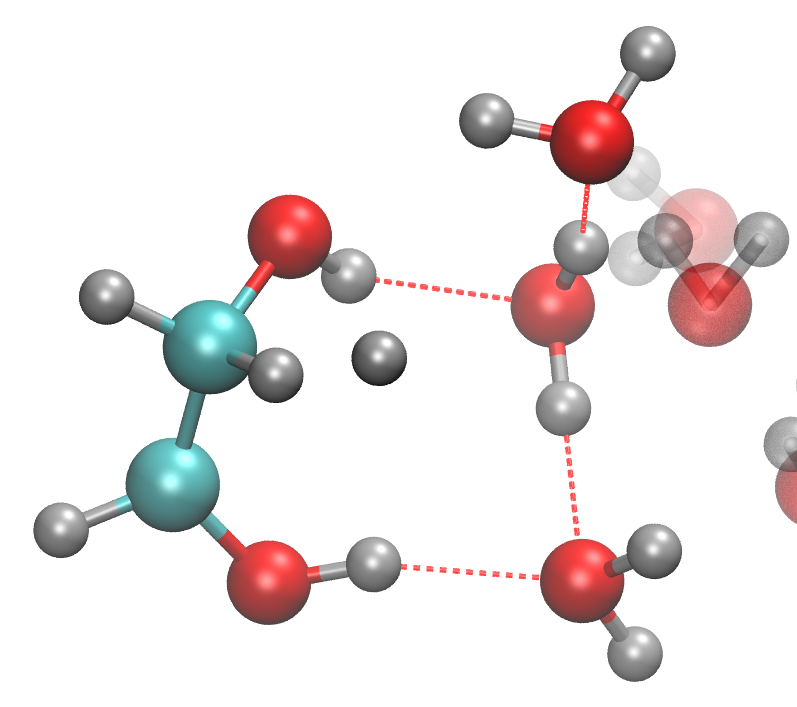} 
    \includegraphics[width=0.40\linewidth]{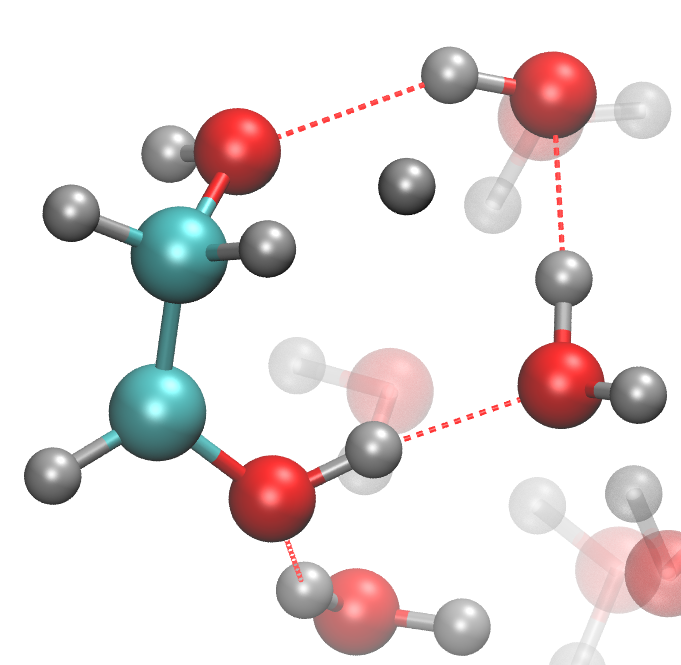} \\
    \vspace{0.1cm}
    \begin{tabular}{lll}
    dO  & \hspace{0.4\linewidth}  dH \\
    \end{tabular} \\
    \hrulefill \\[1ex]
    { \bfseries \ce{(E)-HOC2H2OH} ((E)-1,2-ethenediol)} \\
    \includegraphics[width=0.40\linewidth]{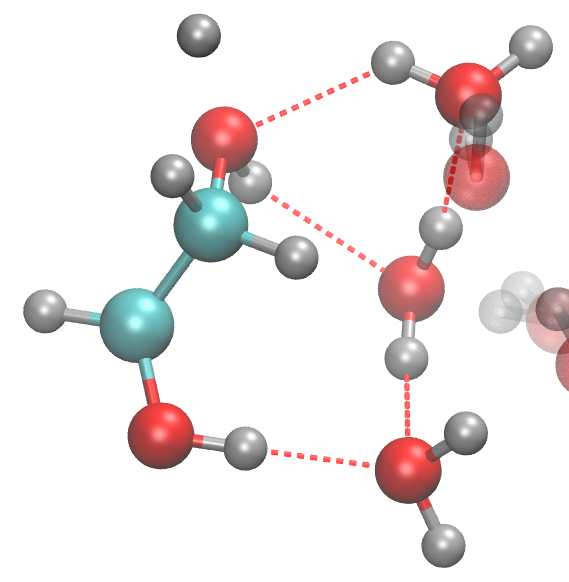} 
    \includegraphics[width=0.40\linewidth]{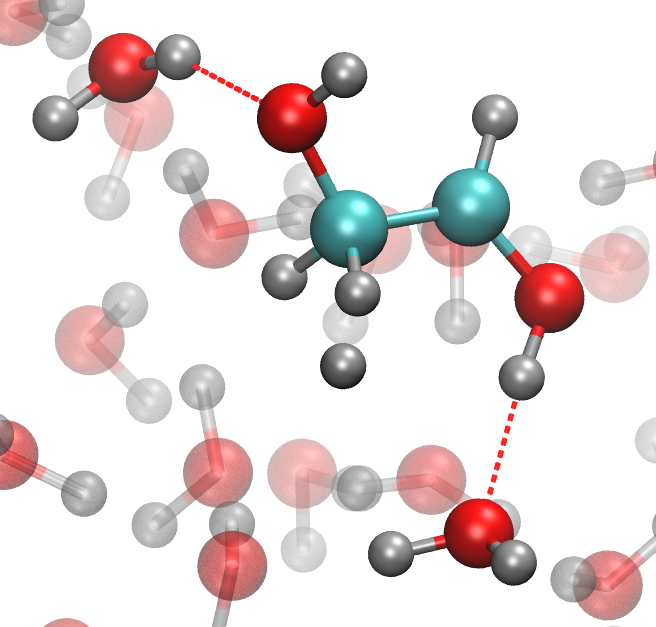} \\
    \vspace{0.1cm}
    \begin{tabular}{lll}
    dO  & \hspace{0.4\linewidth}  Pocket \\
    \end{tabular} \\
    \hrulefill \\[2ex]
    \caption{Close-up snapshots of the H-abstraction process leading to different stereoisomers on selected binding sites. Top panel corresponding to \ce{(Z)-HOC2H2OH}, bottom panel corresponding to \ce{(E)-HOC2H2OH}.}
    \label{fig:snapshots}
\end{figure}

\begin{figure*}[hbt] 
    \centering
    \hrulefill \\[1ex]
    {\large \bfseries \ce{HOC2H4OH} (1,2-ethanediol, ethylene glycol)} \\
    \includegraphics[width=0.30\linewidth]{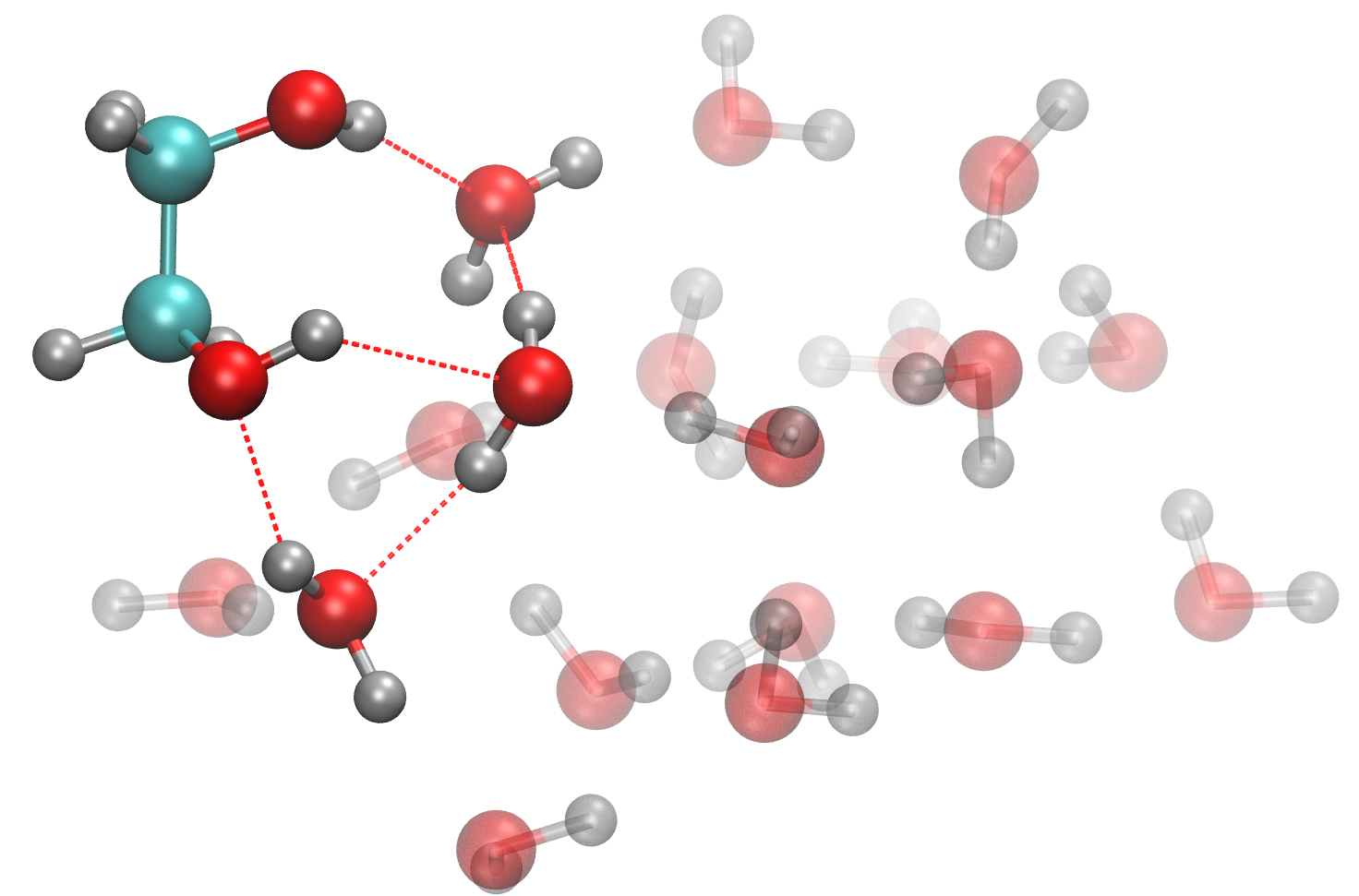} 
    \includegraphics[width=0.30\linewidth]{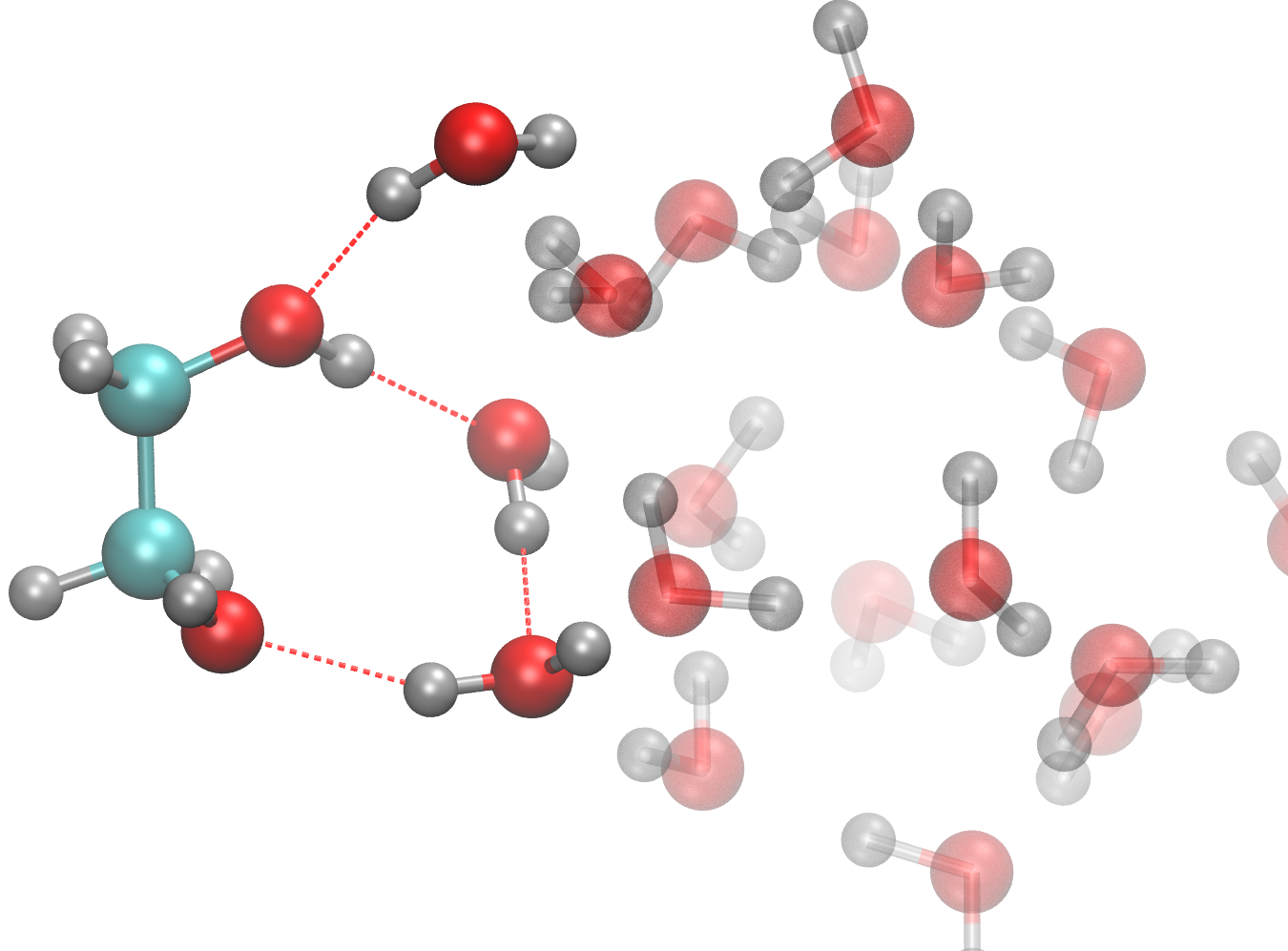}
    \includegraphics[width=0.30\linewidth]{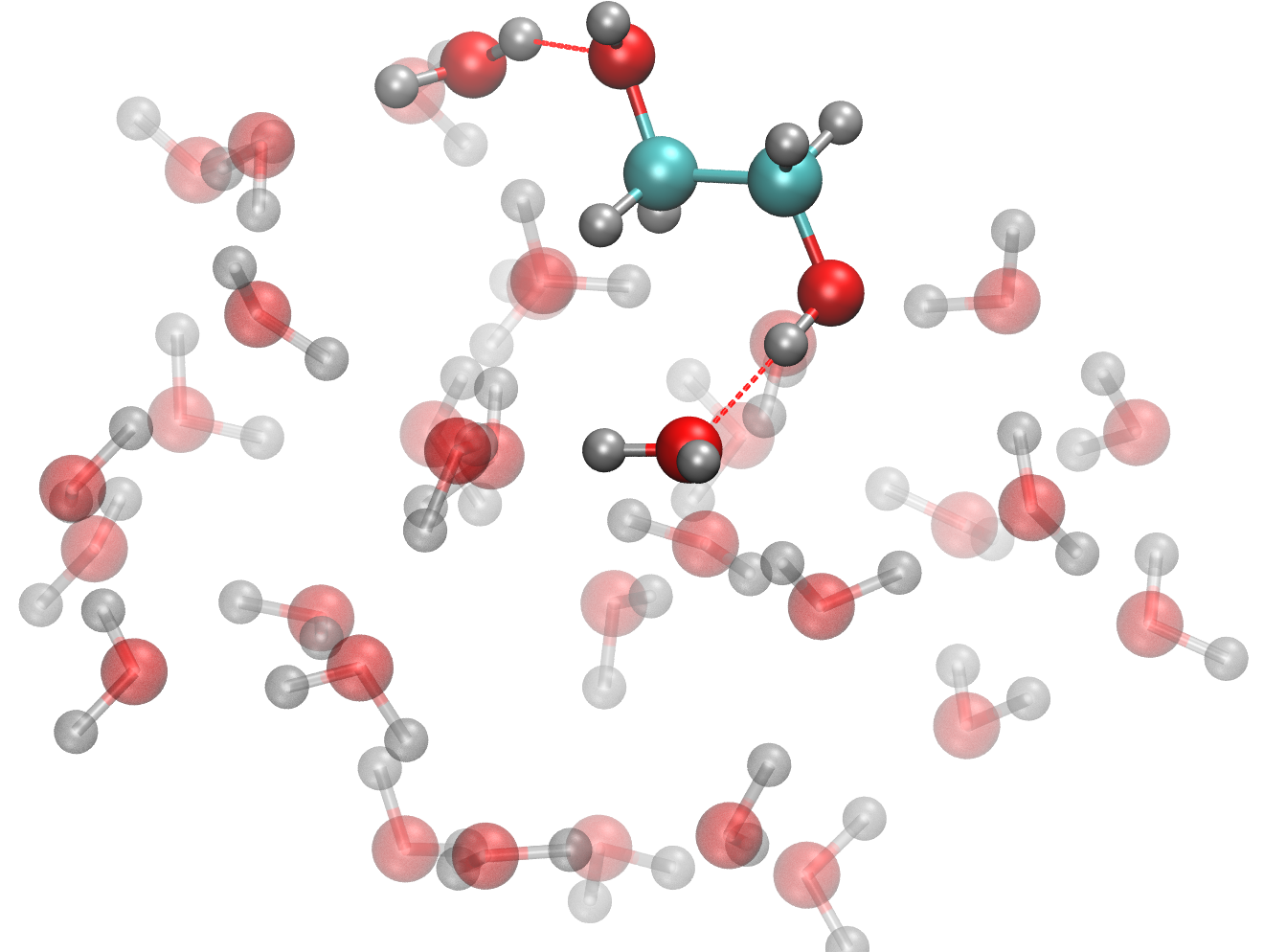} \\
    \vspace{0.1cm}
    \begin{tabular}{lll}
    dO  & \hspace{0.3\linewidth}  dH  & \hspace{0.3\linewidth}  Pocket \\
    \end{tabular} \\
    \hrulefill \\[1ex]
    {\large \bfseries \ce{C2H4(OH)2} (1,1-ethanediol)} \\
    \includegraphics[width=0.30\linewidth]{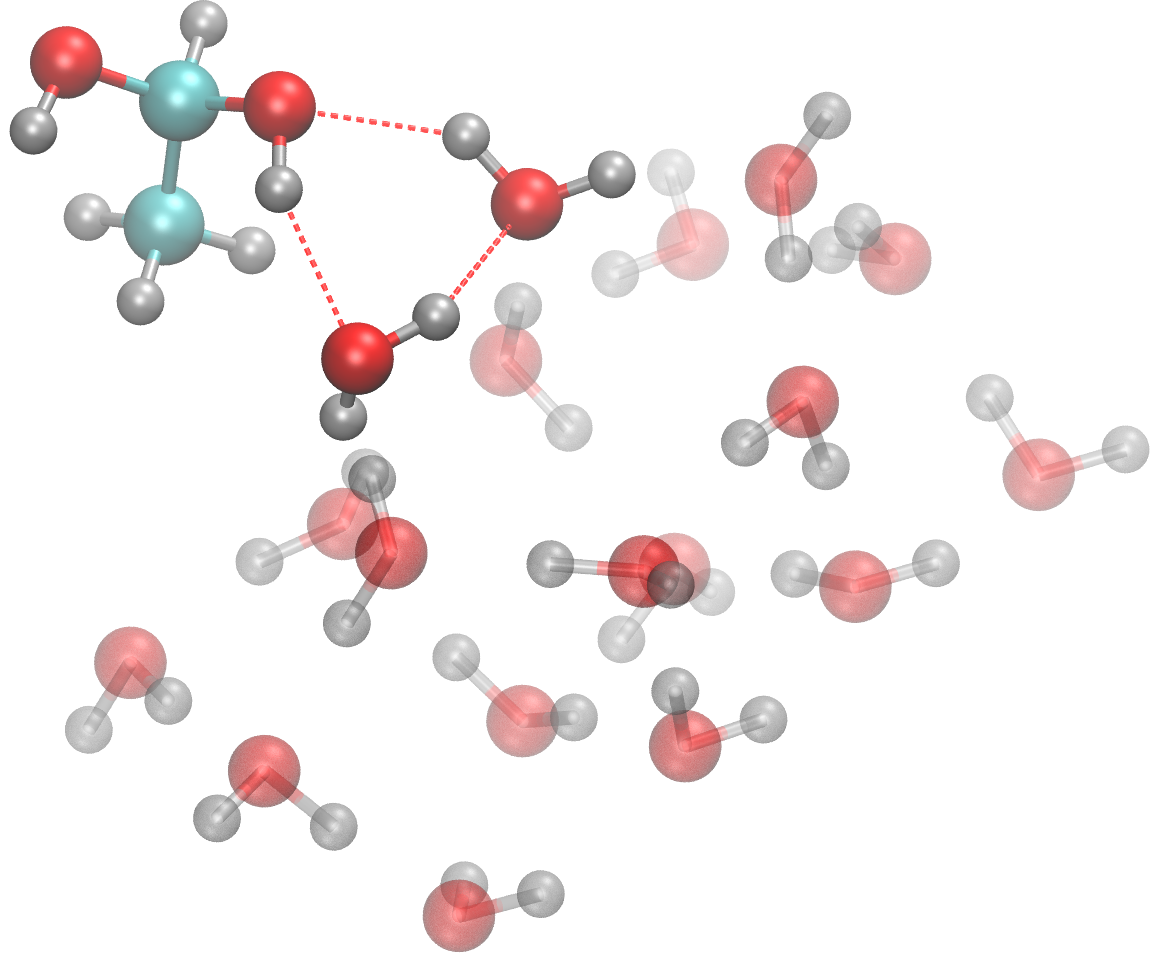} 
    \includegraphics[width=0.30\linewidth]{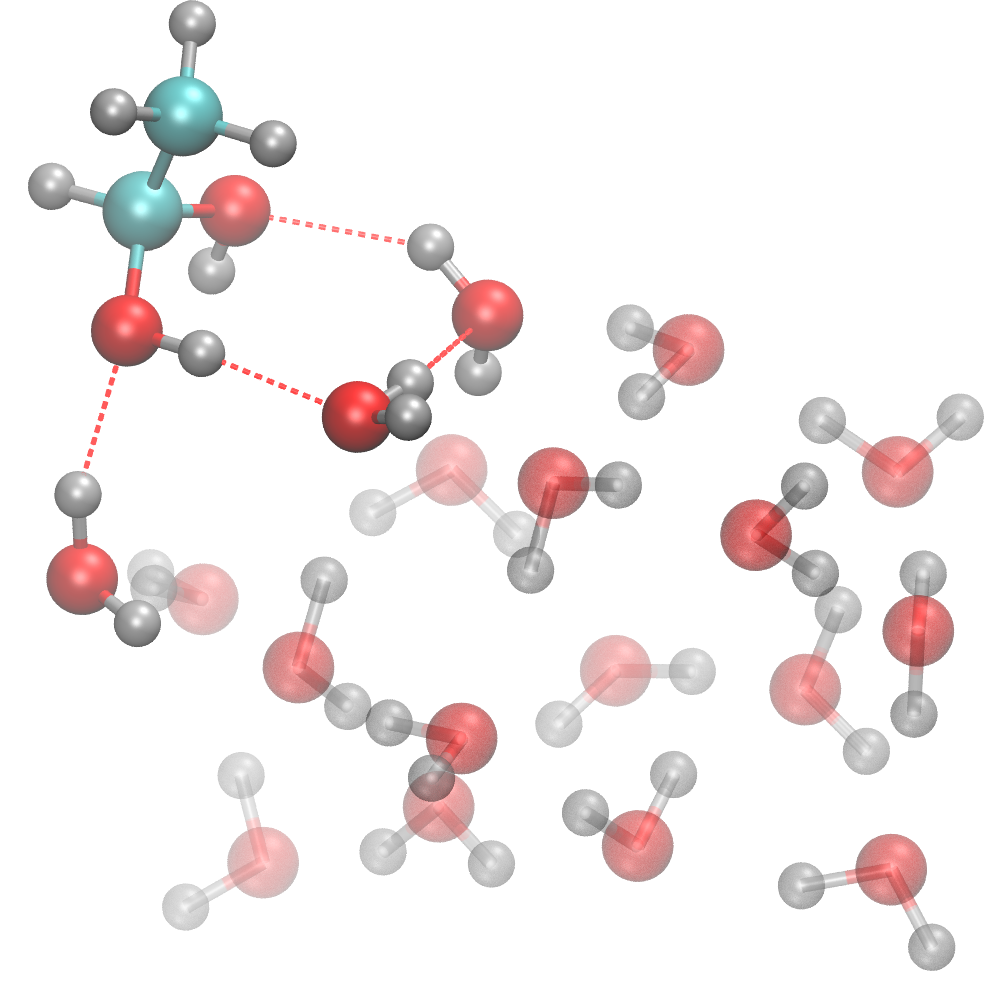}
    \includegraphics[width=0.30\linewidth]{figures/17cav11ethanediol.png} \\
    \vspace{0.1cm}
    \begin{tabular}{lll}
    dO  & \hspace{0.3\linewidth}  dH  & \hspace{0.3\linewidth}  Pocket \\
    \end{tabular} \\
    \hrulefill \\[2ex]
    \caption{Adsorption geometries of 1,2-ethanediol (\textbf{ethylene glycol, }\ce{HOC2H4OH}) (Top panel) and 1,1-ethanediol (\ce{C2H4(OH)2}) (Lower panel). From left to right in each panel we show the different binding sites considered in this work, namely dO, dH and the pocket site. The highlighted atoms in the figure indicate the adsorbate atoms and directly interacting water molecules.}
    \label{fig:Hadditionproducts}
\end{figure*}

\begin{deluxetable*}{clc|cc}[bt!]
\label{tab:OHaddition}
\tablecaption{Summary of the reaction energies, $\Delta H^{\circ, \text{R}}$, and activation energies, $\Delta H^{\circ, \ddagger}$ for the \ce{C2H3OH + OH reaction}. Energies are ZPVE corrected. In parentheses, electronic energies, e.g. without ZPVE. All values are reported in kcal mol$^{-1}$. BL stands for barrierless. As a reference, the activation energies in the gas phase, from the Prc1 complex, reported in \citet{Ballotta23} are 1.0 kcal mol$^{-1}$ for reaction \ref{eq:OH1} and 1.9 kcal mol$^{-1}$ for reaction \ref{eq:OH2}}
\tablehead{
\colhead{Label} &
\colhead{ Reaction        } &
\colhead{ Binding Site } &
\colhead{ $\Delta H^{\circ, \text{R}}$ } &
\colhead{ $\Delta H^{\circ, \ddagger}$ } 
}
\startdata
\ref{eq:OH1} & \ce{C2H3OH + OH -> HOC2H3OH} & dO & -27.5 (-31.6) & BL \\
\ref{eq:OH1} & \ce{C2H3OH + OH -> HOC2H3OH} & dH & -25.4 (-28.2) & 0.3 (0.5)\\
\ref{eq:OH1} & \ce{C2H3OH + OH -> HOC2H3OH} & Pocket & -22.6 (-27.6) & BL  \\
\ref{eq:OH2} & \ce{C2H3OH + OH -> C2H3(OH)2} & dO & -26.3 (-28.4) & 0.9 (0.4) \\
\ref{eq:OH2} & \ce{C2H3OH + OH -> C2H3(OH)2} & dH & -25.1 (-26.4) & 2.1 (2.6) \\
\ref{eq:OH2} & \ce{C2H3OH + OH -> C2H3(OH)2} & Pocket & -23.5 (-25.5) & 1.4 (1.0)
\enddata
\vspace{-2em}
\end{deluxetable*}

From an analysis of the BE for \ce{C2H3OH} (first item in Table \ref{tab:BE} and top panel of Figure \ref{fig:OHadditionproducts}) we infer that, as expected, the BE strongly depends on the binding geometry (see Figure \ref{fig:BEgraph}). The binding energy we found for \ce{C2H3OH} is on par, but somewhat lower than the one found for \ce{CH3OH} in the literature \citep{Wakelam2017, Ferrero2020}. Taking the study by \cite{Ferrero2020}, the authors found that BE for \ce{CH3OH} can be found between 3770--8618 K as extremal values, and our values fall within this range, only slightly lower for the case of the dO binding site. The similarity in BE between \ce{C2H3OH} and \ce{CH3OH} is expected, as both molecules are very similar. Looking at the geometric arrangements of \ce{C2H3OH} on the different sites, we observe that in the dH sites, the H atom of the OH group from vinyl alcohol points inwards the cluster, facilitating an arrangement of the \ce{H2O} molecules participating in the bond (top-middle panel of Figure \ref{fig:OHadditionproducts}). This effect fosters a unique environment at the dH sites, making them conducive to dangling hydrogen (dH) and dangling oxygen (dO) interactions. In the case of dO we do not observe such an effect, and the original site motif is maintained.  Finally, it is worth remarking that, while on dO and dH sites the adsorption geometry is normal to the surface, on the pocket site the alcohol covers the cavity. 

The energetic parameters of the reactions of OH with \ce{C2H3OH} on the binding sites described above can be found in Table \ref{tab:OHaddition}. In general, we observe that the values of $\Delta H^{\circ, \ddagger}$, which is the most important quantity for reactivity are rather in line with the gas phase values \citep{Ballotta23} both for the formation of the \ce{HOC2H3OH} or \ce{C2H3(OH)2} radicals, e.g., the \ce{H2O} surface only plays a minor catalytic role. However, and because all reactions have very low $\Delta H^{\circ, \ddagger}$ in the gas phase, we find that this minor catalytic role can make some pathways totally barrierless for the reaction \ref{eq:OH1}. This is not the case for the reaction \ref{eq:OH2}, where $\Delta H^{\circ, \ddagger}$ varies in the 0.9--2.1 kcal mol$^{-1}$ range. This finding can be important when considering competitive reaction channels when OH thermally diffuses on the surface, at T$\geq$36 K \citep{miyazaki_direct_2022}, but in this work we are more interested in the role of suprathermal OH, where the internal energy of the radical should easily surpass $\Delta H^{\circ, \ddagger}$. We discuss this fact in more detail in Section \ref{sec:discussion}, and here it suffices to say that there should not be a preferential reaction path when suprathermal OH is considered.

The structural arrangements between the adsorbate and surface produced by the reaction will eventually determine the stereoselectivity of the whole reaction path. Vinyl alcohol is a planar molecule, and so the addition of the OH radical produces a pyramidalization of the given carbon center (C sp$^{2}$ $\rightarrow$ C sp$^{3}$). The 1,2 OH addition on \ce{C2H3OH} in the dO and dH sites leaves the initial H-bonds unchanged, to which the newly formed OH group adds two and one new hydrogen bonds respectively (see the middle panels of Figure \ref{fig:OHadditionproducts}). Restructuring \ce{HOC2H3OH} for the 1,2 addition on the pocket binding site is more complicated. The addition distorts the molecule, so that the initial hydrogen bond bonded by the dH site breaks, and a new dH site hydrogen bond is formed through the newly added OH group. We observe that the two OH groups are staggered and that one hydrogen atom bonded to the sp$^{3}$ carbon is forced to rotate into the pocket. This is essential for the following H-abstraction reaction, as it leads to only (E)-1,2-ethenediol. 

Formation of \ce{C2H3(OH)2} radical (geminal addition to \ce{C2H3OH}) always proceeds with the formation of new hydrogen bonds. In the dO site, we observe a cleavage of the original H-bond to form a new H-bond with a neighbour \ce{H2O} molecule, followed by a restructuring of the dO water site, with the -OH moeity of \ce{C2H3(OH)2} acting both as H-donor and acceptor (Lower left corner of Figure \ref{fig:OHadditionproducts}). Besides, the newly added OH group does not interact with the ice slab. Meanwhile, in the dH and pocket sites, the new OH group interacts with the ice, furnishing one new H-bond in each case. Furthermore, the interactions within the pocket force, as in the \ce{HOC2H3OH} case presented above, a motion in the hydrogen atom bonded to the sp$^{3}$ carbon inwards the cluster, with important consequences for the subsequent stereoselectivity of the H-abstraction. Besides, the torsional angle HCCH is shifted, making the gas phase and the adsorbed-phase minimum energy conformations diverge. This will have further implications in Section \ref{section:Habs}.

%\textbf{REMOVED: Regarding the BE of the newly formed radicals (second and third elements in Table \ref{tab:BE}) we observe, in general, a systematic increase in BE with respect to the \ce{C2H3OH} molecule. We find the explanation of this trend pretty straightforward since, as we mentioned in the previous paragraphs, the addition of OH is associated in most cases with an increase in the number of H-bonds between the adsorbate and the cluster. This explanation is also behind the decrease in BE for \ce{HOC2H3OH} at the pocket site or the milder increase for \ce{C2H3(OH)2} at the dO site, because in both cases we find the breaking of hydrogen bonds already formed (and the formation of new ones) to accommodate the geometry of the radicals produced in the reaction. REMOVED}

\textbf{The variations in the BE of the newly formed radicals are related to the creation or breaking of hydrogen bonds due to the addition of OH as the molecules accommodate on the surface.} In absolute terms, the BE of radicals are on par with other molecules susceptible to forming H bonds with the surface, such as \ce{CH3OH} \citep{Ferrero2020}, \ce{HCOOH} \citep{molpeceres_hydrogen_2022} or \ce{NH2OH} \citep{molpeceres_processing_2023}, e.g. close to the BE of \ce{H2O} on the same substrate \citep{tinacci_theoretical_2023}.

%Vynil alcohol is totally planar and so, the addition of the OH radical produces a pyramidalization of the given carbon center. The 1,2 OH addition on vinyl alcohol in the dO and dH sites leaves the initial cleavages unchanged, though it enables the formation of two and one new hydrogen bonds respectively through the other end of the 1,2-ethenediol radical, accounting for three hydrogen bonds in total in both cases. On the pocket nevertheless, the addition distorts the molecule so that the initial hydrogen bond through the dH breaks and a new dH site hydrogen bond is formed through the newly added OH group. We observe that the two OH groups are staggered and one hydrogen atom bonded to the $sp^{3}$ carbon is forced to get rotated into the pocket. This is essential for the following H-abstraction reaction as it leads only to (E)-1,2-ethenediol.

\subsection{H-abstraction and addition reactions from OH addition products}
\label{section:Habs}

The hydrogen abstraction from OH addition products (Section \ref{section:OHadd}) leads to the exothermal formation of unsaturated diols. From \ce{HOC2H3OH} we may obtain \ce{(E)-HOC2H2OH} and \ce{(Z)-HO2C2H2OH} depending on the binding site. Starting from \ce{C2H3(OH)2}, we can produce only \ce{C2H2(OH)2} (e.g. no stereoisomerism).

A summary of the energies for the reaction with H is shown in Table \ref{tab:Hreact}. From it, we observe that the formation of the 1,1 products is consistently more exothermic than that of the 1,2 products regardless of the binding site. Moreover, H-abstraction reactions on the dO site are more exothermic than those on dH sides. We found all H abstraction and addition reactions barrierless, except 
reaction \ref{eq:H2}, which proceeds with a small barrier of $\Delta H^{\circ, \ddagger}$=1.0 kcal mol$^{-1}$. 

The formation of \ce{(Z)-HOC2H2OH} is possible on binding sites dH and dO, and \ce{(E)-HOC2H2OH} on dO sites and the pocket site as no barrier is expected (see Table \ref{tab:Hreact}). We also observe that the sites where \ce{(Z)-HOC2H2OH} is formed are consistent with those where the binding energies of the precursor \ce{HOC2H3OH} are highest. On the dH site, the strong hydrogen bonds acting on both OH groups in \ce{HOC2H3OH} forces them towards the ice surface throughout the H-abstraction process as observed in Figure \ref{fig:Habstractionproducts}, so that H-abstraction leads to \ce{(Z)-HOC2H2OH} exclusively. The weaker interaction on the dO site means that molecules here will have more freedom to rearrange, yielding both \ce{(Z)-HOC2H2OH} or \ce{(E)-HOC2H2OH}. The case of the pocket is straight-forward as the OH groups in \ce{HOC2H3OH} are already in a staggered position and only the H atom on top may be abstracted leading to the \ce{(Z)-HOC2H2OH}. These facts can be also observed in Figure \ref{fig:snapshots} that shows a close up view of the reactions.

%\textbf{REMOVED: Regarding the binding energies shown on Table \ref{tab:BE}, we do not infer any common trend for species \ce{(Z)-HOC2H2OH} or \ce{(E)-HOC2H2OH}. One should highlight, however, the substantial changes in the BE on dO sites and the pocket from \ce{HOC2H3OH} to \ce{(E)-HOC2H2OH} due to Reaction \ref{eq:H1}. In the fist case, the number of H bonds is maintained though the interaction is much weaker as the new C=C double bond makes \ce{(E)-HOC2H2OH} stiff and thus difficult to be accommodated in the original binding sites, weakening the hydrogen bonds and decreasing the BE (around 4.9 kcal/mol). However, in the pocket a new hydrogen bond on a dH site is formed, consequently making the BE rise (around 7.5 kcal/mol). REMOVED}

We acknowledge that in gas phase \ce{(Z)-HOC2H2OH} can be present as conformer (Z)-(syn,anti) and (Z)-(anti,anti), and \ce{(E)-HOC2H2OH} as (E)-(syn,syn), (E)-(syn,anti) and (E)-(anti,anti) regarding the rotation of the alcohol groups. The energy difference between the first two is in the order of 4.1 kcal mol$^{-1}$, whereas the differences between (E)-(syn,syn), (E)-(syn,anti) and (E)-(syn,anti), (E)-(anti,anti) are 0.5 and 0.2 kcal/mol respectively \citep{kleimeier_identification_2021}. In our study we observed isomer (E)-(syn,anti) on a dO site and (E)-(syn,syn) in the pocket, whereas only the (Z)-(syn,anti) was observed regardless of the binding site (see Figure \ref{fig:Habstractionproducts}). The low energy difference in the conformations of the (E)-isomer makes us consider unimolecular isomerization once released to the gas phase as a possible outcome on astronomical timescales, which can affect the detectability of the molecule. In the case of the (Z)-conformer, the high activation energy precludes this interconversion, making the conformer desorbing from the grain the one to be detectable. This is an important aspect to consider when envisaging new astronomical detections of these unsaturated species. 

Reaction \ref{eq:H3} is barrierless and exothermic in all binding sites as shown in Table \ref{tab:Hreact}. Moreover, there is a generalized increase of the BE comparing those of \ce{C2H3(OH)2} and \ce{C2H2(OH)2} except on the dH site (as seen in Figure \ref{fig:BEgraph}). 

%\textbf{REMOVED: A close examination of Figure \ref{fig:Habstractionproducts} provides an explanation. In the dO site, an intramolecular H-bond that leads to a cooperative effect with the one binding \ce{C2H2(OH)2} and the ice slab is formed. For the pocket, the H-bonds are kept in the same positions, the H-abstraction is produced on a sterically hindered atom in \ce{C2H3(OH)2}, which allows the product of Reaction \ref{eq:H3} to adapt better to the ice pocket. The decrease of the BE in the dH site, as previously discussed in the context of Reaction \ref{eq:H1}, is attributed to the inherent rigidity of the C=C bond within \ce{C2H2(OH)2}. This stiffness affects the molecular structure, hindering the accommodation of \ce{(E)-HOC2H2OH} at its original binding sites. Consequently, even though the number of hydrogen bonds is maintained, the interaction becomes significantly weaker. REMOVED}

%\subsection{H-addition reactions of OH addition products} \label{section:Hadd}

The hydrogen addition to \ce{HOC2H3OH} and \ce{C2H3(OH)2} takes place at the formal sp$^{2}$ carbon atom, yielding \ce{HOC2H4OH}, \textbf{ethylene glycol}, and \ce{C2H4(OH)2}, respectively. All reactions are exothermic and barrierless. Reaction \ref{eq:H5} is more exothermic than Reaction \ref{eq:H4} in all binding sites. As we mentioned in the previous subsection, Reaction \ref{eq:H3} is also more exothermic than Reactions \ref{eq:H1} or \ref{eq:H2}. Moreover, in the general picture we observe very small energy differences in the different binding sites, with slightly higher values in dH sites.

%\textbf{REMOVED: The binding energies of species \ce{HOC2H4OH} and \ce{C2H4(OH)2} are higher compared to those of \ce{HOC2H3OH} and \ce{C2H3(OH)2} in all binding sites. Also, the number of hydrogen bonds and the binding site has been conserved during the H-addition reactions as presented in Figures \ref{fig:BEgraph} and \ref{fig:Hadditionproducts}. The rise in the binding energies corresponding to Reaction \ref{eq:H4} and \ref{eq:H5} is on par with a structure change of the adsorbate, so that they stick better to the surface without creating any new hydrogen bond. In the pocket binding site, we observe that the most stable configuration of \ce{HOC2H4OH} as adsorbate and in the gas phase is different. Our binding energies are calculated using as reference the lowest-energy species in the gas phase. Thus, the BE on the surface will also recover the effects from the torsional change between gas phase and adsorbed molecules. REMOVED}

\begin{deluxetable*}{clc|cc}[bt!]
\label{tab:Hreact}
\tablecaption{Summary of the reaction energies, $\Delta H^{\circ, \text{R}}$, and activation energies, $\Delta H^{\circ, \ddagger}$ for the second hydrogenation step studied in this work. Energies are ZPVE corrected. In parentheses, electronic energies, e.g. without ZPVE. All values are reported in kcal mol$^{-1}$. BL means barrierless.}
\tablehead{
\colhead{ Label } &
\colhead{ Reaction        } &
\colhead{ Binding Site } &
\colhead{ $\Delta H^{\circ, \text{R}}$ } &
\colhead{ $\Delta H^{\circ, \ddagger}$ }
}
\startdata
\ref{eq:H1} & \ce{HOC2H3OH + H -> (E)-HOC2H2OH + H2} & dO & -60.7 (-61.2) & BL (0.6)\\
\ref{eq:H2} & \ce{HOC2H3OH + H -> (Z)-HOC2H2OH + H2} & dO & -66.2 (-66.8) & BL\\
\ref{eq:H4} & \ce{HOC2H3OH + H -> HOC2H4OH} & dO & -84.5 (-94.2) & BL\\
\ref{eq:H3} & \ce{C2H3(OH)2 + H -> C2H2(OH)2 + H2} & dO & -82.4 (-80.8) & BL\\
\ref{eq:H5} & \ce{C2H3(OH)2 + H -> C2H4(OH)2} & dO & -97.7 (-107.3) & BL\\
\ref{eq:H2} & \ce{HOC2H3OH + H -> (Z)-HOC2H2OH + H2} & dH & -61.0 (-62.0) & 1.0 (0.9)\\
\ref{eq:H4} & \ce{HOC2H3OH + H -> HOC2H4OH} & dH & -85.5 (-94.4) & BL\\
\ref{eq:H3} & \ce{C2H3(OH)2 + H -> C2H2(OH)2 + H2} & dH & -72.1 (-72.5) & BL (0.5)\\
\ref{eq:H5} & \ce{C2H3(OH)2 + H -> C2H4(OH)2} & dH & -99.6 (-108.0) & BL\\
\ref{eq:H1} & \ce{HOC2H3OH + H -> (E)-HOC2H2OH + H2}$^{*}$ & Pocket & -64.8 (-64.0) & BL (1.4)\\
\ref{eq:H4} & \ce{HOC2H3OH + H -> HOC2H4OH}$^{*}$ & Pocket & -85.1 (-93.8) &  BL\\
\ref{eq:H3} & \ce{C2H3(OH)2 + H -> C2H2(OH)2 + H2}$^{*}$ & Pocket & -73.7 (-75.5) & BL\\
\ref{eq:H4} & \ce{C2H3(OH)2 + H -> C2H4(OH)2}$^{*}$ & Pocket & -98.3 (-107.3) & BL\\
\enddata
\begin{minipage}{\linewidth}
\centering
\vspace{0.5em}
\footnotesize
(*) Ice slab with fixed atoms. See Section \ref{sec:methods} (**) \textbf{Computed with broken-symmetry DFT. A larger uncertainty in $\Delta H^{\circ, \text{R}}$ and $\Delta H^{\circ, \ddagger}$ is expected with respect reactions \ref{eq:OH1} and \ref{eq:OH2}}
\end{minipage}
\end{deluxetable*}

\section{Discussion} \label{sec:discussion}

\subsection{Isomerism} \label{sec:isomerism}

Our calculations show that the formation of \ce{(Z)-HOC2H3OH} from \ce{C2H3OH} is indeed \textbf{feasible} through the action of OH and H radicals on the surface of interstellar dust grains, and that the barriers that need to be overcome are, in all cases, lower than the corresponding thermal diffusion barriers of the reactants involved.\footnote{The diffusion energies of \ce{C2H3OH} and related species are not known, but in light of the high $\Delta H^{\circ}_\text{bin}$ listed in Table \ref{tab:BE}, it is safe to assume that they will be immobile.} \textbf{We thus expect the formation of isomers (E) and \ce{(Z)-HOC2H2OH} as well as the H-addition products (\ce{HOC2H4OH}), the saturated and unsaturated geminal diols (e.g. \ce{R-(OH)2}})

\textbf{The most important point of our work is validating a postulated chemical reaction pathway for the formation \ce{(Z)-HOC2H3OH}, as its recent detection \citep{rivilla_precursors_2022} prompted some questions around chemical isomerism in space. Quantum tunneling mediated unimolecular interconversion has recently been able to explain the thermodynamic isomeric ratio of imines \citep{garcia_de_la_concepcion_origin_2021}, though failing to explain the case of formic acid (HCOOH) \citep{GarciadelaConcepcion2022}. The case of \ce{(E)-HOC2H2OH <=> (Z)-HOC2H2OH} requires an unfeasible torsion through a double bond, making this mechanism inoperative though leaving the door open to photon mediated processes \citep{cuadrado_trans-cis_2016} or chemical conversion in gas or grains \citep{Shingledecker2020, molpeceres_hydrogen_2022, garcia_de_la_concepcion_sequential_2023}}

\textbf{As we already mentioned, the formation of \ce{(Z)-HOC2H2OH} is possible but further reactions are competitive. Hence, a more detailed discussion is required in order to facilitate subsequent astronomical searches. The occurrence of one or another reaction depends mostly on the binding site on ASW, that, as we found in earlier works \citep{Molpeceres2021b, molpeceres_hydrogen_2022}, modulates the stereochemistry based on the H-bond topology of the adsorbate (\ce{C2H3OH} as a precursor in this case), with the surface. Another plausible competitive reaction would be the hydrogenation of \ce{C2H3OH} leading to the formation of ethanol. This was previously studied by \citep{Perrero2022} in a very similar fashion to that of our study by employing ASW clusters. Both possible additions, either on the aliphatic carbon or the OH bearing one, were reported to proceed with small barriers that could be overcome by quantum tunneling. These reaction will compete with the first step of our proposed scheme, and due to the flux of H atoms impinging on interstellar ASW, possibly dominate. This is coherent with the expected abundances of ethanol, an ice molecule recently detected by JWST observations \citep{rocha_jwst_2024}, that should be much more abundant than our title molecule. The presence of competitive hydrogenation routes at low temperatures, however, does not preclude our proposed route. Our route at low temperatures is also enhanced by H accretion on ices, for example through a three body mechanism where a suprathermal OH radical can diffuse and react with a nearby \ce{C2H3OH} molecule \citep{Garrod2011, Jin2020, Garrod2022}, with evidence for this process to occur in the presence of available hydrogenation channels \citep{Chang2018, Ishibashi2021}.}

We emphasize again that our discussion so far pertains to the chemistry at low temperatures (e.g. 10 K) where the OH radical participating in the first reaction \ce{C2H3OH + OH}, is excited by different mechanisms, such as, for example, water photolysis \citep{Ishibashi2021} or transient heating after a reactive event \ce{O + H -> OH$^{*}$} \citep[see, for example][]{Garrod2011}. Because of the attenuation of the primary \textbf{interstellar} UV field, photolysis is not expected to contribute a lot to the generation of OH radicals in molecular clouds. Making this physical regime suitable for the reactions studied in this work. Assuming that the kinetic barriers found in Table \ref{tab:OHaddition} can be overcome suprathermally as well, we see that all products of reactions \ref{eq:OH1} and \ref{eq:OH2} must be equally possible. We do not observe any preferential chemistry, because suprathermal rate constants are barrier and temperature independent, see Equation \ref{eq:suprathermal} and \citep{shingledecker_cosmic-ray-driven_2018}. A posterior reaction with H also shows no or minute barriers (easily passed \emph{via} quantum tunneling) for concluding the reaction (reactions \ref{eq:H1}--\ref{eq:H5}; Table \ref{tab:Hreact}). Although in this case we do observe preferential stereochemistry based on the H-bond topology of the preadsorbed reactant, a mixture of all products of reaction shown in reactions \ref{eq:H1}--\ref{eq:H5} can be possible, with specific abundances depending on the distribution of binding sites on the surface. 

A second regime starts when two conditions are satisfied. First, H atoms population on the surface is negligible due to thermal evaporation (at $\sim$18 K) and OH radicals start to be mobile (36 K, \cite{miyazaki_direct_2022}). The evaporation of H atoms inhibits the formation of suprathermal OH$^{*}$ and reaction \ref{eq:H1}--\ref{eq:H5}, making thermal diffusion the limiting step for the reaction and H-abstraction reactions to proceed \emph{via} alternative reactants (i.e. OH, \ce{NH2}; not considered in this work). Under this regime, we expect that the barrierless or low barrier channel to form the 1,2 product (Reaction \ref{eq:OH1}) dominates. This second regime is \textit{temperature dependent}, because both reactions \ref{eq:OH1} and \ref{eq:OH2} now follow Equation \ref{eq:thermal}. 

In light of this discussion, we consider that, in the absence a direct isomerization mechanism in the gas, that is, \cite{cuadrado_trans-cis_2016}, \cite{garcia_de_la_concepcion_origin_2021} or \cite{garcia_de_la_concepcion_sequential_2023} or a clearly favored reaction route for the stereochemistry of the reaction \citep{Molpeceres2021b} it is very complicated to determine the isomerism of a gray case like the one tackled in this article. The specific isomeric excess depends on the binding site and the temperature window. This is the second important point extracted from this work. Sophisticated chemical modelling efforts are required to quantify the importance of every isomer, structural and spatial, in the total scheme of the reaction. We postpone such a study for future works. A similar speciation of possible isomers was found in the electron irradiation experiments of \citet{kleimeier_identification_2021}, with the inclusion of even further products, e.g. 1-3-oxyranol, due to the high energy dose inoculated in their experiments.

\subsection{Astrochemical rationale: New molecules proposed for detection} \label{sec:astro}

So far, the molecule under scrutiny in this work, (Z)-\ce{HOC2H2OH} has been detected only in the giant molecular cloud G+0.693-0.027, at the center of our galaxy, contrary to the lower energy isomer, glycolaldehyde, that has been detected in more than one source (\cite{coutens14,Jorgensen_2012,Beltran_2009, hollis00}). In their work, \citet{rivilla_precursors_2022} suggest five different routes for the formation of (Z)-\ce{HOC2H2OH}, namely the \ce{H2CO}/\ce{CH2OH} pair, dimerization of hydroxycarbene (HCOH), hydrogenation of glyoxal \ce{HCOCHO}, enolisation of glycolaldehyde (\ce{HCOCH2OH}) and the route that we considered in this work, hydroxylation of vinyl alcohol (\ce{C2H3OH}) followed by a hydrogen abstraction reaction. From all the possible reaction routes, we believe that the one considered in this work is the most plausible one, owing to chemical considerations affecting the alternatives that we enumerate below.

Beginning with the dimerization of HCOH, we recently showed that this species undergoes quick conversion to \ce{H2CO} on polar ices through efficient proton relay mechanisms \citep{Molpeceres2021carbon}. While we found this phenomenon to be binding site dependent, we find it difficult to consider that, given this reactivity, to have two close HCOH molecules together on a dust grain under ISM conditions. By contrast, reactions of HCOH in the gas phase, where reaction with the ice mantle is not possible, are reported to be crucial in the synthesis of sugars \citep{eckhardt_gas-phase_2018}. Second, the \ce{HCOCHO + 2 H} route was already studied in the literature, finding significantly easier hydrogenation at the C position towards the \ce{H2COCHO} radical \citep{Alvarez-Barcia2018}, precluding the formation of \ce{HOC2H2OH}. Third, the enolisation of glycolaldehyde is a unimolecular activated and endothermic process, not feasible under ISM conditions. Therefore, only \ce{H2CO + CH2OH + 2 H} and our proposed route \ce{C2H3OH + OH + H} remain viable candidates. We carried out additional gas phase calculations (at the same MPWB1K-D3(BJ)/def2-TZVP level) on the \ce{H2CO + CH2OH} reaction, finding a barrier of $\sim$ 5.0 kcal mol$^{-1}$, i.e. a sizable barrier. Following the arguments given in this article, such a barrier might be overcome suprathermally, although in this case suprathermal diffusion is a competitive process. Comparing \ce{H2CO} (or \ce{CH2OH}) with OH, the higher number of degrees of freedom of these polyatomic molecules does not allow for similarly efficient excitation of translational degrees of freedom. In fact, while \citet{Ishibashi2021} reported the presence of suprathermal OH, a similar behavior has not been found for bigger molecules. 

Overall, based on pure chemical arguments, our route is the most promising one, although \ce{C2H3OH} is the least abundant of the \ce{C2H4O} family of molecules, at least in Sgr-B2 \citep{belloche_complex_2013}, IRAS 16293-2422 \citep{lykke_alma-pils_2017} or astrochemical models \citep{Garrod2022}. However, recent theoretical calculations show new and efficient routes for the formation of vynil alcohol \citep{Perrero2022, molpeceres_radical_2022} summed to studies addressing the limited resilience of \ce{C2H3OH} in the gas phase \citep{Ballotta23} suggest that the abundance of \ce{C2H3OH} in grains can be larger than expected, serving as feedstock to more complex prebiotic COMs. In this work, we anticipate some of these COMs as not only (Z)-\ce{HOC2H2OH}, but also (E)-\ce{HOC2H2OH}, \ce{HOC2H4OH}, \ce{C2H4(OH)2} and \ce{C2H2(OH)2}. In the absence of destruction routes operating in these molecules, that are beyond the scope of this work, we expect these molecules to be detectable in the ISM and encourage laboratory work informing about the necessary spectroscopic parameters conducive to their detection.

Finally, it remains to understand why (Z)-\ce{HOC2H2OH} has so far only been detected in G+0.693-0.027. Although giving a definitive explanation is extremely difficult, we believe that the distinctive chemistry operative in this astronomical object is a very important factor. So far, the chemistry of G+0.693-0.027 hints to an important contribution of surface chemistry to the molecular inventory of the cloud, not only with (Z)-\ce{HOC2H2OH} but also with i.e. with HC(O)SH \citep{Rodriguez-Almeida2021}, \ce{NH2OH} \citep{Rivilla2020} or very recently \ce{HOCOOH} \citep{sanz-novo_discovery_2023} or \ce{HNSO} \citep{sanz-novo_discovery_2024}. Some of these products' chemistry can be explained by hydrogenations \citep{Molpeceres2021b, molpeceres_processing_2023} but some others necessitate of the diffusion of heavier radicals to proceed. While some of these radicals, like the N or O atoms might diffuse on ASW at $\sim$10-15 K \citep{Minissale2013, Minissale2016a, Zaverkin2021} others are immobile (OH as an example in this work) and non-thermal diffusion and reaction needs to be invoked. G+0.693-0.027, with an extremely high flux of cosmic rays \citep{Goto2013} or cloud-cloud collisions \citep{zeng_cloudcloud_2020} provide the necessary energy to put these energetic phenomena at the forefront of the observed chemical complexity.  Further investigation of unconventional chemical routes that can operate in this source is in our immediate plans. The particular conditions of G+0.693-0.027 are not only ideal for our proposed chemical route, but also for the radiolytic route presented of \citet{kleimeier_identification_2021}, reinforcing the view of G+0.693-0.027 as a factory of prebiotic COMs.

\section{Conclusions}

The current study sheds light on the reactivity of vinyl alcohol on ASW surfaces and its role on the synthesis of the sugar precursor, \ce{(Z)-HOC2H2OH}, highlighting the effect of the binding configurations on the stereoselectivity of the reactions. Our quantum chemical calculations show that all the reactions participating in the chemical scheme under consideration are barrierless or possess rather low barriers, with a couple of exceptions on particular binding sites. Even for exceptions, all activation energies are lower than the diffusion barrier for the OH radical, hence confirming that suprathermal OH must readily react with \ce{C2H3OH}. Our simulations also provide valuable information on the BE of all the species involved in ourt reaction scheme, as well as variations during the reaction and reaction energies. These quantities can be directly included into astrochemical models looking to reproduce the abundances of (Z)-\ce{HOC2H2OH} in interstellar environments, an endeavor that we will pursue with further, dedicated works.

%All sites enable H-addition reactions (\ref{eq:H4} and \ref{eq:H5}) as well as reaction (\ref{eq:H3}). These processes were found exothermic and barrierless. However, as explained in section \ref{section:Habs}, reactions \ref{eq:H1} and \ref{eq:H2} are specific to some adsorption sites and always exothermal. \ce{(Z)-HOC2H2OH} can be only obtained in dO and dH sites. In the first position, the process proceeds barrierles, whereas on the second one, a small barrier is found. Reaction \ref{eq:H2} occurs only in dO sites and the pocket and is also barrierless for both. 

%This set of reactions produces the formation and breakage of different hydrogen bonds that bind the molecular species to the ASW. Thus, we observe changes on the binding energies (BE) as exposed in Figure \ref{tab:BE} during the course of the reactions.

In addition to the finding of kinetic barriers or lack thereof for our postulated chemical scheme, our work also focuses on the isomerism of the considered reactions. Through the analysis of the binding of the reactants to the surface, we are able to determine that the formation of (Z)-\ce{\ce{HOC2H2OH}} is not isomerically pure. In fact, the extreme reactivity exhibited by all the reactants and intermediates considered makes us conclude that all possible products of the reaction, involving stereoisomers (e.g. (E)-\ce{HOC2H2OH}), structural isomers (\ce{C2H2(OH)2}) or addition products (molecules with formula \ce{C2H6O2}). We suggest that these molecules are good candidates for spectroscopic and interstellar search, at least in sources with a low dust temperature (10-18 K, See section \ref{sec:isomerism})) 

In conclusion, our research offers extensive data on the formation of sugar precursors in molecular clouds. We underscore the importance of the molecular interactions with the ASW surface as they dictate the selectivity in the reactions. Likewise, we encourage further experimental, theoretical, and modeling studies to keep growing the astrochemical knowledge on reactions with the OH radical, once its mobility has been proven due to non-thermal effects \citep{Ishibashi2021}.

%% Also note that the akcnowlodgment environment does not support long amounts of text. If you have a lot of people and institutions to acknowledge, do not use this command. Instead, create a new \section{Acknowledgments}.
\begin{acknowledgments}
This work was funded by Deutsche Forschungsgemeinschaft (DFG, German Research Foundation) under Germany's Excellence Strategy - EXC 2075 – 390740016. We acknowledge the support by the Stuttgart Center for Simulation Science (SimTech). G.M acknowledges the support of the grant RYC2022-035442-I funded by MCIN/AEI/10.13039/501100011033 and ESF+. G.M. also received support from the Japan Society for the Promotion of Science through its international fellow program (Grant: P22013) and the Grant-in-aid JP22F22013. P.R thanks the financial support from the Spanish Ministerio de Ciencia e Innovaci\'on (PID2020-117742GB-I00).
The authors acknowledge the support of the state of Baden-Württemberg through bwHPC and the German Research Foundation (DFG) through grant no INST 40/575-1 FUGG (JUSTUS 2 cluster) and the Research Center for Computational Science, Okazaki, Japan (Projects: 22-IMS-C301, 23-IMS-C128). G.M. also received support from project 20245AT016 (Proyectos Intramurales CSIC).
\end{acknowledgments}


\vspace{5mm}
\facilities{bwHPC Justus Cluster (DE), RCCS (JP)}


\software{ Orca, v.5.0.4 \citep{Neese2020} (Quantum chemical calculations), VMD, v.1.9.4 \citep{HUMP96,STON1998} (Visualization and figure preparation), Chemcraft (Chemical structure manipulation and analysis) and Matplotlib v.3.6.3 \citep{thomas_a_caswell_2023_7527665} (Article's plot preparation)}



\appendix

\section{Benchmark of DFT methods} \label{sec:appendix_methods}

Tables \ref{tab:synnormal} and \ref{tab:simgem} show the activation energies of the 1,2 and 1,1 OH-addition to vinyl alcohol for the syn isomer (syn-anti) in gas phase using different DFT functionals (reactions \ref{eq:OH1} and \ref{eq:OH2}). These were obtained after a full geometry optimization and corrected using the zero-point energy. We benchmarked our methods referencing the results for the same reaction in \citep{Ballotta23}. In their work, gas phase reactions on both isomers were performed. In this work, for computational saving reasons, we restrict our research to the case of syn-\ce{C2H3OH}. 

\begin{deluxetable}{l c}[bt!]
\label{tab:synnormal}
\tablecaption{Activation Energies $\Delta H^{\circ, \ddagger}$ for the OH addition to \ce{C2H3OH} leading to \ce{HOC2H3OH} using different methods (kcal mol$^{-1}$).}
\tablehead{
\colhead{Method} &
\colhead{ $\Delta H^{\circ, \ddagger}$ }
}
\startdata
Reference \citep{Ballotta23} & 1.0 \\
BB1K/def2-TZVP & 0.5 \\
BHLYP-D3(BJ)/def2-TZVP & 2.7 \\
DSD-PBEP86-D3(BJ)/def2-TZVP & 2.8 \\
M06-2X-D3/def2-TZVP & 1.6 \\
M08-HX/def2-TZVP & 1.1 \\
MPWB1K-D3(BJ)/def2-TZVP & 0.6 \\
MPWKCIS1K/def2-TZVP & 0.6 \\
$\omega$B97M-V/def2-TZVP & 1.4 \\
\enddata
\end{deluxetable}

\begin{deluxetable}{l c}[bt!]
\label{tab:simgem}
\tablecaption{Activation Energies $\Delta H^{\circ, \ddagger}$ for the OH addition to \ce{C2H3OH} leading to \ce{C2H3(OH)2} using different methods (kcal mol$^{-1}$).}
\tablehead{
\colhead{Method} &
\colhead{ $\Delta H^{\circ, \ddagger}$ }
}
\startdata
Reference \citep{Ballotta23} & 1.9 \\
BB1K/def2-TZVP & 1.8 \\
BHLYP-D3(BJ)/def2-TZVP & 4.1 \\
DSD-PBEP86-D3(BJ)/def2-TZVP & 3.9 \\
M06-2X-D3/def2-TZVP & 3.4 \\
M08-HX/def2-TZVP & 2.6 \\
MPWB1K-D3(BJ)/def2-TZVP & 1.8 \\
MPWKCIS1K/def2-TZVP & 2.3 \\
$\omega$B97M-V/def2-TZVP & 1.2 \\
\enddata
\end{deluxetable}

In \citep{Ballotta23} calculations are performed using Truhlar's calendar basis set Jun-cc-pVTZ \citep{papajak11} which adds diffuse functions to the cc-pVTZ basis except to H, He and removes the highest angular momentum diffuse function from all other atoms. We opted for the 3$\zeta$ basis set def2-TZVP \citep{Weigend2005}. We use the def2/J \citep{Weigend_2006} and def2-TZVP/c (in the case of double hybrids, see below) auxiliary basis set for the RI-approximation \citep{Dunlap_Connolly_Sabin_1979,Whitten_1973}. 

Functionals were chosen to cover several existing families. We employed functionals designed for main-group thermochemistry, thermochemical kinetics and non-covalent interactions as well as more advanced hybrid functional approaches. The set is composed by BB1K \citep{Zhao_Lynch_Truhlar_2004}, BHLYP \citep{Becke_1993BHLYP}, DSD-PBEP86 \citep{Kozuch_Martin_2011}, M06-2X \citep{Zhao2007}, M08-HX \citep{Zhao_Truhlar_2008}, MPWKCIS1K \citep{Zhao_González-García_Truhlar_2005}, MPWB1K \citep{Zhao_Truhlar_2004MPWB1K}, $\omega$B97M-V \citep{Mardirossian2016}. We used dispersion corrections for functionals where such correction was parametrized, namely D3 \citep{Grimme2010}, D3(BJ) \citep{Grimme_Ehrlich_Goerigk_2011} or the correction by non-local correlation (NL; \cite{Vydrov2010}, included in $\omega$B97M-V). The specific functionals for which we apply either correction or no correction at all can be consulted in Tables \ref{tab:synnormal} and \ref{tab:simgem}.

Considering the size of the ASW clusters, we chose the functional with the best cost/accuracy ratio. We discarded BHLYP-D3(BJ), DSD-PBEP86-D3(BJ) and M062-2X-D3 according to the constant overestimation of the barriers of both reactions. From the remaining functionals, we chose MPWB1K-D3(BJ)/def2-TZVP as the method for this study due to its good prediction for the barrier in Table \ref{tab:simgem}. Though it may exhibit a relatively small error compared to the reference in Table \ref{tab:synnormal}, we considered it acceptable for the specific requirements of our study. The computational efficiency makes it overall a well-balanced choice considering the manageable computational cost associated with this method within our system.

%\section{Calculation of molecular properties for gas phase reactivity}

%\textcolor{red}{A rellenar por mi cuando esten los modelos}

%\section{Miscellaneous reactions added to the chemical reaction network}

%\textcolor{red}{A rellenar por mi cuando esten los modelos}


\bibliography{sample631}{}

\begin{thebibliography}{}
\expandafter\ifx\csname natexlab\endcsname\relax\def\natexlab#1{#1}\fi
\providecommand{\url}[1]{\href{#1}{#1}}
\providecommand{\dodoi}[1]{doi:~\href{http://doi.org/#1}{\nolinkurl{#1}}}
\providecommand{\doeprint}[1]{\href{http://ascl.net/#1}{\nolinkurl{http://ascl.net/#1}}}
\providecommand{\doarXiv}[1]{\href{https://arxiv.org/abs/#1}{\nolinkurl{https://arxiv.org/abs/#1}}}

\bibitem[{Agúndez {et~al.}(2021)Agúndez, Marcelino, Tercero, Cabezas,
  De~Vicente, \& Cernicharo}]{agundez_o-bearing_2021}
Agúndez, M., Marcelino, N., Tercero, B., {et~al.} 2021, Astronomy \&
  Astrophysics, 649, L4, \dodoi{10.1051/0004-6361/202140978}

\bibitem[{Andrés {et~al.}(2024)Andrés, Rivilla, Colzi, Jiménez-Serra,
  Martín-Pintado, Megías, López-Gallifa, Martínez-Henares, Massalkhi, Zeng,
  Sanz-Novo, Tercero, de~Vicente, Martín, Requena-Torres, Molpeceres, \& de~la
  Concepción}]{andres_first_2024}
Andrés, D.~S., Rivilla, V.~M., Colzi, L., {et~al.} 2024,
  \dodoi{10.48550/ARXIV.2404.03334}

\bibitem[{Ballotta {et~al.}(2023)Ballotta, Martínez-Núñez, Rampino, \&
  Barone}]{Ballotta23}
Ballotta, B., Martínez-Núñez, E., Rampino, S., \& Barone, V. 2023, ACS Earth
  and Space Chemistry, 7, 1467–1477,
  \dodoi{10.1021/acsearthspacechem.3c00110}

\bibitem[{Becke(1993)}]{Becke_1993BHLYP}
Becke, A.~D. 1993, The Journal of Chemical Physics, 98, 1372–1377,
  \dodoi{10.1063/1.464304}

\bibitem[{Belloche {et~al.}(2019)Belloche, Garrod, Müller, Menten, Medvedev,
  Thomas, \& Kisiel}]{belloche_re-exploring_2019}
Belloche, A., Garrod, R.~T., Müller, H. S.~P., {et~al.} 2019, Astronomy \&
  Astrophysics, 628, A10, \dodoi{10.1051/0004-6361/201935428}

\bibitem[{Belloche {et~al.}(2013)Belloche, Müller, Menten, Schilke, \&
  Comito}]{belloche_complex_2013}
Belloche, A., Müller, H. S.~P., Menten, K.~M., Schilke, P., \& Comito, C.
  2013, Astronomy \& Astrophysics, 559, A47,
  \dodoi{10.1051/0004-6361/201321096}

\bibitem[{{Belloche, A.} {et~al.}(2008){Belloche, A.}, {Menten, K. M.},
  {Comito, C.}, {Müller, H. S. P.}, {Schilke, P.}, {Ott, J.}, {Thorwirth, S.},
  \& {Hieret, C.}}]{belloche08}
{Belloche, A.}, {Menten, K. M.}, {Comito, C.}, {et~al.} 2008, A\&A, 492,
  769–773, \dodoi{10.1051/0004-6361:20079203e}

\bibitem[{Beltrán {et~al.}(2008)Beltrán, Codella, Viti, Neri, \&
  Cesaroni}]{Beltran_2009}
Beltrán, M.~T., Codella, C., Viti, S., Neri, R., \& Cesaroni, R. 2008, The
  Astrophysical Journal, 690, L93, \dodoi{10.1088/0004-637X/690/2/L93}

\bibitem[{Butlerow(1861)}]{butlerow_bildung_1861}
Butlerow, A. 1861, Justus Liebigs Annalen der Chemie, 120, 295,
  \dodoi{10.1002/jlac.18611200308}

\bibitem[{Caselli {et~al.}(1999)Caselli, Walmsley, Tafalla, Dore, \&
  Myers}]{Caselli1999}
Caselli, P., Walmsley, C.~M., Tafalla, M., Dore, L., \& Myers, P.~C. 1999, The
  Astrophysical Journal, 523, L165, \dodoi{10.1086/312280}

\bibitem[{Caswell {et~al.}(2023)Caswell, Lee, Droettboom, de~Andrade, Hoffmann,
  Klymak, Hunter, Firing, Stansby, Varoquaux, Nielsen, Root, May, Elson,
  Seppänen, Lee, Dale, Gustafsson, hannah, McDougall, Straw, Hobson, Lucas,
  Gohlke, Vincent, Yu, Ma, Silvester, Moad, \&
  Kniazev}]{thomas_a_caswell_2023_7527665}
Caswell, T.~A., Lee, A., Droettboom, M., {et~al.} 2023, matplotlib/matplotlib:
  REL: v3.6.3, v3.6.3,  Zenodo, \dodoi{10.5281/zenodo.7527665}

\bibitem[{Cernicharo {et~al.}(2021)Cernicharo, Agúndez, Cabezas, Tercero,
  Marcelino, Pardo, \& De~Vicente}]{Cernicharo2021}
Cernicharo, J., Agúndez, M., Cabezas, C., {et~al.} 2021, Astronomy and
  Astrophysics, 649, \dodoi{10.1051/0004-6361/202141156}

\bibitem[{Cernicharo {et~al.}(2022)Cernicharo, Fuentetaja, Agúndez, Kaiser,
  Cabezas, Marcelino, Tercero, Pardo, \& de~Vicente}]{fulvenallene}
Cernicharo, J., Fuentetaja, R., Agúndez, M., {et~al.} 2022, Astronomy \&
  Astrophysics, 663, L9, \dodoi{10.1051/0004-6361/202244399}

\bibitem[{{Chang} \& {Herbst}(2016)}]{Chang2018}
{Chang}, Q., \& {Herbst}, E. 2016, \apj, 819, 145,
  \dodoi{10.3847/0004-637X/819/2/145}

\bibitem[{{Coutens, A.} {et~al.}(2015){Coutens, A.}, {Persson, M. V.},
  {Jørgensen, J. K.}, {Wampfler, S. F.}, \& {Lykke, J. M.}}]{coutens14}
{Coutens, A.}, {Persson, M. V.}, {Jørgensen, J. K.}, {Wampfler, S. F.}, \&
  {Lykke, J. M.} 2015, A\&A, 576, A5, \dodoi{10.1051/0004-6361/201425484}

\bibitem[{Cuadrado {et~al.}(2016)Cuadrado, Goicoechea, Roncero, Aguado,
  Tercero, \& Cernicharo}]{cuadrado_trans-cis_2016}
Cuadrado, S., Goicoechea, J.~R., Roncero, O., {et~al.} 2016, Astronomy \&
  Astrophysics, 596, L1, \dodoi{10.1051/0004-6361/201629913}

\bibitem[{Dunlap {et~al.}(1979)Dunlap, Connolly, \&
  Sabin}]{Dunlap_Connolly_Sabin_1979}
Dunlap, B.~I., Connolly, J.~W., \& Sabin, J.~R. 1979, The Journal of Chemical
  Physics, 71, 3396–3402, \dodoi{10.1063/1.438728}

\bibitem[{Eckhardt {et~al.}(2018)Eckhardt, Linden, Wende, Bernhardt, \&
  Schreiner}]{eckhardt_gas-phase_2018}
Eckhardt, A.~K., Linden, M.~M., Wende, R.~C., Bernhardt, B., \& Schreiner,
  P.~R. 2018, Nature Chemistry, 10, 1141, \dodoi{10.1038/s41557-018-0128-2}

\bibitem[{Etim {et~al.}(2017)Etim, Gorai, Das, \&
  Arunan}]{Etim_Gorai_Das_Arunan_2017}
Etim, E.~E., Gorai, P., Das, A., \& Arunan, E. 2017, Astrophysics and Space
  Science, 363, \dodoi{10.1007/s10509-017-3226-5}

\bibitem[{Ferrero {et~al.}(2023)Ferrero, Pantaleone, Ceccarelli, Ugliengo,
  Sodupe, \& Rimola}]{ferrero_where_2023}
Ferrero, S., Pantaleone, S., Ceccarelli, C., {et~al.} 2023, The Astrophysical
  Journal, 944, 142, \dodoi{10.3847/1538-4357/acae8e}

\bibitem[{Ferrero {et~al.}(2020)Ferrero, Zamirri, Ceccarelli, Witzel, Rimola,
  \& Ugliengo}]{Ferrero2020}
Ferrero, S., Zamirri, L., Ceccarelli, C., {et~al.} 2020, ApJ, 904, 11,
  \dodoi{10.3847/1538-4357/abb953}

\bibitem[{García De La~Concepción {et~al.}(2021)García De La~Concepción,
  Jiménez-Serra, Carlos~Corchado, Rivilla, \&
  Martín-Pintado}]{garcia_de_la_concepcion_origin_2021}
García De La~Concepción, J., Jiménez-Serra, I., Carlos~Corchado, J.,
  Rivilla, V.~M., \& Martín-Pintado, J. 2021, The Astrophysical Journal
  Letters, 912, L6, \dodoi{10.3847/2041-8213/abf650}

\bibitem[{García De La~Concepción {et~al.}(2023)García De La~Concepción,
  Jiménez-Serra, Corchado, Molpeceres, Martínez-Henares, Rivilla, Colzi, \&
  Martín-Pintado}]{garcia_de_la_concepcion_sequential_2023}
García De La~Concepción, J., Jiménez-Serra, I., Corchado, J.~C., {et~al.}
  2023, Astronomy \& Astrophysics, 675, A109,
  \dodoi{10.1051/0004-6361/202243966}

\bibitem[{García de~la Concepción {et~al.}(2022)García de~la Concepción,
  Colzi, Jiménez-Serra, Molpeceres, Corchado, Rivilla, Martín-Pintado,
  Beltrán, \& Mininni}]{GarciadelaConcepcion2022}
García de~la Concepción, J., Colzi, L., Jiménez-Serra, I., {et~al.} 2022,
  Astronomy \& Astrophysics, 658, A150, \dodoi{10.1051/0004-6361/202142287}

\bibitem[{Garrod {et~al.}(2022)Garrod, Jin, Matis, Jones, Willis, \&
  Herbst}]{Garrod2022}
Garrod, R.~T., Jin, M., Matis, K.~A., {et~al.} 2022, The Astrophysical Journal
  Supplement Series, 259, 1, \dodoi{10.3847/1538-4365/ac3131}

\bibitem[{Garrod \& Pauly(2011)}]{Garrod2011}
Garrod, R.~T., \& Pauly, T. 2011, Astrophysical Journal, 735, 15,
  \dodoi{10.1088/0004-637X/735/1/15}

\bibitem[{Goto(2014)}]{Goto2013}
Goto, M. 2014, Proceedings of the International Astronomical Union, 9, 429,
  \dodoi{10.1017/S1743921314001070}

\bibitem[{Grimme {et~al.}(2010)Grimme, Antony, Ehrlich, \& Krieg}]{Grimme2010}
Grimme, S., Antony, J., Ehrlich, S., \& Krieg, H. 2010, Journal of Chemical
  Physics, 132, 154104, \dodoi{10.1063/1.3382344}

\bibitem[{Grimme {et~al.}(2011)Grimme, Ehrlich, \&
  Goerigk}]{Grimme_Ehrlich_Goerigk_2011}
Grimme, S., Ehrlich, S., \& Goerigk, L. 2011, Journal of Computational
  Chemistry, 32, 1456–1465, \dodoi{10.1002/jcc.21759}

\bibitem[{Hasegawa \& Herbst(1993)}]{hasegawa_three-phase_1993}
Hasegawa, T.~I., \& Herbst, E. 1993, Monthly Notices of the Royal Astronomical
  Society, 263, 589, \dodoi{10.1093/mnras/263.3.589}

\bibitem[{Herbst \& van Dishoeck(2009)}]{Herbst2009}
Herbst, E., \& van Dishoeck, E.~F. 2009, Annual Review of Astronomy and
  Astrophysics, 47, 427, \dodoi{10.1146/annurev-astro-082708-101654}

\bibitem[{Hollis {et~al.}(2000)Hollis, Lovas, \& Jewell}]{hollis00}
Hollis, J.~M., Lovas, F.~J., \& Jewell, P.~R. 2000, The Astrophysical Journal,
  540, L107

\bibitem[{Hollis {et~al.}(2002)Hollis, Lovas, Jewell, \&
  Coudert}]{hollis_interstellar_2002}
Hollis, J.~M., Lovas, F.~J., Jewell, P.~R., \& Coudert, L.~H. 2002, The
  Astrophysical Journal, 571, L59, \dodoi{10.1086/341148}

\bibitem[{Humphrey {et~al.}(1996)Humphrey, Dalke, \& Schulten}]{HUMP96}
Humphrey, W., Dalke, A., \& Schulten, K. 1996, Journal of Molecular Graphics,
  14, 33

\bibitem[{Ioppolo {et~al.}(2021)Ioppolo, Fedoseev, Chuang, Cuppen, Clements,
  Jin, Garrod, Qasim, Kofman, van Dishoeck, \& Linnartz}]{Ioppolo2020}
Ioppolo, S., Fedoseev, G., Chuang, K.~J., {et~al.} 2021, Nature Astronomy, 5,
  197, \dodoi{10.1038/s41550-020-01249-0}

\bibitem[{Ishibashi {et~al.}(2021)Ishibashi, Hidaka, Oba, Kouchi, \&
  Watanabe}]{Ishibashi2021}
Ishibashi, A., Hidaka, H., Oba, Y., Kouchi, A., \& Watanabe, N. 2021, The
  Astrophysical Journal Letters, 921, L13, \dodoi{10.3847/2041-8213/ac3005}

\bibitem[{Jiménez-Serra {et~al.}(2022)Jiménez-Serra, Rodríguez-Almeida,
  Martín-Pintado, Rivilla, {Melosso, Mattia}, {Zeng, Shaoshan}, {Colzi,
  Laura}, {Kawashima, Yoshiyuki}, {Hirota, Eizi}, {Puzzarini, Cristina},
  {Tercero, Belén}, {de Vicente, Pablo}, {Rico-Villas, Fernando},
  {Requena-Torres, Miguel A.}, \& {Martín, Sergio}}]{npropanol}
Jiménez-Serra, I., Rodríguez-Almeida, L., Martín-Pintado, J., {et~al.} 2022,
  A\&A, 663, A181, \dodoi{10.1051/0004-6361/202142699}

\bibitem[{Jin \& Garrod(2020)}]{Jin2020}
Jin, M., \& Garrod, R.~T. 2020, The Astrophysical Journal Supplement Series,
  249, 26, \dodoi{10.3847/1538-4365/ab9ec8}

\bibitem[{Jørgensen {et~al.}(2012)Jørgensen, Favre, Bisschop, Bourke, van
  Dishoeck, \& Schmalzl}]{Jorgensen_2012}
Jørgensen, J.~K., Favre, C., Bisschop, S.~E., {et~al.} 2012, The Astrophysical
  Journal, 757, L4, \dodoi{10.1088/2041-8205/757/1/l4}

\bibitem[{Kleimeier {et~al.}(2021)Kleimeier, Eckhardt, \&
  Kaiser}]{kleimeier_identification_2021}
Kleimeier, N.~F., Eckhardt, A.~K., \& Kaiser, R.~I. 2021, Journal of the
  American Chemical Society, 143, 14009, \dodoi{10.1021/jacs.1c07978}

\bibitem[{Kozuch \& Martin(2011)}]{Kozuch_Martin_2011}
Kozuch, S., \& Martin, J.~M. 2011, Physical Chemistry Chemical Physics, 13,
  20104, \dodoi{10.1039/c1cp22592h}

\bibitem[{Kruse \& Grimme(2012)}]{Kruse_Grimme_2012}
Kruse, H., \& Grimme, S. 2012, The Journal of Chemical Physics, 136,
  \dodoi{10.1063/1.3700154}

\bibitem[{Landau \& Lifshitz(1976)}]{Landau1976Mechanics}
Landau, L.~D., \& Lifshitz, E.~M. 1976, Mechanics, Third Edition: Volume 1
  (Course of Theoretical Physics), 3rd edn. (Butterworth-Heinemann).
\newblock \url{http://www.worldcat.org/isbn/0750628960}

\bibitem[{Lykke {et~al.}(2017)Lykke, Coutens, Jørgensen, Van Der~Wiel, Garrod,
  Müller, Bjerkeli, Bourke, Calcutt, Drozdovskaya, Favre, Fayolle, Jacobsen,
  Öberg, Persson, Van~Dishoeck, \& Wampfler}]{lykke_alma-pils_2017}
Lykke, J.~M., Coutens, A., Jørgensen, J.~K., {et~al.} 2017, Astronomy \&
  Astrophysics, 597, A53, \dodoi{10.1051/0004-6361/201629180}

\bibitem[{Mardirossian \& Head-Gordon(2016)}]{Mardirossian2016}
Mardirossian, N., \& Head-Gordon, M. 2016, J. Chem. Phys., 144, 214110,
  \dodoi{10.1063/1.4952647}

\bibitem[{McGuire(2018)}]{mcguire_2018_2018}
McGuire, B.~A. 2018, The Astrophysical Journal Supplement Series, 239, 17,
  \dodoi{10.3847/1538-4365/aae5d2}

\bibitem[{McGuire(2022)}]{McGuire2022}
---. 2022, The Astrophysical Journal Supplement Series, 259, 30,
  \dodoi{10.3847/1538-4365/ac2a48}

\bibitem[{McGuire {et~al.}(2018)McGuire, Burkhardt, Kalenskii, Shingledecker,
  Remijan, Herbst, \& McCarthy}]{c6h5cn}
McGuire, B.~A., Burkhardt, A.~M., Kalenskii, S., {et~al.} 2018, Science, 359,
  202–205, \dodoi{10.1126/science.aao4890}

\bibitem[{McGuire {et~al.}(2021)McGuire, Loomis, Burkhardt, Lee, Shingledecker,
  Charnley, Cooke, Cordiner, Herbst, Kalenskii, Siebert, Willis, Xue, Remijan,
  \& McCarthy}]{pah}
McGuire, B.~A., Loomis, R.~A., Burkhardt, A.~M., {et~al.} 2021, Science, 371,
  1265–1269, \dodoi{10.1126/science.abb7535}

\bibitem[{Melosso {et~al.}(2022)Melosso, Bizzocchi, Gazzeh, Tonolo, Guillemin,
  Alessandrini, Rivilla, Dore, Barone, \& Puzzarini}]{melosso}
Melosso, M., Bizzocchi, L., Gazzeh, H., {et~al.} 2022, Chem. Commun., 58,
  2750–2753, \dodoi{10.1039/D1CC06919E}

\bibitem[{Minissale {et~al.}(2016)Minissale, Congiu, \&
  Dulieu}]{Minissale2016a}
Minissale, M., Congiu, E., \& Dulieu, F. 2016, Astronomy and Astrophysics, 585,
  \dodoi{10.1051/0004-6361/201526702}

\bibitem[{Minissale {et~al.}(2013)Minissale, Congiu, Baouche, Chaabouni,
  Moudens, Dulieu, Accolla, Cazaux, Manicó, \& Pirronello}]{Minissale2013}
Minissale, M., Congiu, E., Baouche, S., {et~al.} 2013, Physical Review Letters,
  111, 053201, \dodoi{10.1103/PhysRevLett.111.053201}

\bibitem[{Miyazaki {et~al.}(2022)Miyazaki, Tsuge, Hidaka, Nakai, \&
  Watanabe}]{miyazaki_direct_2022}
Miyazaki, A., Tsuge, M., Hidaka, H., Nakai, Y., \& Watanabe, N. 2022, The
  Astrophysical Journal Letters, 940, L2, \dodoi{10.3847/2041-8213/ac9d30}

\bibitem[{Molpeceres {et~al.}(2021{\natexlab{a}})Molpeceres, García de~la
  Concepción, \& Jiménez-Serra}]{Molpeceres2021b}
Molpeceres, G., García de~la Concepción, J., \& Jiménez-Serra, I.
  2021{\natexlab{a}}, The Astrophysical Journal, 923, 159,
  \dodoi{10.3847/1538-4357/ac2ebc}

\bibitem[{Molpeceres {et~al.}(2021{\natexlab{b}})Molpeceres, Kästner,
  Fedoseev, Qasim, Schömig, Linnartz, \& Lamberts}]{Molpeceres2021carbon}
Molpeceres, G., Kästner, J., Fedoseev, G., {et~al.} 2021{\natexlab{b}},
  Journal of Physical Chemistry Letters, 12, 10854,
  \dodoi{10.1021/acs.jpclett.1c02760}

\bibitem[{Molpeceres \& Rivilla(2022)}]{molpeceres_radical_2022}
Molpeceres, G., \& Rivilla, V.~M. 2022, Astronomy \& Astrophysics, 665, A27,
  \dodoi{10.1051/0004-6361/202243892}

\bibitem[{Molpeceres {et~al.}(2023{\natexlab{a}})Molpeceres, Rivilla, Furuya,
  Kästner, Maté, \& Aikawa}]{molpeceres_processing_2023}
Molpeceres, G., Rivilla, V.~M., Furuya, K., {et~al.} 2023{\natexlab{a}},
  Monthly Notices of the Royal Astronomical Society, 521, 6061,
  \dodoi{10.1093/mnras/stad892}

\bibitem[{Molpeceres {et~al.}(2023{\natexlab{b}})Molpeceres, Zaverkin, Furuya,
  Aikawa, \& Kästner}]{molpeceres_reaction_2023}
Molpeceres, G., Zaverkin, V., Furuya, K., Aikawa, Y., \& Kästner, J.
  2023{\natexlab{b}}, Astronomy \& Astrophysics, 673, A51,
  \dodoi{10.1051/0004-6361/202346073}

\bibitem[{Molpeceres {et~al.}(2022)Molpeceres, Jiménez-Serra, Oba, Nguyen,
  Watanabe, de~la Concepción, Maté, Oliveira, \&
  Kästner}]{molpeceres_hydrogen_2022}
Molpeceres, G., Jiménez-Serra, I., Oba, Y., {et~al.} 2022, Astronomy \&
  Astrophysics, 663, A41, \dodoi{10.1051/0004-6361/202243366}

\bibitem[{Mullikin {et~al.}(2021)Mullikin, Anderson, O’Hern, Farrah,
  Arumainayagam, Van~Dishoeck, Gerakines, Vasyunin, Majumdar, Caselli, \&
  Shingledecker}]{mullikin_new_2021}
Mullikin, E., Anderson, H., O’Hern, N., {et~al.} 2021, The Astrophysical
  Journal, 910, 72, \dodoi{10.3847/1538-4357/abd778}

\bibitem[{Neese {et~al.}(2020)Neese, Wennmohs, Becker, \&
  Riplinger}]{Neese2020}
Neese, F., Wennmohs, F., Becker, U., \& Riplinger, C. 2020, Journal of Chemical
  Physics, 152, 224108, \dodoi{10.1063/5.0004608}

\bibitem[{Neese {et~al.}(2009)Neese, Wennmohs, Hansen, \& Becker}]{NEESE200998}
Neese, F., Wennmohs, F., Hansen, A., \& Becker, U. 2009, Chemical Physics, 356,
  98–109, \dodoi{10.1016/j.chemphys.2008.10.036}

\bibitem[{Nonne {et~al.}(2024)Nonne, Melosso, Tonolo, Bizzocchi, Alessandrini,
  Guillemin, Dore, \& Puzzarini}]{nonne_tracing_2024}
Nonne, M., Melosso, M., Tonolo, F., {et~al.} 2024, The Journal of Physical
  Chemistry A, 128, 4850, \dodoi{10.1021/acs.jpca.4c02533}

\bibitem[{Pantaleone {et~al.}(2021)Pantaleone, Enrique-Romero, Ceccarelli,
  Ferrero, Balucani, Rimola, \& Ugliengo}]{Pantaleone2021}
Pantaleone, S., Enrique-Romero, J., Ceccarelli, C., {et~al.} 2021, The
  Astrophysical Journal, 917, 49, \dodoi{10.3847/1538-4357/ac0142}

\bibitem[{Pantaleone {et~al.}(2020)Pantaleone, Enrique-Romero, Ceccarelli,
  Ugliengo, Balucani, \& Rimola}]{Pantaleone_2020}
---. 2020, The Astrophysical Journal, 897, 56, \dodoi{10.3847/1538-4357/ab8a4b}

\bibitem[{Papajak {et~al.}(2011)Papajak, Zheng, Xu, Leverentz, \&
  Truhlar}]{papajak11}
Papajak, E., Zheng, J., Xu, X., Leverentz, H.~R., \& Truhlar, D.~G. 2011,
  Journal of Chemical Theory and Computation, 7, 3027–3034,
  \dodoi{10.1021/ct200106a}

\bibitem[{Perrero {et~al.}(2022)Perrero, Enrique-Romero, Martínez-Bachs,
  Ceccarelli, Balucani, Ugliengo, \& Rimola}]{Perrero2022}
Perrero, J., Enrique-Romero, J., Martínez-Bachs, B., {et~al.} 2022, ACS Earth
  and Space Chemistry, 6, 496, \dodoi{10.1021/acsearthspacechem.1c00369}

\bibitem[{Rimola {et~al.}(2018)Rimola, Skouteris, Balucani, Ceccarelli,
  Enrique-Romero, Taquet, \& Ugliengo}]{Rimola2018}
Rimola, A., Skouteris, D., Balucani, N., {et~al.} 2018, ACS Earth and Space
  Chemistry, 2, 720, \dodoi{10.1021/acsearthspacechem.7b00156}

\bibitem[{Rivilla {et~al.}(2020)Rivilla, Martín-Pintado, Jiménez-Serra,
  Martín, Rodríguez-Almeida, Requena-Torres, Rico-Villas, Zeng, \&
  Briones}]{Rivilla2020}
Rivilla, V.~M., Martín-Pintado, J., Jiménez-Serra, I., {et~al.} 2020, The
  Astrophysical Journal, 899, L28, \dodoi{10.3847/2041-8213/abac55}

\bibitem[{Rivilla {et~al.}(2021)Rivilla, Jiménez-Serra, Martín-Pintado,
  Briones, Rodríguez-Almeida, Rico-Villas, Tercero, Zeng, Colzi, De~Vicente,
  Martín, \& Requena-Torres}]{Rivilla2021}
Rivilla, V.~M., Jiménez-Serra, I., Martín-Pintado, J., {et~al.} 2021,
  Proceedings of the National Academy of Sciences of the United States of
  America, 118, e2101314118, \dodoi{10.1073/pnas.2101314118}

\bibitem[{Rivilla {et~al.}(2022)Rivilla, Colzi, Jiménez-Serra,
  Martín-Pintado, Megías, Melosso, Bizzocchi, López-Gallifa,
  Martínez-Henares, Massalkhi, Tercero, de~Vicente, Guillemin, García de~la
  Concepción, Rico-Villas, Zeng, Martín, Requena-Torres, Tonolo,
  Alessandrini, Dore, Barone, \& Puzzarini}]{rivilla_precursors_2022}
Rivilla, V.~M., Colzi, L., Jiménez-Serra, I., {et~al.} 2022, The Astrophysical
  Journal Letters, 929, L11, \dodoi{10.3847/2041-8213/ac6186}

\bibitem[{Rocha {et~al.}(2024)Rocha, Van~Dishoeck, Ressler, Van~Gelder,
  Slavicinska, Brunken, Linnartz, Ray, Beuther, Caratti O~Garatti, Geers,
  Kavanagh, Klaassen, Justtanont, Chen, Francis, Gieser, Perotti, Tychoniec,
  Barsony, Majumdar, Le~Gouellec, Chu, Lew, Henning, \&
  Wright}]{rocha_jwst_2024}
Rocha, W. R.~M., Van~Dishoeck, E.~F., Ressler, M.~E., {et~al.} 2024, Astronomy
  \& Astrophysics, 683, A124, \dodoi{10.1051/0004-6361/202348427}

\bibitem[{Rodler(1985)}]{RODLER198523}
Rodler, M. 1985, Journal of Molecular Spectroscopy, 114, 23–30,
  \dodoi{10.1016/0022-2852(85)90332-7}

\bibitem[{Rodríguez-Almeida {et~al.}(2021)Rodríguez-Almeida, Jiménez-Serra,
  Rivilla, Martín-Pintado, Zeng, Tercero, de~Vicente, Colzi, Rico-Villas,
  Martín, \& Requena-Torres}]{Rodriguez-Almeida2021}
Rodríguez-Almeida, L.~F., Jiménez-Serra, I., Rivilla, V.~M., {et~al.} 2021,
  The Astrophysical Journal Letters, 912, L11, \dodoi{10.3847/2041-8213/abf7cb}

\bibitem[{Ruaud {et~al.}(2016)Ruaud, Wakelam, \& Hersant}]{ruaud_gas_2016}
Ruaud, M., Wakelam, V., \& Hersant, F. 2016, Monthly Notices of the Royal
  Astronomical Society, 459, 3756, \dodoi{10.1093/mnras/stw887}

\bibitem[{Saito(1976)}]{SAITO1976399}
Saito, S. 1976, Chemical Physics Letters, 42, 399–402,
  \dodoi{10.1016/0009-2614(76)80638-0}

\bibitem[{Sanz-Novo {et~al.}(2023)Sanz-Novo, Rivilla, Jiménez-Serra,
  Martín-Pintado, Colzi, Zeng, Megías, López-Gallifa, Martínez-Henares,
  Massalkhi, Tercero, De~Vicente, Martín, Andrés, \&
  Requena-Torres}]{sanz-novo_discovery_2023}
Sanz-Novo, M., Rivilla, V.~M., Jiménez-Serra, I., {et~al.} 2023, The
  Astrophysical Journal, 954, 3, \dodoi{10.3847/1538-4357/ace523}

\bibitem[{Sanz-Novo {et~al.}(2024)Sanz-Novo, Rivilla, Müller, Jiménez-Serra,
  Martín-Pintado, Colzi, Zeng, Megías, López-Gallifa, Martínez-Henares,
  Tercero, de~Vicente, Andrés, Martín, \&
  Requena-Torres}]{sanz-novo_discovery_2024}
Sanz-Novo, M., Rivilla, V.~M., Müller, H. S.~P., {et~al.} 2024,
  \dodoi{10.48550/ARXIV.2404.01044}

\bibitem[{Shapiro(1988)}]{shapiro_prebiotic_1988}
Shapiro, R. 1988, Origins of life and evolution of the biosphere, 18, 71,
  \dodoi{10.1007/BF01808782}

\bibitem[{Shingledecker {et~al.}(2020)Shingledecker, Molpeceres, Rivilla,
  Majumdar, \& Kästner}]{Shingledecker2020}
Shingledecker, C., Molpeceres, G., Rivilla, V., Majumdar, L., \& Kästner, J.
  2020, The Astrophysical Journal, 897, 158, \dodoi{10.3847/1538-4357/ab94b5}

\bibitem[{Shingledecker {et~al.}(2018)Shingledecker, Tennis, Gal, \&
  Herbst}]{shingledecker_cosmic-ray-driven_2018}
Shingledecker, C.~N., Tennis, J., Gal, R.~L., \& Herbst, E. 2018, The
  Astrophysical Journal, 861, 20, \dodoi{10.3847/1538-4357/aac5ee}

\bibitem[{Shingledecker {et~al.}(2019)Shingledecker, Vasyunin, Herbst, \&
  Caselli}]{shingledecker_simulating_2019}
Shingledecker, C.~N., Vasyunin, A., Herbst, E., \& Caselli, P. 2019, The
  Astrophysical Journal, 876, 140, \dodoi{10.3847/1538-4357/ab16d5}

\bibitem[{Stone(1998)}]{STON1998}
Stone, J. 1998, Master's thesis, Computer Science Department, University of
  Missouri-Rolla

\bibitem[{Tinacci {et~al.}(2023)Tinacci, Germain, Pantaleone, Ceccarelli,
  Balucani, \& Ugliengo}]{tinacci_theoretical_2023}
Tinacci, L., Germain, A., Pantaleone, S., {et~al.} 2023, The Astrophysical
  Journal, 951, 32, \dodoi{10.3847/1538-4357/accae8}

\bibitem[{Vydrov \& {Van Voorhis}(2010)}]{Vydrov2010}
Vydrov, O.~A., \& {Van Voorhis}, T. 2010, Journal of Chemical Physics, 133,
  244103, \dodoi{10.1063/1.3521275}

\bibitem[{Wakelam {et~al.}(2017)Wakelam, Loison, Mereau, \&
  Ruaud}]{Wakelam2017}
Wakelam, V., Loison, J.~C., Mereau, R., \& Ruaud, M. 2017, Molecular
  Astrophysics, 6, 22, \dodoi{10.1016/j.molap.2017.01.002}

\bibitem[{Weigend(2006)}]{Weigend_2006}
Weigend, F. 2006, Physical Chemistry Chemical Physics, 8, 1057,
  \dodoi{10.1039/b515623h}

\bibitem[{Weigend \& Ahlrichs(2005)}]{Weigend2005}
Weigend, F., \& Ahlrichs, R. 2005, Physical Chemistry Chemical Physics, 7,
  3297, \dodoi{10.1039/b508541a}

\bibitem[{Whitten(1973)}]{Whitten_1973}
Whitten, J.~L. 1973, The Journal of Chemical Physics, 58, 4496–4501,
  \dodoi{10.1063/1.1679012}

\bibitem[{Zaverkin {et~al.}(2022)Zaverkin, Molpeceres, \&
  Kästner}]{Zaverkin2021}
Zaverkin, V., Molpeceres, G., \& Kästner, J. 2022, Monthly Notices of the
  Royal Astronomical Society, 510, 3063, \dodoi{10.1093/mnras/stab3631}

\bibitem[{Zeng {et~al.}(2019)Zeng, Quénard, Jiménez-Serra, Martín-Pintado,
  Rivilla, Testi, \& Martín-Doménech}]{zeng19}
Zeng, S., Quénard, D., Jiménez-Serra, I., {et~al.} 2019, Monthly Notices of
  the Royal Astronomical Society: Letters, 484, L43,
  \dodoi{10.1093/mnrasl/slz002}

\bibitem[{Zeng {et~al.}(2020)Zeng, Zhang, Jiménez-Serra, Tercero, Lu,
  Martín-Pintado, de Vicente, Rivilla, \& Li}]{zeng_cloudcloud_2020}
Zeng, S., Zhang, Q., Jiménez-Serra, I., {et~al.} 2020, Monthly Notices of the
  Royal Astronomical Society, 497, 4896, \dodoi{10.1093/mnras/staa2187}

\bibitem[{Zeng {et~al.}(2021)Zeng, Jiménez-Serra, Rivilla, Martín-Pintado,
  Rodríguez-Almeida, Tercero, de~Vicente, Rico-Villas, Colzi, Martín, \&
  Requena-Torres}]{Zeng_2021}
Zeng, S., Jiménez-Serra, I., Rivilla, V.~M., {et~al.} 2021, The Astrophysical
  Journal Letters, 920, L27, \dodoi{10.3847/2041-8213/ac2c7e}

\bibitem[{Zhao {et~al.}(2004)Zhao, Lynch, \& Truhlar}]{Zhao_Lynch_Truhlar_2004}
Zhao, Y., Lynch, B.~J., \& Truhlar, D.~G. 2004, The Journal of Physical
  Chemistry A, 108, 2715–2719, \dodoi{10.1021/jp049908s}

\bibitem[{Zhao \& Truhlar(2004)}]{Zhao_Truhlar_2004MPWB1K}
Zhao, Y., \& Truhlar, D.~G. 2004, The Journal of Physical Chemistry A, 108,
  6908–6918, \dodoi{10.1021/jp048147q}

\bibitem[{Zhao \& Truhlar(2005)}]{Zhao2005}
---. 2005, J. Phys. Chem. A, 109, 5656, \dodoi{10.1021/jp050536c}

\bibitem[{Zhao \& Truhlar(2008{\natexlab{a}})}]{Zhao2007}
---. 2008{\natexlab{a}}, Theor. Chem. Acc., 120, 215,
  \dodoi{10.1007/s00214-007-0310-x}

\bibitem[{Zhao \& Truhlar(2008{\natexlab{b}})}]{Zhao_Truhlar_2008}
---. 2008{\natexlab{b}}, Journal of Chemical Theory and Computation, 4,
  1849–1868, \dodoi{10.1021/ct800246v}

\bibitem[{Álvarez Barcia {et~al.}(2018)Álvarez Barcia, Russ, Kästner, \&
  Lamberts}]{Alvarez-Barcia2018}
Álvarez Barcia, S., Russ, P., Kästner, J., \& Lamberts, T. 2018, Monthly
  Notices of the Royal Astronomical Society, 479, 2007,
  \dodoi{10.1093/MNRAS/STY1478}

\end{thebibliography}
\bibliographystyle{aasjournal}



\end{document}